\pdfoutput=1
\documentclass[11pt,twoside,a4paper,cmspaper,final,collab]{cms-tdr}
\def\svnVersion{c4fe84c}\def\svnDate{2023/09/26}\def\cmsCernNoTag{CERN-EP-2022-169}\def\cmsCernDate{\today}\def\cmsMessage{Published in the Journal of High Energy Physics as \href{http://dx.doi.org/10.1007/JHEP09(2023)149}{\doi{10.1007/JHEP09(2023)149}.}}
\begin{document}\cmsNoteHeader{SUS-21-007}

\newcommand{\cmsTable}[1]{\resizebox{\textwidth}{!}{#1}}
\newlength{\cmsTabSkip}\setlength\cmsTabSkip{0.4ex}
\newcommand{\pp}{\ensuremath{\Pp\Pp}\xspace}
\renewcommand{\MET}{\ensuremath{\ptmiss}\xspace}
\newcommand{\LT}{\ensuremath{L_{\mathrm{T}}}\xspace}
\newcommand{\DF}{\ensuremath{\Delta\phi}\xspace}
\newcommand{\DFz}{\ensuremath{\DF_0}\xspace}
\newcommand{\Rcs}{\ensuremath{R^{\mathrm{CS}}}\xspace}
\newcommand{\Lp}{\ensuremath{L_{\mathrm{P}}}\xspace}
\newcommand{\nbtag}{\ensuremath{n_{\PQb}}\xspace}
\newcommand{\ntop}{\ensuremath{n_{\PQt}}\xspace}
\newcommand{\nwtag}{\ensuremath{n_{\PW}}\xspace}
\newcommand{\njet}{\ensuremath{n_{\text{jet}}}\xspace}
\newcommand{\njetav}{\ensuremath{\langle\njet\rangle}\xspace}
\newcommand{\kappab}{\ensuremath{\kappa_{\PQb}}\xspace}
\newcommand{\kappaW}{\ensuremath{\kappa_{\PW}}\xspace}
\newcommand{\kappatt}{\ensuremath{\kappa_{\ttbar}}\xspace}
\newcommand{\kappaew}{\ensuremath{\kappa_{\text{EW}}}\xspace}
\newcommand{\wDL}{\ensuremath{w_{\text{DL}}}\xspace}

\newcommand{\mGlu}{\ensuremath{m_{\PSg}}\xspace}
\newcommand{\mLSP}{\ensuremath{m_{\PSGczDo}}\xspace}

\newcommand{\Fratio}{\ensuremath{F_{\mathrm{SA}}}\xspace}
\newcommand{\Nsel}{\ensuremath{N_{\text{QCD selected}}}\xspace}
\newcommand{\Nanti}{\ensuremath{N_{\text{QCD anti-selected}}}\xspace}

\newcommand{\multib}{\ensuremath{\text{multi-\PQb}}\xspace}
\newcommand{\zerob}{\ensuremath{\text{zero-\PQb}}\xspace}
\newcommand{\Wjets}{\ensuremath{\PW\text{+jets}}\xspace}
\newcommand{\WToLNuPlusJets}{\ensuremath{\PW\to \ell \nu\text{+jets}}\xspace}
\newcommand{\Wother}{\ensuremath{\PW\text{+other}}\xspace}
\newcommand{\ttV}{\ensuremath{\ttbar\PV}\xspace}

\newcommand{\Imini}{\ensuremath{I_{\text{mini}}}\xspace}
\newcommand{\Irel}{\ensuremath{I_{\text{rel}}}\xspace}
\newcommand{\abseta}{\ensuremath{\abs{\eta}}\xspace}

\newcommand{\WW}{\ensuremath{\PW\PW}\xspace}
\newcommand{\WZ}{\ensuremath{\PW\PZ}\xspace}
\newcommand{\ptvecW}{\ensuremath{\ptvec^{\PW}}\xspace}
\newcommand{\ptvecL}{\ensuremath{\ptvec^{\Pell}}\xspace}
\newcommand{\thetast}{\ensuremath{\theta^\ast}\xspace}

\newcommand{\NMBSRpred}{\ensuremath{N^{\text{MB,SR}}_{\text{Pred}}}\xspace}
\newcommand{\NMBSRpredew}{\ensuremath{N^{\text{MB,SR}}_{\text{Pred,EW}}}\xspace}
\newcommand{\NMBSRpredqcd}{\ensuremath{N^{\text{MB,SR}}_{\text{Pred,QCD}}}\xspace}
\newcommand{\NMBSRpredw}{\ensuremath{N^{\text{MB,SR}}_{\text{Pred,\PW}}}\xspace}
\newcommand{\NMBSRpredtt}{\ensuremath{N^{\text{MB,SR}}_{\text{Pred,\ttbar}}}\xspace}
\newcommand{\NMBSRmcother}{\ensuremath{N^{\text{MB,SR}}_{\text{MC,other}}}\xspace}
\newcommand{\NSBSRmc}[1][]{\ensuremath{N^\text{SB#1,SR}_{\text{MC}}}\xspace}
\newcommand{\NSBSRdata}[1][]{\ensuremath{N^\text{SB#1,SR}_{\text{data}}}\xspace}
\newcommand{\NSBSRpredqcd}[1][]{\ensuremath{N^\text{SB#1,SR}_{\text{Pred,QCD}}}\xspace}
\newcommand{\NSBCRmc}[1][]{\ensuremath{N^\text{SB#1,CR}_{\text{MC}}}\xspace}
\newcommand{\NSBCRdata}[1][]{\ensuremath{N^\text{SB#1,CR}_{\text{data}}}\xspace}
\newcommand{\NSBCRpredqcd}[1][]{\ensuremath{N^\text{SB#1,CR}_{\text{Pred,QCD}}}\xspace}
\newcommand{\NMBCR}{\ensuremath{N^{\text{MB,CR}}}\xspace}
\newcommand{\NMBCRdata}{\ensuremath{\NMBCR_{\text{data}}}\xspace}
\newcommand{\RCSdata}{\ensuremath{\Rcs_{\text{data}}}\xspace}
\newcommand{\RCSdatacorr}{\ensuremath{R^{\text{CS,corr}}_{\text{data}}}\xspace}
\newcommand{\RCSmc}{\ensuremath{\Rcs_{\text{MC}}}\xspace}
\newcommand{\RCSmcew}{\ensuremath{\Rcs_{\text{MC,EW}}}\xspace}
\newcommand{\RCStt}{\ensuremath{\Rcs_{\ttbar}}\xspace}
\newcommand{\RCSw}{\ensuremath{\Rcs_{\PW}}\xspace}
\newcommand{\ftt}{\ensuremath{f_{\ttbar}}\xspace}
\newcommand{\fttSBCR}{\ensuremath{\ftt^{\text{SB,CR}}}\xspace}

\newcommand{\threefourjets}{\ensuremath{\njet\in[3,4]}\xspace}
\newcommand{\fourfivejets}{\ensuremath{\njet\in[4,5]}\xspace}
\newcommand{\jetsasinmb}{\ensuremath{\njet\text{ as in MB}}\xspace}
\newcommand{\onebtag}{\ensuremath{\nbtag\geq1}\xspace}
\newcommand{\zerobtag}{\ensuremath{\nbtag=0}\xspace}
\newcommand{\onemuon}{\ensuremath{n_{\PGm}=1}\xspace}
\newcommand{\medminmax}{\ensuremath{\text{median [min, max] [\%]}}\xspace}

\cmsNoteHeader{SUS-21-007}

\title{Search for supersymmetry in final states with a single electron or muon using angular correlations and heavy-object identification in proton-proton collisions at \texorpdfstring{$\sqrt{s}=13\TeV$}{sqrt(s)=13 TeV}}

\date{\today}

\abstract{A search for supersymmetry is presented in events with a single charged lepton, electron or muon, and multiple hadronic jets. The data correspond to an integrated luminosity of 138\fbinv of proton-proton collisions at a center-of-mass energy of 13\TeV, recorded by the CMS experiment at the CERN LHC. The search targets gluino pair production, where the gluinos decay into final states with the lightest supersymmetric particle (LSP) and either a top quark-antiquark (\ttbar) pair, or a light-flavor quark-antiquark (\qqbar) pair and a virtual or on-shell \PW boson. The main backgrounds, \ttbar pair and \Wjets production, are suppressed by requirements on the azimuthal angle between the momenta of the lepton and of its reconstructed parent \PW boson candidate, and by top quark and \PW boson identification based on a machine-learning technique. The number of observed events is consistent with the expectations from standard model processes. Limits are evaluated on supersymmetric particle masses in the context of two simplified models of gluino pair production. Exclusions for gluino masses reach up to 2120 (2050)\GeV at 95\% confidence level for a model with gluino decay to a \ttbar pair (a \qqbar pair and a \PW boson) and the LSP. For the same models, limits on the mass of the LSP reach up to 1250 (1070)\GeV.}

\hypersetup{%
pdfauthor={CMS Collaboration},%
pdftitle={Search for supersymmetry in final states with a single electron or muon using angular correlations and heavy-object identification in proton-proton collisions at sqrt(s)=13 TeV},%
pdfsubject={CMS},%
pdfkeywords={CMS, supersymmetry}}

\maketitle

\section{Introduction}

Supersymmetry (SUSY)~\cite{Wess:1973kz, Fayet:1976cr, Barbieri:1982eh, Nilles:1983ge, Haber:1984rc, Martin:1997ns} is an appealing extension of the standard model (SM) of particle physics, which is able to address several shortcomings of the SM by introducing a new symmetry that predicts superpartners to the existing bosons and fermions.
The supersymmetric partner of the gluon is the gluino (\PSg).
The superpartners of the electroweak gauge bosons and the Higgs bosons mix to form mass eigenstates called neutralinos (\PSGcz) and charginos (\PSGcpm).
In SUSY models that conserve $R$-parity~\cite{Farrar:1978xj}, the SUSY particles have to be produced in pairs and the lightest SUSY particle (LSP) is stable, providing a possible dark matter candidate, which can in certain models explain the dark matter content of the universe.

The search in this paper targets final states containing a single lepton (electron or muon), missing transverse momentum, and large hadronic activity.
The proton-proton (\pp) collision data at $\sqrt{s}=13\TeV$ recorded by the CMS experiment at the CERN LHC during 2016--2018 and corresponding to an integrated luminosity of 138\fbinv are used.
Search regions (SRs) with and without \PQb tagging requirements are defined, so that the search is sensitive to the strong production of superpartners with different decays to lighter states and different mass splittings.
The sensitivity is further enhanced by using a large number of SR bins defined by several variables characterizing the event topology and kinematical properties.
The results are interpreted in terms of two simplified SUSY models~\cite{Arkani-Hamed:2007gys, Alwall:2008ag, Alwall:2008va, LHCNewPhysicsWorkingGroup:2011mji, CMS:2013wdg}.

\begin{figure}[!htb]
\centering
\includegraphics[width=0.4\textwidth]{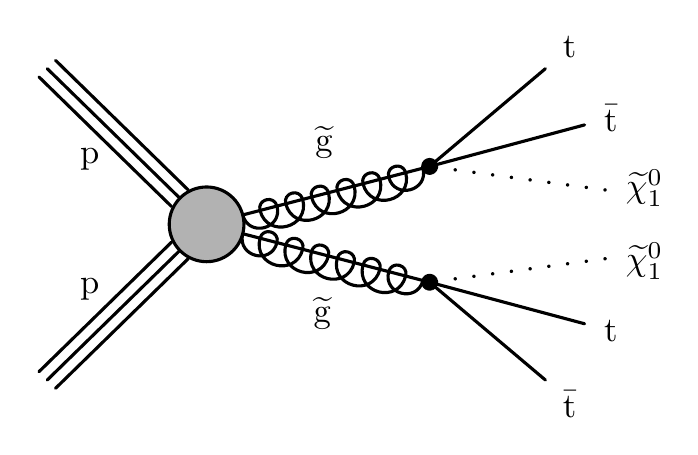}
\hspace{.05\textwidth}
\includegraphics[width=0.4\textwidth]{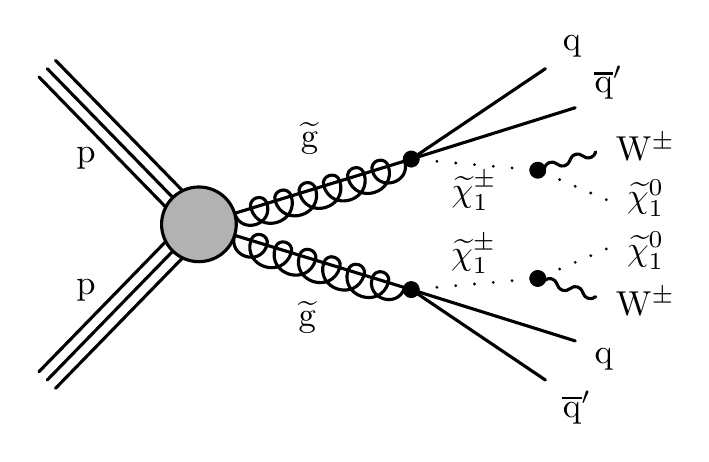}
\caption{Diagrams showing the simplified SUSY models T1tttt (left) and T5qqqqWW (right).}
\label{fig:feynman}
\end{figure}

The diagrams of the specific $R$-parity conserving models of gluino pair production that are used to interpret the results are shown in Fig.~\ref{fig:feynman}.
The results of the search with at least one \PQb-tagged jet, referred to as ``\multib analysis'', are interpreted in terms of the simplified model, labeled as ``T1tttt'' (left), where the gluino always decays to a top quark-antiquark pair (\ttbar) and the lightest neutralino (\PSGczDo), which is the LSP.
The top quarks will decay into a bottom quark (\PQb quark) and a \PW boson, which further decays hadronically or into a lepton and a neutrino.
The observations in the SR bins with no \PQb-tagged jets, referred to as ``\zerob analysis'', are interpreted in the model labeled as ``T5qqqqWW'' (right).
In this model, each gluino decays to a light-flavor quark-antiquark pair of different quark flavors ($\qqbar^\prime$) and the lightest chargino (\PSGcpmDo), which then decays further to a \PSGczDo and a \PW boson, which finally decays hadronically or into a lepton and a neutrino. In T5qqqqWW, the \PW boson can be virtual, depending on the mass difference between the lightest chargino (\PSGcpmDo) and the lightest neutralino (\PSGczDo).
The mass of the \PSGcpmDo is fixed at the value halfway between the masses of the \PSg and the \PSGczDo.

Searches targeting gluino pair production in the single-lepton final state have been performed by both the ATLAS~\cite{ATLAS:2021twp, ATLAS:2017xvf, ATLAS:2022ihe,ATLAS:2016maz, ATLAS:2016gty} and CMS~\cite{CMS:2020cur, CMS:2017qth, CMS:2016muu, CMS:2016krz, CMS:2016muu, CMS:2018rst, CMS:2017umd} Collaborations.
The investigated models have also been tested by ATLAS~\cite{ATLAS:2020syg, ATLAS:2017mjy, ATLAS:2015gky, ATLAS:2020xgt} and CMS~\cite{CMS:2017tec, CMS:2013xio, CMS:2016ohy, CMS:2017abv, CMS:2019ybf, CMS:2019zmd, CMS:2017qxu, CMS:2017okm} in other final states.
The results presented in this paper supersede the CMS search presented in Ref.~\cite{CMS:2017qth}, which follows a similar strategy and uses data recorded in 2016, corresponding to an integrated luminosity of 35.9\fbinv.
Improvements stem not only from the larger analyzed data set, but also from significantly reduced SM background contributions in the SR.
This is achieved by requiring at least one jet or jet cluster to be consistent with a hadronically decaying top quark (\PW boson) in the \multib (\zerob) final states as determined by the multivariate classifiers described in Section~\ref{sec:obj}.

Tabulated results are provided in the HEPData record for this analysis~\cite{hepdata}.

\section{The CMS detector}

{\tolerance=800
The CMS apparatus is a multipurpose, nearly hermetic detector, designed to trigger on and identify electrons, muons, photons, and charged and neutral hadrons~\cite{CMS:2020uim, CMS:2018rym, CMS:2014pgm}.
The detector comprises an all-silicon inner tracker and by the crystal electromagnetic calorimeter (ECAL) and brass-scintillator hadron calorimeters (HCAL), operating inside a 3.8\unit{T} superconducting solenoid, with data from the gas-ionization muon detectors embedded in the flux-return yoke outside the solenoid.
Each of these parts of the detector is composed of a cylindrical barrel section and two endcap sections.
The pseudorapidity ($\eta$) coverage of the barrel and endcap detectors is extended by forward calorimeters that lie very close to the LHC beam line.
Outside the solenoid, the returning magnetic flux is guided through a steel return yoke.
Gas-ionization detectors are sandwiched in between the layers of the return yoke and are used to detect muons.
\par}

The events used in the search were collected using a two-tiered trigger system.
The first level, composed of custom hardware processors, uses information from the calorimeters and muon detectors to select events at a rate of around 100\unit{kHz} within a fixed latency of about 4\mus~\cite{CMS:2020cmk}.
The second level, known as the high-level trigger (HLT), consists of a farm of processors running a version of the full event reconstruction software optimized for fast processing, and reduces the event rate to around 1\unit{kHz} before data storage~\cite{CMS:2016ngn}.
The CMS detector is described in more detail, along with the coordinate system and basic kinematic variables, in Ref.~\cite{CMS:2008xjf}.

\section{Simulation}
\label{sec:simulation}

Simulated background events are used to optimize the event selection and calculate correction factors for the background estimation, which is mainly based on control samples in data.
The SM processes are simulated with different Monte Carlo (MC) event generators:
Events for \ttbar, \Wjets, and Drell--Yan (DY) production, as well as for the background from SM events composed uniquely of jets produced through the strong interaction, referred to as quantum chromodynamics (QCD) multijet events, are simulated using the \MGvATNLO event generator at leading order (LO) (versions 2.2.2 for 2016 and 2.4.2 for 2017 and 2018)~\cite{Alwall:2014hca}.
The \ttbar events are generated with up to three additional partons in the matrix-element calculations, while the \Wjets and DY events are generated with up to four additional partons.
Single top quark events produced through the $s$ channel; events containing a \ttbar pair produced in association with a \PZ boson, a \PW boson, or a photon; and rare events such as those containing multiple electroweak or Higgs bosons (\PW, \PZ, \PGg, and \PH) are generated with \MGvATNLO at next-to-LO (NLO)~\cite{Frederix:2012ps}.
Events containing a single top quark produced through the $t$ channel and $\PQt\PW$ production, as well as \WW and $\ttbar\PH$ events, are calculated at NLO with the \POWHEG v1 (v2)~\cite{Nason:2004rx, Frixione:2007vw, Alioli:2010xd, Alioli:2009je, Re:2010bp, Melia:2011tj, Nason:2013ydw, Hartanto:2015uka} program for 2016 (2017 and 2018).
The $\PZ\PZ$ events are generated at NLO with either \POWHEG or \MGvATNLO, depending on the decay mode, while \WZ production is simulated at LO with \PYTHIA8.226 (8.230)~\cite{Sjostrand:2014zea} for 2016 (2017 and 2018).
The normalization of the simulated background samples is performed using the most accurate cross section calculations available~\cite{Alioli:2009je, Re:2010bp, Alwall:2014hca, Gavin:2012sy, Gavin:2010az, Li:2012wna, Gehrmann:2014fva, Campbell:2011bn, Campbell:2011bn, Campbell:2015qma, Beneke:2011mq, Cacciari:2011hy, Baernreuther:2012ws, Czakon:2012zr, Czakon:2012pz, Czakon:2013goa, Czakon:2011xx}, which typically correspond to NLO or next-to-NLO (NNLO) accuracy.

Simulated signal events are used to optimize the event selection and to estimate the signal acceptance and selection efficiency.
They are generated using \MGvATNLO at LO including up to two additional partons in the matrix-element calculations.
The production cross sections are determined with approximate NNLO plus next-to-next-to-leading logarithmic (NNLL) corrections~\cite{Beenakker:2016lwe, Borschensky:2014cia, Beenakker:1996ch, Kulesza:2008jb, Kulesza:2009kq, Beenakker:2009ha, Beenakker:2011sf, Beenakker:2013mva, Beenakker:2014sma, Beenakker:1997ut, Beenakker:2010nq, Beenakker:2016gmf}.
The signal events are produced on a two-dimensional grid for different gluino and LSP masses.

{\tolerance=800
The parton showering and hadronization for all simulated samples is performed with the \PYTHIA8.226 (8.230) program for 2016 (2017 and 2018).
For samples that are simulated at NLO with \MGvATNLO, the partons from the matrix-element calculations are matched to those from the parton showers using the FxFx~\cite{Frederix:2012ps} scheme, while for samples simulated at LO the MLM scheme~\cite{Mangano:2006rw} is adopted.
The CUETP8M1~\cite{CMS:2015wcf} \PYTHIA8.226 tune is used for both SM and signal samples for the analysis of the 2016 data.
For 2017 and 2018, the CP5 (CP2)~\cite{CMS:2019csb} tunes are used for the SM background (signal) samples.
Simulated background samples generated at LO (NLO) with the CUETP8M1 tune use the NNPDF3.0LO (NNPDF3.0NLO)~\cite{Ball:2015} sets for the parton distribution functions (PDFs), respectively.
For signal samples, the NNPDF3.1LO was used~\cite{Ball:2017nwa}.
The samples using the CP2 or CP5 tune use the NNPDF3.1LO or NNPDF3.1NNLO sets, respectively.
\par}

Simulated SM events are processed through a \GEANTfour-based~\cite{Agostinelli:2002hh} simulation of the CMS detector, while the simulated signal events are processed through the CMS fast simulation program~\cite{Abdullin:2011zz, Giammanco:2014bza} in order to save computing time.
The results of the fast simulation are found to be generally consistent with the \GEANTfour-based simulation.

All simulated events are generated with nominal distributions of additional \pp interactions per bunch crossing and nearby bunch crossings, referred to as pileup.
Any residual difference between the pileup distribution used in the simulation and the one observed in the data is corrected via a weighting procedure applied to the simulated events.

In order to improve the modeling of additional jets originating mainly from initial-state radiation (ISR) in events containing \ttbar, the \MGvATNLO prediction is compared to data in a \ttbar-enriched dileptonic control region, and scale factors (SF) are extracted that are applied to the \ttbar simulation for the year 2016, and to the SUSY signal simulation for the years 2016--2018.
The \ttbar simulation of the years 2017 and 2018 is performed with an updated tune resulting in a good agreement between simulation and data, such that no SF are needed.
The values of the reweighting factors are about 90\% for most events and reach 50\% at the high tails of the distribution in the number of ISR jets.

\section{Object reconstruction}
\label{sec:obj}

The ``particle-flow'' (PF) algorithm~\cite{CMS:2017yfk} aims to reconstruct and identify each particle in an event, with an optimized combination of information from the various elements of the CMS detector, and classifies each either as a photon, electron, muon, charged hadron, or a neutral hadron.
The energy of charged hadrons is determined from a combination of their momentum measured in the tracker and the matching ECAL and HCAL energy deposits, corrected for the response function of the calorimeters to hadronic showers.
The energy of neutral hadrons is obtained from the corresponding corrected ECAL and HCAL energies.
The primary vertex (PV) is taken to be the vertex corresponding to the hardest scattering in the event, evaluated using tracking information alone, as described in Section 9.4.1 of Ref.~\cite{CMS-TDR-15-02}.

The energy and momentum of electrons is determined from a combination of the electron momentum at the PV as determined by the tracker, the energy of the corresponding ECAL cluster, and the energy sum of all bremsstrahlung photons spatially compatible with originating from the electron track~\cite{CMS:2020uim}.
For electrons with transverse momentum $\pt\approx45\GeV$ from $\PZ\to\EE$ decays, the momentum resolution ranges from 1.7 to 4.5\%.
It is generally better in the barrel region than in the endcaps, and also depends on the bremsstrahlung energy emitted by the electron as it traverses the material in front of the ECAL.

The energy of muons is obtained from the curvature of the corresponding track.
Muons are measured in the range $\abseta<2.4$, with detection planes made using three technologies: drift tubes, cathode strip chambers, and resistive plate chambers.
Matching muons to tracks measured in the tracker results in a relative \pt resolution of 1\% in the barrel and 3\% in the endcaps for muons with \pt up to 100\GeV, and of better than 7\% in the barrel for muons with \pt up to 1\TeV ~\cite{CMS:2018rym}.

A relative isolation variable is defined as the \pt sum of all PF objects within a cone around the lepton candidate (excluding the candidate itself), divided by the lepton \pt.
This analysis uses the so-called mini-isolation variable (\Imini), which is an optimized version~\cite{CMS:2016krz} of the originally proposed mini-isolation in Ref.~\cite{Rehermann:2010vq}.
The cone size $R$, referring to the distance in the $\phi$--$\eta$ plane (where $\phi$ is the azimuthal angle), depends on the \pt of the lepton: for $\pt<50\GeV$, $R=0.2$; for $50<\pt<200\GeV$, $R=10\GeV/\pt$; and for $\pt>200\GeV$, $R=0.05$.

Two categories of leptons are defined, denoted by ``veto leptons'' and ``good leptons'', in the range of $\abseta<2.4$ and with a minimum \pt threshold of 10 and 25\GeV, respectively.
Muons that fulfill the ``loose'' working point (WP) of the standard muon identification (ID) criteria~\cite{CMS:2018rym} are defined as ``veto muons'', while muons that fulfill the ``medium'' WP are defined as ``good muons''.
The dedicated ``veto'' WP of the standard electron ID criteria~\cite{CMS:2020uim} is used to define ``veto electrons'', and the ``tight'' WP is used to define ``good electrons''.
We use a common requirement of $\Imini<0.4$ for all veto leptons, whereas for good muons (electrons) $\Imini<0.2$ (0.1) is required.
The use of \Imini enhances the selection efficiency of signal events that contain a large amount of hadronic energy compared to an isolation definition with a fixed cone size.
The efficiency for reconstructing a veto muon exceeds 99\% and is equal to 95\% for a veto electron.
The efficiency to select a good muon (electron) is more than 98 (70)\%.
A conversion veto and the requirement of zero lost hits in the tracker are applied for good electrons to reject converted photons.

Jets are clustered with the anti-\kt algorithm~\cite{Cacciari:2008gp, Cacciari:2011ma} with a distance parameter $R$ of 0.4 (AK4), or, in order to identify large-radius jets, with $R=0.8$ (AK8).
The jet \pt is determined from the vectorial \ptvec sum of all PF objects in the jet, and is found from simulation to be, on average, within 5 to 10\% of the true \pt over the whole \pt spectrum and detector acceptance.
Pileup interactions contribute additional tracks and calorimetric energy depositions, increasing the apparent jet momentum.
To mitigate this effect, different strategies are applied.
For AK4 jets, tracks identified to be originating from pileup vertices are discarded and a correction for remaining contributions is applied~\cite{CMS:2020ebo}.
For AK8 jets, the pileup per particle identification algorithm~\cite{CMS:2020ebo,Bertolini:2014bba} is used, which makes use of local shape information, event pileup properties, and tracking information.
A local shape variable is defined, which distinguishes between collinear and soft diffuse distributions of other particles surrounding the particle under consideration.
The former is attributed to particles originating from the hard scatter and the latter to particles originating from pileup interactions.
Charged particles identified to be originating from pileup vertices are discarded.
For each neutral particle, a local shape variable is computed using the surrounding charged particles compatible with the primary vertex within the tracker acceptance ($\abseta<2.5$), and using both charged and neutral particles in the region outside of the tracker coverage.
The momenta of the neutral particles are then rescaled according to their probability to originate from the primary interaction vertex deduced from the local shape variable, superseding the need for jet-based pileup corrections~\cite{CMS:2020ebo}.
In the following, ``jet'' will refer to AK4 jets, unless specified otherwise.

We apply jet energy corrections, derived from simulation studies, to match the average measured energy of jets to that of particle-level jets, clustered from all stable particles excluding neutrinos.
In situ measurements of the momentum balance in dijet, photon+jet, {\PZ}+jet, and multijet events are used to determine any residual differences between the jet energy scale in data and in simulation, and appropriate corrections are applied in the analysis~\cite{CMS:2016lmd}.
For the fast simulation that is used for the signal, dedicated jet energy corrections are applied.

Additional selection criteria are applied to each jet to remove jets potentially affected by instrumental effects or reconstruction failures~\cite{CMS:2020ebo}.
The jets are selected with $\pt>30\GeV$ and $\abseta<2.4$.
The jets that lie within a cone of $R=0.4$ around any good or veto lepton are removed, to avoid double counting.

To identify jets originating from \PQb quarks we use an inclusive deep neural network based combined secondary vertex tagger is used at the medium WP~\cite{CMS:2017wtu}.
The efficiency to identify \PQb jets varies between 50 and 70\%, depending on the jet \pt, with a misidentification probability of 10--15\% for \PQc jets and 1--5\% for light-flavor quark and gluon jets, also depending on the jet \pt.

AK8 jets are selected with $\pt>200\GeV$ and $\abseta<2.4$.
In order to identify hadronic decays of top quarks and \PW bosons with a large Lorentz boost, we apply a dedicated algorithm that is based on convolutional neural networks, the DeepAK8 algorithm~\cite{CMS:2020poo}.
It is a multiclass classifier for top quark, \PW boson, \PZ boson, Higgs boson, and QCD jets, and takes input from all the PF candidates and secondary vertices associated with the AK8 jet.

Hadronically decaying top quarks with $\pt>400\GeV$ are usually merged into one large-radius jet, and are identified using the DeepAK8 algorithm at an efficiency of $\approx$68\% and a mistagging rate of $\approx$8\%, as computed in a \WToLNuPlusJets sample.
The dominant source of the mistagging rate stems from QCD multijet events.
Top quarks with $\pt<400\GeV$ are usually not boosted enough to be caught in one large-radius jet cone, and therefore are identified by a resolved top quark tagging (\PQt tagging) algorithm, as used in Ref.~\cite{CMS:2017mbm}.
It identifies hadronically decaying top quarks whose decay products form three individual jets.
A boosted decision tree is used to distinguish between trijet combinations whose three jets all match the decay products of a top quark versus those that do not.
It uses high-level information such as the invariant mass of the trijet as well as information from each jet.
The resolved \PQt tagger yields an efficiency of $\approx$42\%, while the mistagging rate is $\approx$4\%.

To avoid double counting, a cross cleaning between resolved and merged \PQt tags is performed by first reconstructing the merged top quarks as identified by the DeepAK8 algorithm.
In the next step, resolved top quark candidates that contain any jet within a cone of radius $R=0.8$ of the merged top quark (\ie, the cone of the AK8 jet classified as a top quark by the DeepAK8) are removed.
In the following, \ntop is used to denote the number of identified \PQt tags.

Hadronically decaying \PW bosons are identified with the DeepAK8 algorithm as well.
For the \zerob analysis, this tagging algorithm utilizes AK8 jets to identify hadronically decaying \PW bosons with $\pt>200\GeV$. The efficiency for \PW boson tagging is $\approx$62\%, while the mistagging rate is $\approx$7\%.
The number of identified \PW bosons is denoted as \nwtag.

All mistagging rates for the heavy object taggers were measured in \Wjets data samples, where the \PW decays leptonically.
The efficiencies of these taggers were measured using MC simulation samples of hadronic \PQt and \PW decays.

The missing transverse momentum vector \ptvecmiss is computed as the negative vector \pt sum of all the PF candidates in an event, and its magnitude is denoted as \ptmiss~\cite{CMS:2019ctu}.
The \ptvecmiss is modified to account for corrections to the energy scale of the reconstructed jets in the event.
Anomalous high-\ptmiss events can occur because of a variety of reconstruction failures, detector malfunctions, or noncollision backgrounds.
Such events are rejected by event filters that are designed to identify more than 85--90\% of the spurious high-\ptmiss events with an error rate of less than 0.1\%.

Two kinematic variables are used to describe the energy scale of an event:
The \LT variable is defined as the scalar sum of the lepton \pt and \MET, reflecting the ``leptonic'' energy scale of the event.
The \HT variable reflects the ``hadronic'' energy scale of the event.
It denotes the scalar \pt sum of all selected jets.

In the second half of the 2018 data-taking period, a detector malfunction prevented the readout from a small fraction of the HCAL.
This is taken into account by reweighting simulated events such that it reflects the overall 2018 efficiency of the HCAL in the relevant region.

\section{Baseline event selection}

Events are selected with a combination of HLT paths, relying on the kinematic variables of reconstructed leptons, \HT, \ptmiss, or combinations thereof, to maximize the trigger efficiency.
The main HLT path requires a loosely isolated lepton with $\pt>15\GeV$ and \HT greater than a threshold equal to 350, 400, and 450\GeV for 2016, 2017, and 2018, respectively.
The additional HLT paths require \ptmiss greater than a threshold equal to 100, 110, and 120\GeV for the three data-taking years; isolated electrons with $\pt>27$ (35)\GeV in 2016 (2017 and 2018); isolated muons with $\pt>24\GeV$; or leptons with no isolation requirement and a higher \pt threshold of 105 or 115\GeV for electrons (depending on the year), or 50\GeV for muons.
The trigger efficiency is measured in control samples recorded either with single-lepton triggers or with triggers based on an \HT requirement.
For the electron channel, it is found to be 98, 93, and 97\% in 2016, 2017, and 2018, respectively, while for the muon channel it is 99\% for all three data-taking years.
The inefficiency is mainly caused by the lepton selection in the trigger.
The uncertainty in the measured trigger efficiencies is about 1\%.

For the baseline event selection, one good lepton with $\pt>25\GeV$ is required, and events with additional veto leptons with $\pt>10\GeV$ are removed.
Events with two genuine leptons, of which one is not identified, constitute one of the main backgrounds in the SR bins.
In order to reduce this background contribution, we remove events with an isolated track that fulfills the following criteria.
Charged particle tracks from the PV with $\pt>5\GeV$ are selected, and an isolation variable \Irel is defined as the \pt sum of all tracks within a cone of $R=0.3$ around the track candidate (excluding the candidate itself), divided by the track \pt.
The isolated tracks considered here come from two different sources, one is from isolated leptons that satisfy looser ID criteria than lepton candidates, and the other from isolated charged hadrons.
Charged hadron (lepton) candidates are required to satisfy $\Irel<0.1$ (0.2).
In case of multiple isolated track candidates in an event, the one with the highest \pt that has the opposite charge with respect to the selected lepton is chosen.
Events with such isolated tracks are rejected if the \mTii variable~\cite{Lester:1999tx}, calculated from the momenta of the isolated track and the selected lepton, is below 60 (80)\GeV for isolated tracks associated with charged hadrons (leptons).

Furthermore, we require $\HT>500\GeV$ and $\LT>250\GeV$. The usage of \LT instead of \MET allows the analysis to be not only sensitive to events with high \MET, but also to signal events with very low \MET but higher lepton \pt.
A minimum number of three jets is required for the baseline selection, and the two highest \pt jets are required to fulfill $\pt>80\GeV$.

Events are selected exclusively for the \multib or the \zerob analysis, depending on the number of \PQb-tagged jets in the event.
Events in the \multib analysis are additionally required to contain at least one \PQt tag.
The baseline event selection is summarized in Table~\ref{tab:basiccuts}.

\begin{table}[!htb]
\centering\renewcommand\arraystretch{1.1}
\topcaption{Baseline event selection.}
\label{tab:basiccuts}
\begin{tabular}{l}\hline
One good lepton with $\pt>25\GeV$\\
No additional veto leptons with $\pt>10\GeV$ \\
No isolated tracks with $\pt>5\GeV$ with $\mTii<60$ (80)\GeV for hadronic (leptonic) tracks \\
$\LT>250\GeV$ \\
$\HT>500\GeV$ \\
Number of AK4 jets $\njet\geq3$ \\
At least 2 jets with $\pt>80\GeV$ \\
\onebtag and $\ntop\geq1$ (\multib analysis) or \zerobtag (\zerob analysis) \\
\hline
\end{tabular}
\end{table}

\section{Search strategy and background estimation}

\begin{figure}[!htb]
\centering
\includegraphics[width=.45\textwidth]{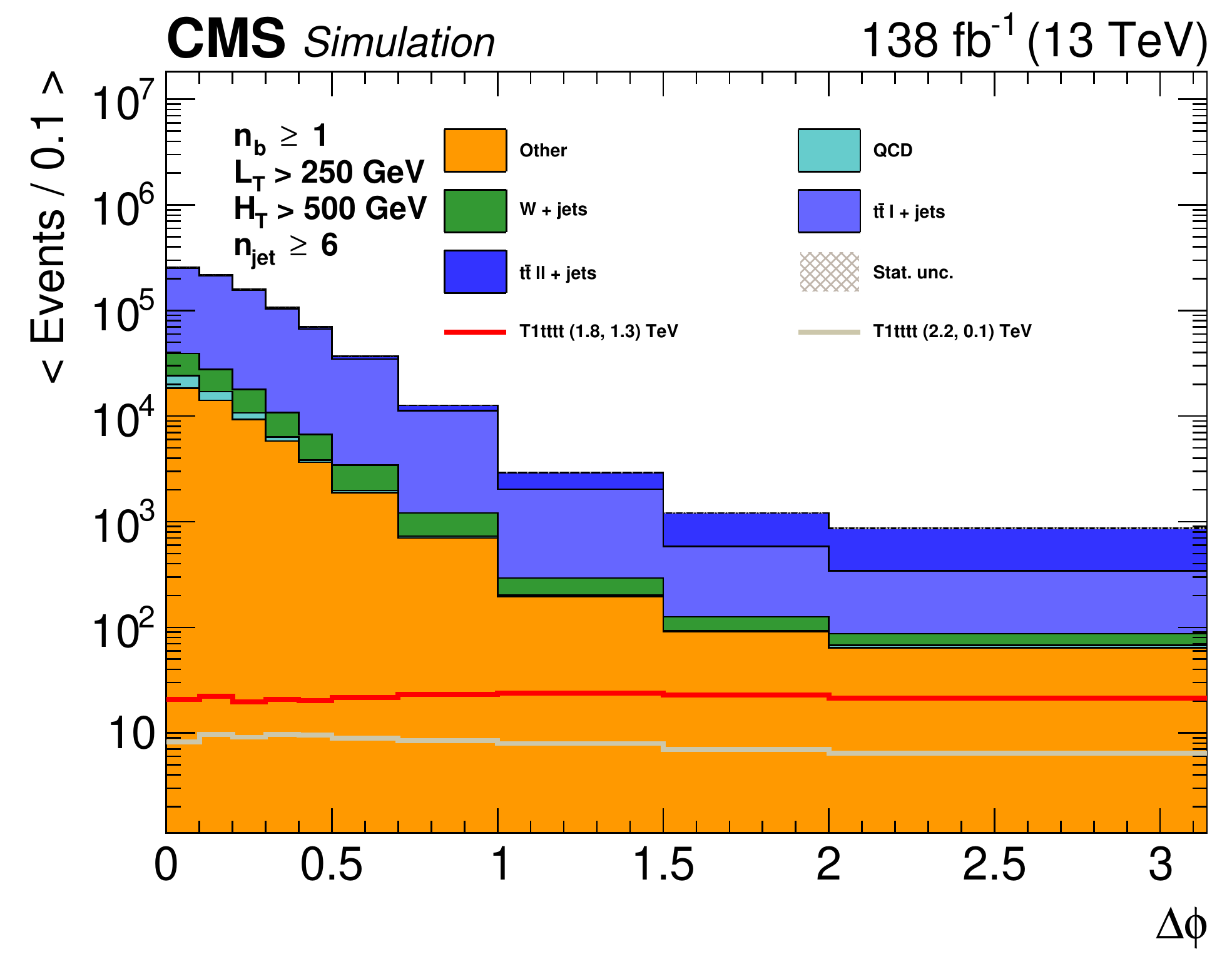}%
\hspace{.05\textwidth}%
\includegraphics[width=.45\textwidth]{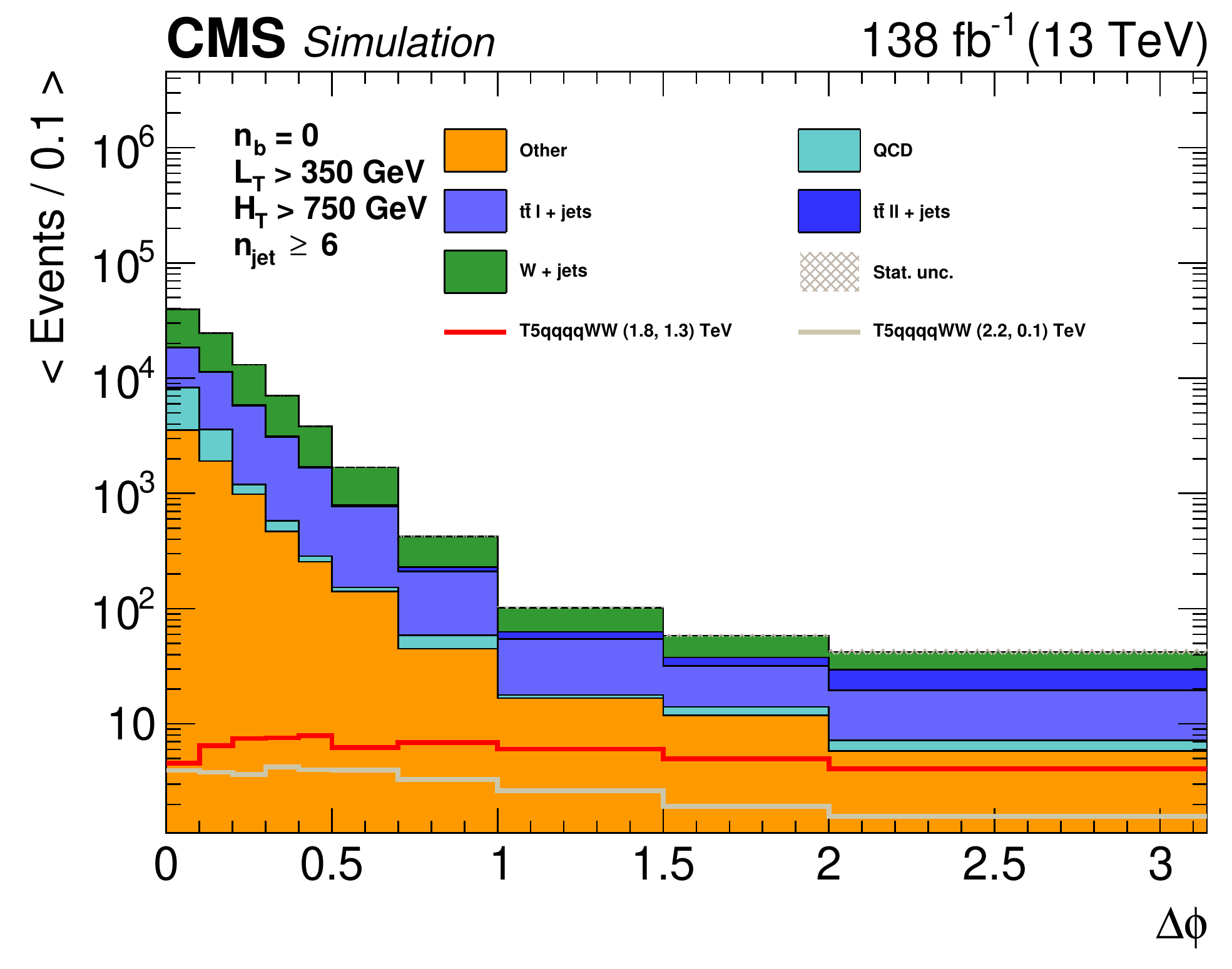}
\caption{Signal and background distributions of the \DF variable, as predicted by simulation, for the \multib analysis, requiring $\njet\geq6$, $\LT>250\GeV$, $\HT>500\GeV$ (left), and the \zerob analysis, requiring $\njet\geq6$, $\LT>350\GeV$, $\HT>750\GeV$ (right).
The predicted signal distributions are also shown for two representative combinations of (gluino, neutralino) masses with large (2.2, 0.1)\TeV and small (1.8, 1.3)\TeV mass differences.}
\label{fig:dPhi}
\end{figure}

The central kinematic variable of this analysis is the absolute value of the azimuthal angle \DF between the \ptvec of a hypothetical \PW boson decaying leptonically and that of the decay lepton
\begin{linenomath}\begin{equation}\label{eq:dphi}
    \DF = \sphericalangle(\ptvecL, \ptvecW),
\end{equation}\end{linenomath}
where the \ptvec of the \PW boson candidate is reconstructed as $\ptvecW=\ptvecL+\ptvecmiss$.
After the baseline event selection, the main backgrounds are events containing one lepton and jets from \ttbar or \Wjets decays.
These backgrounds contain both one prompt lepton and one neutrino from the \PW boson decay in the final state.
Since the neutrino and lepton are boosted in the direction of the momentum of the \PW boson and the neutrino is the only source of \ptvecmiss, a small \DF value is expected.
On the other hand, the SUSY models with two neutralinos in the final state break this correlation, because the neutralinos cannot be detected in the CMS detector and both of them will contribute to the \ptvecmiss, randomizing its direction.
This behavior is indeed observed in Fig.~\ref{fig:dPhi} showing simulated \DF distributions. While most background contributions are at low \DF values, the signal is almost flat over the whole range.

\begin{figure}[!htb]
\centering
\includegraphics[width=.45\textwidth]{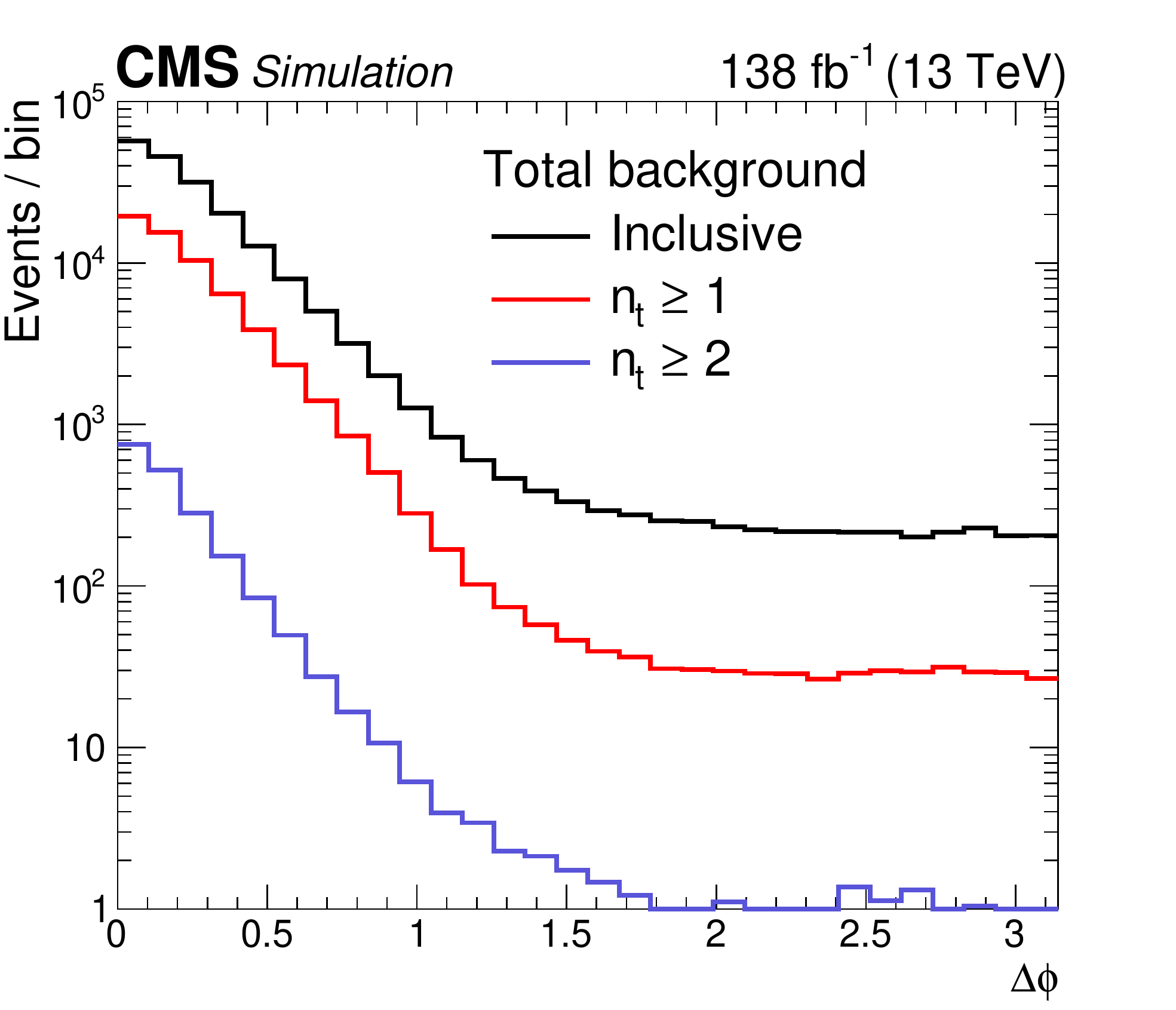}%
\hspace{.05\textwidth}%
\includegraphics[width=.45\textwidth]{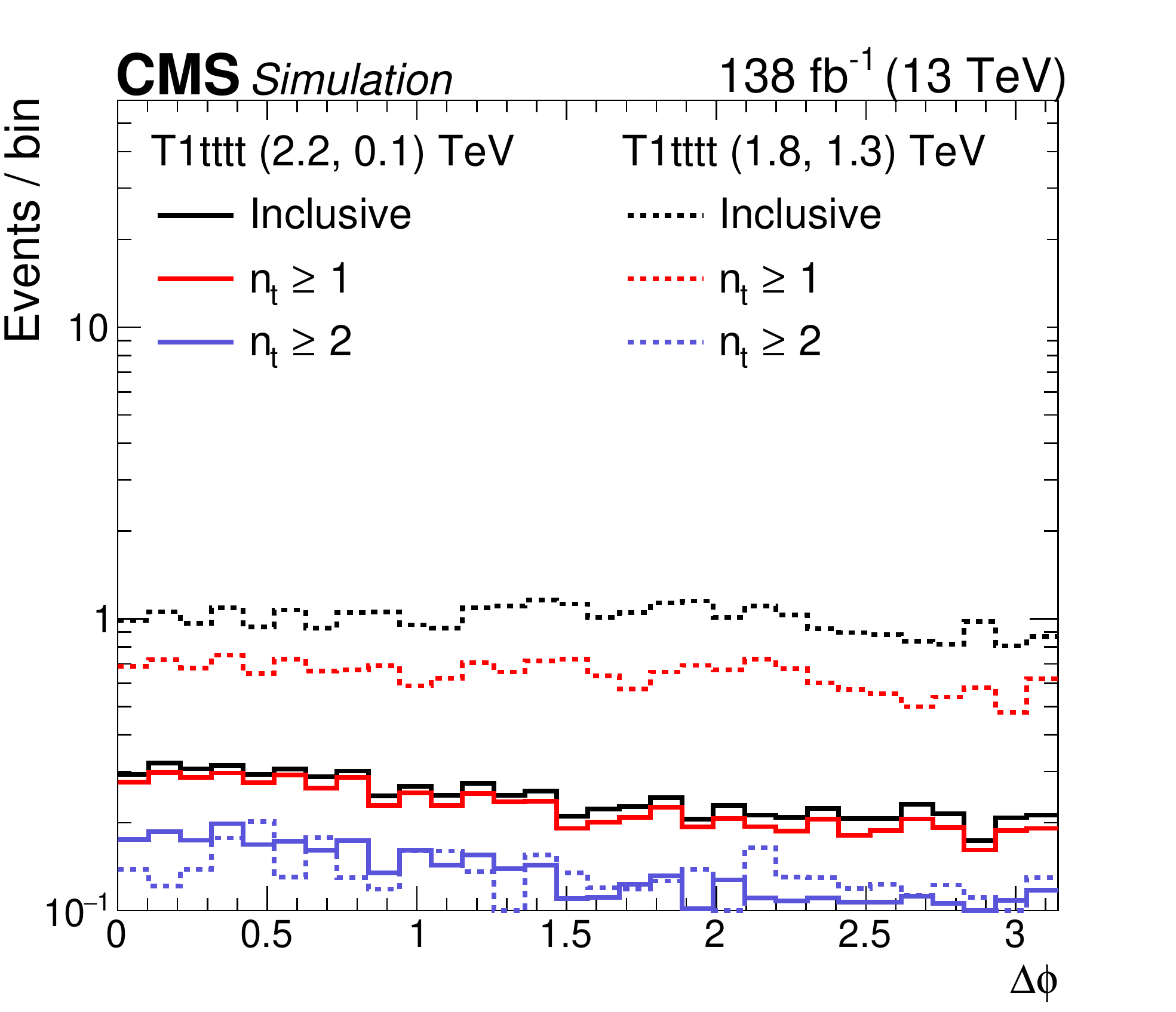}
\caption{Distributions of \DF as obtained from simulation, requiring various \PQt tag multiplicities for the total background (left) and for the signal in two representative combinations of (gluino, neutralino) masses with large (2.2, 0.1)\TeV and small (1.8, 1.3)\TeV mass difference (right).}
\label{fig:top-tagging}
\end{figure}

The effect of the \PQt tagging in the \multib analysis is shown in Fig.~\ref{fig:top-tagging} for the sum of all background contributions and for two representative signal models with a (gluino, neutralino) mass of (2.2, 0.1)\TeV and (1.8, 1.3)\TeV.
While the background in the SR at high \DF is reduced by an order of magnitude when requiring one \PQt tag, the signal yield is only slightly decreased.
Therefore, one \PQt tag is always required, and a few SR bins are defined to have two or more.
When applying \PW boson tagging at the ``tight'' WP in the \zerob analysis, the signal is reduced by about 40\%, while the \Wjets background is reduced by more than 90\%.
The SR bins are split to contain events with either $\nwtag=0$ or $\nwtag\geq1$.

The \DF variable is used to further suppress the background contributions.
The regions with large \DF above a threshold value \DFz are defined as SRs, while those with small \DF values are used as control regions (CRs).
For the \multib analysis, the SR is defined by $\DF>0.75$, whereas in the \zerob analysis the \DF threshold depends on \LT and ranges between 0.5 and 1.
This accounts for a possible higher boost of the \PW boson and correspondingly smaller \DF at larger values of \LT.
The SR is split into bins of \njet, \nbtag, \LT, and \HT, and further categorized by \ntop (\nwtag) for the \multib (\zerob) analysis.
The different SRs must provide good sensitivity for the different signal models and signal parameters, while ensuring sufficient statistical accuracy in CRs to predict the background in the corresponding SR.

The principal tool to estimate the background contributions in the SR bins is a transfer factor, called \Rcs, from CR to SR, which is measured in data with lower jet multiplicity, for each SR bin separately.
For this estimation, we split the regions into a low-\njet region, which is called the sideband (SB), and a high-\njet region, which is called the mainband (MB).
Both of these bands are further divided by \DF into a CR (with $\DF<\DFz$) and an SR (with $\DF>\DFz$) as described above.
This method can be considered as a factorization approach in \DF and jet multiplicity with four regions indexed by pairs of CR or SR and SB or MB.
We note that signal contamination in the SB SR, SB CR, and MB CR is small, typically $<$0.5\% for both \multib and \zerob analysis, and is taken into account in the final fit.

To account for possible deviations from the factorization assumption in the extrapolation from SB to MB, we define multiplicative correction factors $\kappa$, determined from simulations, as described in Sections~\ref{sec:background_mb} and \ref{sec:background_zb}.

In the \multib analysis, the background is dominated by \ttbar events.
In regions with one \PQb-tagged jet and four or five jets, about 80\% \ttbar events and 15 to 20\% \Wjets and single top quark events are expected, with small contributions from QCD multijet events.
In all other \multib regions, the \ttbar background contribution is completely dominant.
With only one SM process dominating the background contribution, a single \Rcs factor is defined in the \multib analysis for each SR bin, after having subtracted the small QCD multijets contribution in the SB.
The background estimation is performed for each year and explained in detail in Section~\ref{sec:background_mb}.

In the \zerob analysis, backgrounds from \ttbar production are suppressed and contributions from \Wjets production are found to be of the same size.
Here, an extension of the \multib strategy is employed, which takes into account differences in the \Rcs values for these two backgrounds, for all years combined, as detailed in Section~\ref{sec:background_zb}.

\begin{figure}[!htb]
\centering
\includegraphics[width=.8\textwidth]{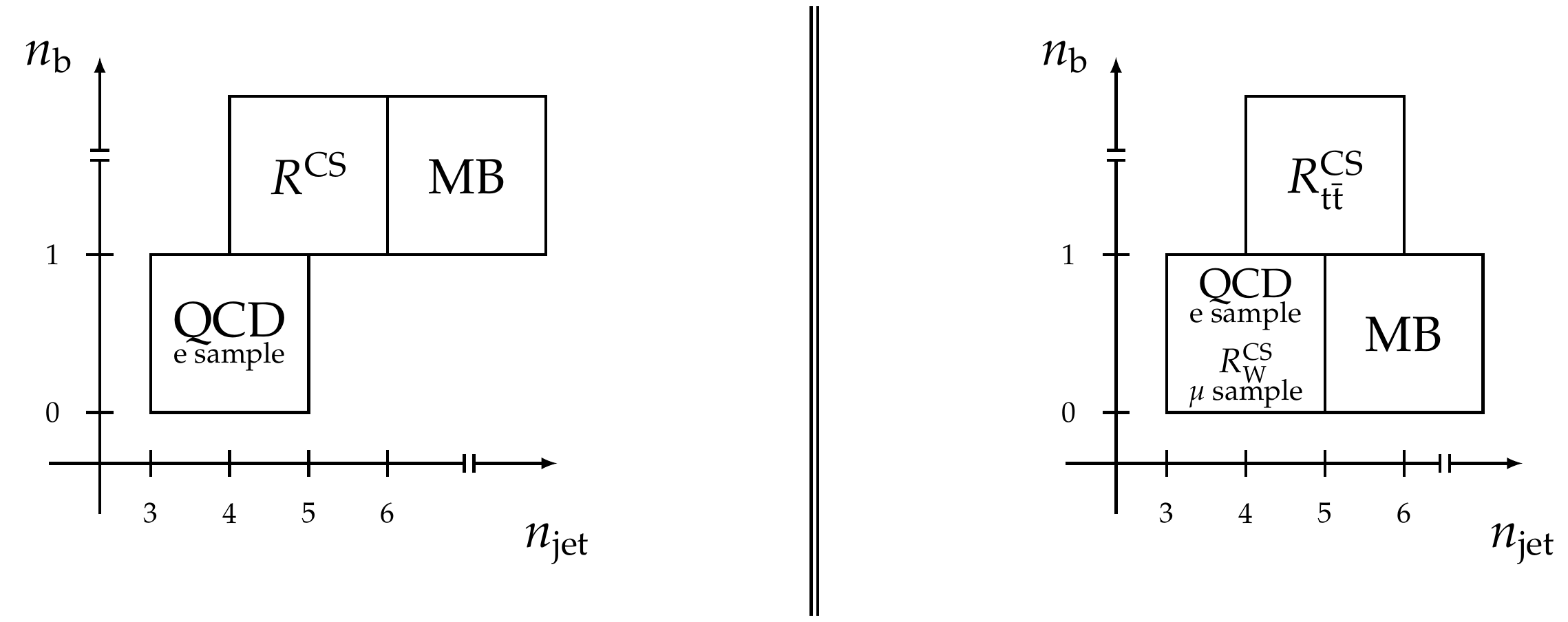}
\caption{Overview of the regions used to calculate \Rcs for the \multib (left) and \zerob (right) analysis.
For the multijet (QCD) fit, the electron (\Pe) sample is used, while the muon (\PGm) sample is used for the determination of \RCSw.}
\label{fig:Regions}
\end{figure}

An overview of the (\njet, \nbtag) regions used in this analysis is given in Fig.~\ref{fig:Regions}.
The \multib and the \zerob analysis share SB regions, but their MB SR bins are exclusive and are never used simultaneously, since the results are interpreted in different simplified models.

To enhance the sensitivity of the search, we further split the MB SR into SRs using \LT, \HT, \nbtag, \njet, and \ntop (\nwtag), as defined in Tables~\ref{tab:searchbins_mb} and~\ref{tab:searchbins_zb} for the \multib and \zerob analysis, respectively.
For each of the search bins in Tables~\ref{tab:searchbins_mb} and~\ref{tab:searchbins_zb}, we define a CR by inverting the \DF cut, and a side band by selecting the low \njet region defined in Sections~\ref{sec:background_mb} and \ref{sec:background_zb}.

\begin{table}[!phtb]
\centering\renewcommand\arraystretch{1.1}
\topcaption{Observed number of events in the MB SR bins of the \multib analysis, together with the predicted yields for background and two T1tttt (\mGlu, \mLSP) signal points.
All bins are defined with $\DF>0.75$.}
\cmsTable{\begin{tabular}{cccccccccc}
\hline
\multirow{2}{*}{\njet} & \multirow{2}{*}{\nbtag} & \multirow{2}{*}{\LT [{\GeVns}]} & \multirow{2}{*}{\HT [{\GeVns}]} & \multirow{2}{*}{\ntop} & \multirow{2}{*}{Bin name} & \multicolumn{2}{c}{T1tttt signal events} & Predicted & Observed \\
& & & & & & (1.8, 1.3)\TeV & (2.2, 0.1)\TeV & background events & events \\ \hline
[6, 8]
    & 1       & [250, 450] & [500, 1500] & 1       & A1a &$1.31 \pm 0.11$ & $<$0.01         & $576 \pm 29$    & 570 \\
    &         &            &             & $\geq$2 & A1b &$0.08 \pm 0.03$ & $<$0.01         & $13 \pm 2$      & 14  \\[\cmsTabSkip]
    &         &            & $>$1500     & 1       & A2a &$<$0.01         & $0.01 \pm 0.01$ & $47 \pm 7$      & 42  \\
    &         &            &             & $\geq$2 & A2b &$<$0.01         & $0.04 \pm 0.01$ & $5 \pm 1$       & 3   \\[\cmsTabSkip]
    &         & [450, 600] & $>$500      & 1       & A3a &$0.56 \pm 0.08$ & $0.05 \pm 0.01$ & $31 \pm 6$      & 16  \\
    &         &            &             & $\geq$2 & A3b &$0.04 \pm 0.02$ & $0.07 \pm 0.01$ & $1.0 \pm 0.3$   & 1   \\[\cmsTabSkip]
    &         & $>$600     & $>$500      & 1       & A4a &$0.24 \pm 0.05$ & $0.51 \pm 0.03$ & $7 \pm 2$       & 8   \\
    &         &            &             & $\geq$2 & A4b &$<$0.01         & $0.59 \pm 0.03$ & $1.0 \pm 0.5$   & 0   \\[\cmsTabSkip]
    & 2       & [250, 450] & [500, 1500] & 1       & B1a &$2.69 \pm 0.15$ & $0.01 \pm 0.01$ & $532 \pm 26$    & 586 \\
    &         &            &             & $\geq$2 & B1b &$0.40 \pm 0.06$ & $<$0.01         & $16 \pm 2$      & 19  \\[\cmsTabSkip]
    &         &            & $>$1500     & 1       & B2a &$0.01 \pm 0.01$ & $0.03 \pm 0.01$ & $30 \pm 5$      & 34  \\
    &         &            &             & $\geq$2 & B2b &$<$0.01         & $0.07 \pm 0.01$ & $3.4 \pm 0.8$   & 1   \\[\cmsTabSkip]
    &         & [450, 600] & $>$500      & 1       & B3a &$1.16 \pm 0.10$ & $0.07 \pm 0.01$ & $27 \pm 6$      & 34  \\
    &         &            &             & $\geq$2 & B3b &$0.18 \pm 0.04$ & $0.11 \pm 0.01$ & $1.1 \pm 0.5$   & 2   \\[\cmsTabSkip]
    &         & $>$600     & $>$500      & 1       & B4a &$0.49 \pm 0.07$ & $0.76 \pm 0.03$ & $6.2 \pm 1.6$   & 6   \\
    &         &            &             & $\geq$2 & B4b &$0.12 \pm 0.03$ & $0.93 \pm 0.03$ & $0.23 \pm 0.08$ & 0   \\[\cmsTabSkip]
    & $\geq$3 & [250, 450] & [500, 1500] & 1       & C1a &$3.69 \pm 0.17$ & $0.01 \pm 0.01$ & $115 \pm 7$     & 105 \\
    &         &            &             & $\geq$2 & C1b &$0.64 \pm 0.07$ & $<$0.01         & $6 \pm 1$       & 3   \\[\cmsTabSkip]
    &         &            & $>$1500     & 1       & C2a &$<$0.01         & $0.04 \pm 0.01$ & $7 \pm 2$       & 10  \\
    &         &            &             & $\geq$2 & C2b &$<$0.01         & $0.08 \pm 0.01$ & $1.0 \pm 0.4$   & 2   \\[\cmsTabSkip]
    &         & [450, 600] & $>$500      & 1       & C3a &$1.25 \pm 0.10$ & $0.07 \pm 0.01$ & $5 \pm 1$       & 4   \\
    &         &            &             & $\geq$2 & C3b &$0.27 \pm 0.05$ & $0.12 \pm 0.01$ & $0.63 \pm 0.43$ & 0   \\[\cmsTabSkip]
    &         & $>$600     & $>$500      & 1       & C4a &$0.52 \pm 0.07$ & $0.70 \pm 0.03$ & $1.4 \pm 0.4$   & 4   \\
    &         &            &             & $\geq$2 & C4b &$0.09 \pm 0.03$ & $0.87 \pm 0.04$ & $0.05 \pm 0.04$ & 0   \\[\cmsTabSkip]
$\geq$9
    & 1       & [250, 450] & [500, 1500] & 1       & D1a &$0.39 \pm 0.06$ & $<$0.01         & $32 \pm 3$      & 26  \\
    &         &            &             & $\geq$2 & D1b &$0.11 \pm 0.03$ & $<$0.01         & $2.1 \pm 0.6$   & 4   \\[\cmsTabSkip]
    &         &            & $>$1500     & 1       & D2a &$0.02 \pm 0.01$ & $0.01 \pm 0.01$ & $6 \pm 1$       & 11  \\
    &         &            &             & $\geq$2 & D2b &$0.02 \pm 0.01$ & $0.02 \pm 0.01$ & $1.0 \pm 0.3$   & 2   \\[\cmsTabSkip]
    &         & [450, 600] & $>$500      & $\geq$1 & D3  &$0.19 \pm 0.05$ & $0.04 \pm 0.01$ & $2.3 \pm 0.6$   & 2   \\[\cmsTabSkip]
    &         & $>$600     & $>$500      & $\geq$1 & D4  &$0.18 \pm 0.04$ & $0.33 \pm 0.02$ & $0.6 \pm 0.3$   & 0   \\[\cmsTabSkip]
    & 2       & [250, 450] & [500, 1500] & 1       & E1a &$1.00 \pm 0.09$ & $<$0.01         & $35 \pm 3$      & 35  \\
    &         &            &             & $\geq$2 & E1b &$0.42 \pm 0.06$ & $<$0.01         & $3.2 \pm 0.7$   & 2   \\[\cmsTabSkip]
    &         &            & $>$1500     & 1       & E2a &$0.04 \pm 0.02$ & $0.01 \pm 0.01$ & $8 \pm 2$       & 6   \\
    &         &            &             & $\geq$2 & E2b &$0.03 \pm 0.02$ & $0.04 \pm 0.01$ & $1.0 \pm 0.4$   & 2   \\[\cmsTabSkip]
    &         & [450, 600] & $>$500      & 1       & E3a &$0.53 \pm 0.06$ & $0.04 \pm 0.01$ & $1.7 \pm 0.5$   & 1   \\
    &         &            &             & $\geq$2 & E3b &$0.14 \pm 0.04$ & $0.04 \pm 0.01$ & $0.2 \pm 0.1$   & 0   \\[\cmsTabSkip]
    &         & $>$600     & $>$500      & 1       & E4a &$0.42 \pm 0.06$ & $0.22 \pm 0.02$ & $0.9 \pm 0.4$   & 1   \\
    &         &            &             & $\geq$2 & E4b &$0.18 \pm 0.04$ & $0.43 \pm 0.02$ & $0.06 \pm 0.04$ & 0   \\[\cmsTabSkip]
    & $\geq$3 & [250, 450] & [500, 1500] & 1       & F1a &$2.25 \pm 0.13$ & $<$0.01         & $13 \pm 2$      & 7   \\
    &         &            &             & $\geq$2 & F1b &$1.09 \pm 0.09$ & $<$0.01         & $2.4 \pm 0.8$   & 2   \\[\cmsTabSkip]
    &         &            & $>$1500     & 1       & F2a &$0.03 \pm 0.02$ & $0.02 \pm 0.01$ & $4 \pm 1$       & 0   \\
    &         &            &             & $\geq$2 & F2b &$0.04 \pm 0.02$ & $0.08 \pm 0.01$ & $0.7 \pm 0.3$   & 0   \\[\cmsTabSkip]
    &         & [450, 600] & $>$500      & $\geq$1 & F3  &$1.39 \pm 0.10$ & $0.12 \pm 0.01$ & $1.1 \pm 0.4$   & 2   \\[\cmsTabSkip]
    &         & $>$600     & $>$500      & $\geq$1 & F4  &$0.89 \pm 0.08$ & $0.96 \pm 0.03$ & $0.24 \pm 0.16$ & 2   \\
     \hline
\end{tabular}}
\label{tab:searchbins_mb}
\end{table}

\begin{table}[!phtb]
\centering\renewcommand\arraystretch{1.1}
\topcaption{Observed number of events in the MB SR bins of the \zerob analysis, together with the predicted yields for background and two T5qqqqWW (\mGlu, \mLSP) signal points.}
\cmsTable{\begin{tabular}{cccccccccc}
\hline
\multirow{2}{*}{\njet} & \multirow{2}{*}{\LT [{\GeVns}]} & \multirow{2}{*}{\HT [{\GeVns}]} & \multirow{2}{*}{\DF} & \multirow{2}{*}{\nwtag} & \multirow{2}{*}{Bin name} & \multicolumn{2}{c}{T5qqqqWW signal events} & Predicted & Observed \\
& & & & & & (1.8, 1.3)\TeV & (2.2, 0.1)\TeV & background events & events \\ \hline
5
    & [250, 350] & [500, 750]  & $>$1    & 0         & G0a & $0.84 \pm 0.23$ & $<$0.01          & $342 \pm 24$  & 333 \\
    &            &             &         & $\geq$1   & G0b & $0.38 \pm 0.12$ & $<$0.01          & $70 \pm 8$    & 77  \\[\cmsTabSkip]
    &            & $>$750      &         & 0         & G1a & $0.23 \pm 0.09$ & $<$0.01          & $292 \pm 22$  & 304 \\
    &            &             &         & $\geq$1   & G1b & $0.15 \pm 0.07$ & $<$0.01          & $69 \pm 10$   & 62  \\[\cmsTabSkip]
    & [350, 450] & [500, 750]  & $>$1    & 0         & G2a & $1.11 \pm 0.29$ & $<$0.01          & $71 \pm 8$    & 63  \\
    &            &             &         & $\geq$1   & G2b & $0.45 \pm 0.15$ & $<$0.01          & $14 \pm 5$    & 25  \\[\cmsTabSkip]
    &            & $>$750      &         & 0         & G3a & $0.34 \pm 0.12$ & $0.01 \pm 0.01$ & $66 \pm 8$    & 44  \\
    &            &             &         & $\geq$1   & G3b & $0.20 \pm 0.09$ & $<$0.01          & $14 \pm 4$    & 13  \\[\cmsTabSkip]
    & [450, 650] & [500, 750]  & $>$0.75 & 0         & G4a & $1.44 \pm 0.34$ & $<$0.01          & $52 \pm 7$    & 45  \\
    &            &             &         & $\geq$1   & G4b & $0.74 \pm 0.19$ & $<$0.01          & $12 \pm 3$    & 9   \\[\cmsTabSkip]
    &            & [750, 1250] &         & 0         & G5a & $0.64 \pm 0.18$ & $<$0.01          & $42 \pm 6$    & 35  \\
    &            &             &         & $\geq$1   & G5b & $0.29 \pm 0.10$ & $<$0.01          & $10 \pm 3$    & 6   \\[\cmsTabSkip]
    &            & $>$1250     &         & 0         & G6a & $<$0.01          & $0.07 \pm 0.02$ & $16 \pm 3$    & 19  \\
    &            &             &         & $\geq$1   & G6b & $<$0.01          & $0.08 \pm 0.02$ & $3 \pm 1$     & 3   \\[\cmsTabSkip]
    & $>$650     & [500, 1250] & $>$0.5  & 0         & G7a & $0.74 \pm 0.20$ & $0.05 \pm 0.01$ & $33 \pm 8$    & 32  \\
    &            &             &         & $\geq$1   & G7b & $0.27 \pm 0.09$ & $0.02 \pm 0.01$ & $7 \pm 2$     & 8   \\[\cmsTabSkip]
    &            & $>$1250     &         & 0         & G8a & $0.14 \pm 0.05$ & $0.70 \pm 0.15$ & $11 \pm 3$    & 8   \\
    &            &             &         & $\geq$1   & G8b & $0.04 \pm 0.02$ & $0.59 \pm 0.13$ & $0.6 \pm 0.4$ & 2   \\[\cmsTabSkip]
[6, 7]
    & [250, 350] & [500, 1000] & $>$1    & 0         & H1a & $1.94 \pm 0.45$ & $<$0.01          & $281 \pm 22$  & 292 \\
    &            &             &         & $\geq$1   & H1b & $0.84 \pm 0.22$ & $<$0.01          & $71 \pm 9$    & 71  \\[\cmsTabSkip]
    &            & $>$1000     &         & 0         & H2a & $0.22 \pm 0.09$ & $0.03 \pm 0.01$ & $121 \pm 11$  & 121 \\
    &            &             &         & $\geq$1   & H2b & $0.10 \pm 0.05$ & $<$0.01          & $29 \pm 5$    & 21  \\[\cmsTabSkip]
    & [350, 450] & [500, 1000] & $>$1    & 0         & H3a & $1.99 \pm 0.45$ & $<$0.01          & $51 \pm 6$    & 71  \\
    &            &             &         & $\geq$1   & H3b & $1.08 \pm 0.26$ & $<$0.01          & $12 \pm 3$    & 15  \\[\cmsTabSkip]
    &            & $>$1000     &         & 0         & H4a & $0.20 \pm 0.08$ & $0.03 \pm 0.01$ & $31 \pm 7$    & 21  \\
    &            &             &         & $\geq$1   & H4b & $0.09 \pm 0.06$ & $0.02 \pm 0.01$ & $6 \pm 2$     & 6   \\[\cmsTabSkip]
    & [450, 650] & [500, 750]  & $>$0.75 & 0         & H5a & $2.08 \pm 0.47$ & $<$0.01          & $19 \pm 4$    & 17  \\
    &            &             &         & $\geq$1   & H5b & $1.13 \pm 0.27$ & $<$0.01          & $5 \pm 2$     & 9   \\[\cmsTabSkip]
    &            & [750, 1250] &         & 0         & H6a & $1.76 \pm 0.40$ & $<$0.01          & $29 \pm 4$    & 18  \\
    &            &             &         & $\geq$1   & H6b & $0.98 \pm 0.24$ & $<$0.01          & $7 \pm 2$     & 4   \\[\cmsTabSkip]
    &            & $>$1250     &         & 0         & H7a & $0.18 \pm 0.07$ & $0.19 \pm 0.05$ & $15 \pm 3$    & 14  \\
    &            &             &         & $\geq$1   & H7b & $0.13 \pm 0.06$ & $0.13 \pm 0.03$ & $3 \pm 1$     & 1   \\[\cmsTabSkip]
    & $>$650     & [500, 1250] & $>$0.5  & 0         & H8a & $1.62 \pm 0.36$ & $0.04 \pm 0.01$ & $13 \pm 3$    & 17  \\
    &            &             &         & $\geq$1   & H8b & $0.60 \pm 0.16$ & $0.03 \pm 0.01$ & $4 \pm 1$     & 4   \\[\cmsTabSkip]
    &            & $>$1250     &         & 0         & H9a & $0.50 \pm 0.13$ & $1.69 \pm 0.35$ & $9 \pm 3$     & 6   \\
    &            &             &         & $\geq$1   & H9b & $0.27 \pm 0.08$ & $1.32 \pm 0.27$ & $2 \pm 1$     & 1   \\[\cmsTabSkip]
$\geq$8
    & [250, 350] & [500, 1000] & $>$1    & 0         & I1a & $0.31 \pm 0.11$ & $<$0.01          & $23 \pm 5$    & 25  \\
    &            &             &         & $\geq$1   & I1b & $0.21 \pm 0.08$ & $<$0.01          & $7 \pm 3$     & 5   \\[\cmsTabSkip]
    &            & $>$1000     &         & 0         & I2a & $0.14 \pm 0.06$ & $0.02 \pm 0.01$ & $22 \pm 5$    & 23  \\
    &            &             &         & $\geq$1   & I2b & $0.06 \pm 0.04$ & $<$0.01          & $8 \pm 2$     & 12  \\[\cmsTabSkip]
    & [350, 450] & [500, 1000] & $>$1    & 0         & I3a & $0.41 \pm 0.12$ & $<$0.01          & $3.0 \pm 0.7$ & 10  \\
    &            &             &         & $\geq$1   & I3b & $0.17 \pm 0.06$ & $<$0.01          & $1.1 \pm 0.4$ & 0   \\[\cmsTabSkip]
    &            & $>$1000     &         & 0         & I4a & $0.23 \pm 0.09$ & $0.03 \pm 0.01$ & $5 \pm 1$     & 5   \\
    &            &             &         & $\geq$1   & I4b & $0.26 \pm 0.09$ & $0.03 \pm 0.01$ & $3 \pm 1$     & 2   \\[\cmsTabSkip]
    & [450, 650] & [500, 1250] & $>$0.75 & 0         & I5a & $0.83 \pm 0.22$ & $<$0.01          & $3.4 \pm 0.9$ & 4   \\
    &            &             &         & $\geq$1   & I5b & $0.61 \pm 0.17$ & $<$0.01          & $0.5 \pm 0.3$ & 1   \\[\cmsTabSkip]
    &            & $>$1250     &         & 0         & I6a & $0.24 \pm 0.09$ & $0.11 \pm 0.03$ & $2.6 \pm 0.8$ & 2   \\
    &            &             &         & $\geq$1   & I6b & $0.12 \pm 0.06$ & $0.07 \pm 0.02$ & $0.5 \pm 0.3$ & 2   \\[\cmsTabSkip]
    & $>$650     & [500, 1250] & $>$0.5  & $\geq$0   & I7  & $0.67 \pm 0.18$ & $<$0.01          & $1.58 \pm 0.63$ & 2   \\
    &            & $>$1250     &         &           & I8  & $0.81 \pm 0.22$ & $1.41 \pm 0.30$ & $1.58 \pm 0.71$ & 1   \\ \hline
\end{tabular}}
\label{tab:searchbins_zb}
\end{table}

\subsection{Background estimate in the \texorpdfstring{\multib}{multi-b} final state}
\label{sec:background_mb}

In the \multib analysis, the predicted number \NMBSRpred of background events in each MB SR bin is given as the sum of the number of background events from \ttbar and electroweak processes \NMBSRpredew and the number of QCD multijet events \NMBSRpredqcd:
\begin{linenomath}\begin{equation}
    \NMBSRpred = \NMBSRpredew + \NMBSRpredqcd\;.
\end{equation}\end{linenomath}
The generic label ``EW'' refers to all backgrounds other than QCD multijet events.
About 10--15\% of the SM background events in the SB CR are expected to be QCD multijet events, while this fraction is significantly smaller in the MB SR.
This background contribution is estimated independently from a fit to data, as described in Section~\ref{sec:qcd}.
The multijet background is subtracted from the number of background events when calculating the transfer factor \RCSdata from data:
\begin{linenomath}\begin{equation}
    \RCSdata\big(\fourfivejets\big) = \frac{\NSBSRdata - \NSBSRpredqcd}{\NSBCRdata - \NSBCRpredqcd}\;.
    \label{eq:Rcs}
\end{equation}\end{linenomath}
Here, \NSBSRdata is the number of events in the SB SR, while \NSBCRdata corresponds to the number of events in the SB CR. The independently estimated number of multijet events for these two regions are \NSBSRpredqcd and \NSBCRpredqcd.

\begin{figure}[!htb]
\centering
\includegraphics[width=\textwidth]{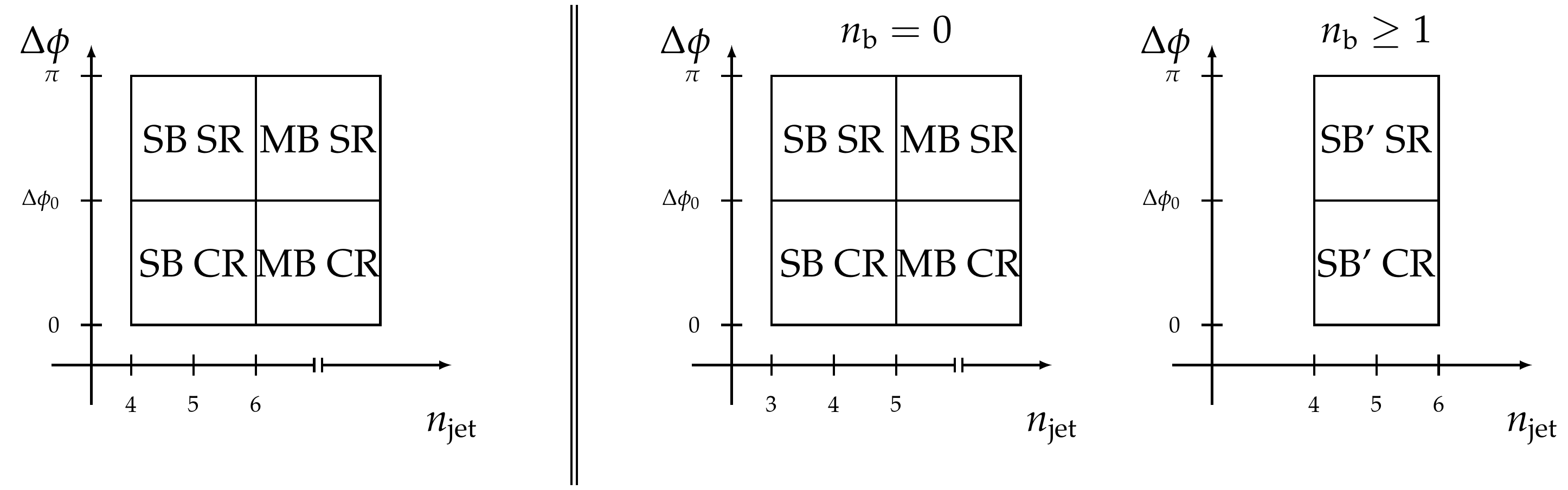}
\caption{Graphical presentation of the regions indexed by pairs of SB or MB and CR or SR: for the \multib (left) and for the \zerob (middle and right) analysis.
The value of \DF separating CR and SR is labeled as \DFz.
It is independent of the SR bin for the \multib analysis with a value of 0.75, but varies from 0.5 to 1 among the \zerob SR bins.}
\label{fig:search_bins}
\end{figure}

The SB region, where \Rcs is determined for each SR bin, is required to have four or five jets, while the MB region must satisfy $\njet\in[6,8]$ or $\njet\geq9$.
This is represented graphically in Fig.~\ref{fig:search_bins}~(left).
The \Rcs factor is calculated separately for each search bin in \LT, \HT, \nbtag, and \ntop.
At very high \HT, \Rcs is determined jointly across all three \nbtag bins to increase the number of events, as the overall uncertainty of the background prediction for several of the search bins is dominated by the statistical uncertainty of the yield in the SB SR.

Small differences in \Rcs between SB and MB are corrected by the additional factor \kappaew, which is determined in simulation as the ratio of the \Rcs for simulated events:
\begin{linenomath}\begin{equation}
    \kappaew = \frac{\RCSmcew\big(\jetsasinmb\big)}{\RCSmcew\big(\fourfivejets\big)}\;.
\end{equation}\end{linenomath}
For the \multib analysis, the label ``\jetsasinmb'' refers to either $\njet\in[6,8]$ or $\njet\geq9$, depending on the specific search bin.
The \kappaew factor is determined separately for each search bin, except that a common \kappaew factor is applied for the $\nbtag\geq2$ search bins with the same \HT and \LT, since the \kappaew factors are found to be nearly independent of \nbtag.
In general, these correction factors are found to be close to unity, within 20--30\%.
With these definitions, the number of predicted EW events in the MB SR is given by:
\begin{linenomath}\begin{equation}
    \NMBSRpredew = \kappaew \RCSdata\big(\fourfivejets\big) \Big(\NMBCRdata - \NMBCR_{\text{Pred,QCD}}\Big).
    \label{eq:prediction}
\end{equation}\end{linenomath}
For events containing one lepton and jets from \ttbar production, \Rcs typically has values of 0.01 to 0.02, depending on the search bin.
Similar values, ranging from 0.01 to 0.04, are found for \Wjets events.
In events with more than one high-\pt neutrino, \eg, in \ttbar events in which both \PW bosons decay leptonically, \Rcs is higher with values of around 0.5.
This is expected, since a large fraction of background events at high \DF is due to the dileptonic \ttbar background, while the low-\DF region is dominated by events with only one neutrino.
A larger \Rcs is also expected for events with three neutrinos, such as $\ttbar\PZ$, when the \ttbar system decays into a lepton and jets and the \PZ boson decays to two neutrinos.

A small fraction of the background arises from dileptonic \ttbar events in which one lepton is undetected.
Having fewer jets than single-lepton \ttbar events, these events tend to populate the SB.
At the same time their \DF distribution is flatter, leading to an overestimate of \Rcs.
Accordingly a separate correction is developed as described in Section 6.3.

\subsection{Background estimate in the \texorpdfstring{\zerob}{zero-b} final state}
\label{sec:background_zb}

Unlike the \multib analysis, where we have only one dominant background in each bin, the SR bins in the \zerob analysis require the prediction of two backgrounds of almost the same size, \ttbar and \Wjets events.
These background contributions are estimated by applying the \Rcs method separately for each of the two components.
Since we split these two contributions in the background estimation, we typically have smaller bin counts compared to the \multib and consequently larger statistical fluctuations.
To guarantee sufficient statistical precision, we perform the prediction for the full data set instead of separate estimations for 2016, 2017, and 2018.

This strategy implies the use of two sidebands enriched in \Wjets (SB) and \ttbar events (SB'), respectively.
We decompose the total background in each bin, for example in the MB SR, as:
\begin{linenomath}\begin{equation}
    \NMBSRpred = \NMBSRpredw + \NMBSRpredtt + \NMBSRmcother + \NMBSRpredqcd\;,
\end{equation}\end{linenomath}
where the numbers of predicted \Wjets and \ttbar events are denoted by \NMBSRpredw and
\NMBSRpredtt, respectively.
We also include \WW and \WZ events, where the \PW boson decays leptonically and the second \PW or the \PZ boson hadronically, as a part of \Wjets estimation, since they have similar kinematic properties and \Rcs values.
All other diboson events are treated as part of the rare backgrounds, which are estimated from simulation and denoted by \NMBSRmcother.
The small contribution of the QCD multijet background is fixed to the yield estimated from data as described in Section~\ref{sec:qcd} and noted as \NMBSRpredqcd.

The \ttbar and the \Wjets contributions are estimated with an \Rcs method in a similar way as described in the previous section.
The \Rcs values for \Wjets and \ttbar events are measured in separate SB regions with different \PQb-tagged jet requirements, as laid out in Fig.~\ref{fig:search_bins} middle and right, respectively.

Similarly to Eq.~\eqref{eq:Rcs}, the value of \Rcs for \ttbar events is calculated in the \multib sideband (SB') with \fourfivejets and \onebtag.
The differences are the definition of \DFz and the requirements that define the corresponding search bins:
\begin{linenomath}\begin{equation}
    \RCSdata\big(\fourfivejets, \onebtag\big) = \frac{\NSBSRdata['] - \NSBSRpredqcd[']}{\NSBCRdata['] - \NSBCRpredqcd[']}\;.
\end{equation}\end{linenomath}
A correction factor \kappab, defined in Eq.~\eqref{eq:kappab} and appearing in Eq.~\eqref{eq:NMBSRpredtt} below, accounts for the difference of \RCStt between samples with zero \PQb-tagged jets and samples with at least one \PQb-tagged jet and also takes into account non-\ttbar background components in the EW category.
It is taken from simulation:
\begin{linenomath}\begin{equation} \label{eq:kappab}
    \kappab = \frac{\RCStt\big(\fourfivejets, \zerobtag\big)}{\Rcs_{\text{EW}}\big(\fourfivejets, \onebtag\big)}
    \text{, where }
    \RCSmc = \frac{\NSBSRmc[']}{\NSBCRmc[']}
    \text{ and }
    \text{MC} \in [\ttbar, \text{EW}].
\end{equation}\end{linenomath}
A second factor \kappatt, defined in Eq.~\eqref{eq:kappatt} and appearing in Eq.~\eqref{eq:NMBSRpredtt} below, corrects for a residual dependence of \RCStt on \njet, in analogy to the \kappaew factor defined in Section~\ref{sec:background_mb}.
It is defined as:
\begin{linenomath}\begin{equation}\label{eq:kappatt}
    \kappatt = \frac{\RCStt\big(\jetsasinmb, \zerobtag\big)}{\RCStt\big(\fourfivejets, \zerobtag\big)}\;.
\end{equation}\end{linenomath}
Similar to the \multib analysis, the number of simulated dilepton \ttbar events in the factor \kappatt is corrected by the slight difference in the \njet shape measured in dilepton and one-lepton CRs, as described in Section~\ref{sec:dileptonnjet}.
The product of both correction factors \kappab and \kappatt has typical values of 0.7 to 1.0 and statistical uncertainties from the simulation are propagated to the predicted yields.

Finally, the fraction of \ttbar and \Wjets events in the MB CR is estimated by a template fit to the \nbtag distribution for each search bin.
The number of QCD events in these fits is consistently fixed to the number of events predicted from data as described in Section~\ref{sec:qcd}, while all other rare backgrounds are taken from simulation and fixed in the fit as well.
The templates are taken from simulation.
Only the number of \ttbar and \Wjets events is adjusted in the fit.
The fractions are:
\begin{linenomath}\begin{equation}
    f_{\text{MC}}^{\text{MB,CR}} = \frac{\NMBCR_{\text{fit,MC}}}{\NMBCRdata},
    \text{ with }
    \text{MC} \in [\ttbar, \Wjets].
\end{equation}\end{linenomath}
The uncertainties in these two components are propagated as systematic uncertainties to the final prediction.

The final \ttbar prediction is:
\begin{linenomath}\begin{equation} \label{eq:NMBSRpredtt}
    \NMBSRpredtt = \underbrace{
        \kappab \kappatt \RCSdata\big(\onebtag, \fourfivejets\big)
    }_{\text{transfer factor}}
    \underbrace{
        \ftt^{\text{MB,CR}} \NMBCRdata
    }_{\text{\makebox[0pt]{\ttbar contribution in the control region}}}\;.
\end{equation}\end{linenomath}
The \Wjets contribution \NMBSRpredw is also estimated using an \Rcs method.
The \zerob SB is chosen with \threefourjets, \zerobtag. With respect to the SB used for the estimate of \RCStt, a lower jet multiplicity is chosen in order to limit the contamination from \ttbar events.
Here we select only events where the lepton is identified as a muon, since this sample has a negligible contamination from QCD multijet events, contrary to the electron channel.
A systematic uncertainty is derived from simulation to cover potential differences between the muon and the combined electron and muon samples.

\begin{figure}[!htb]
\centering
\includegraphics[width=.495\textwidth]{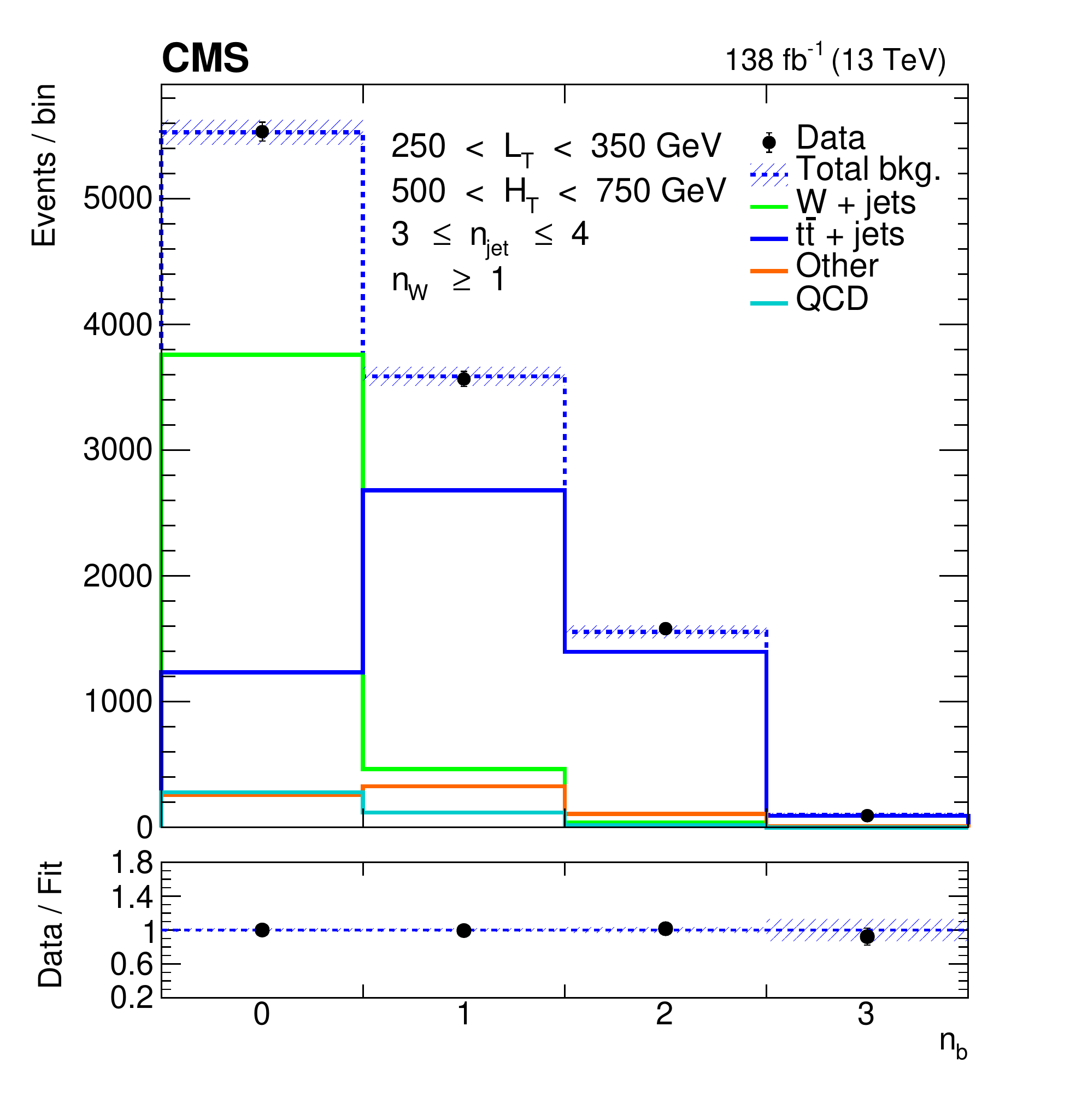}%
\hspace{.0025\textwidth}%
\includegraphics[width=.495\textwidth]{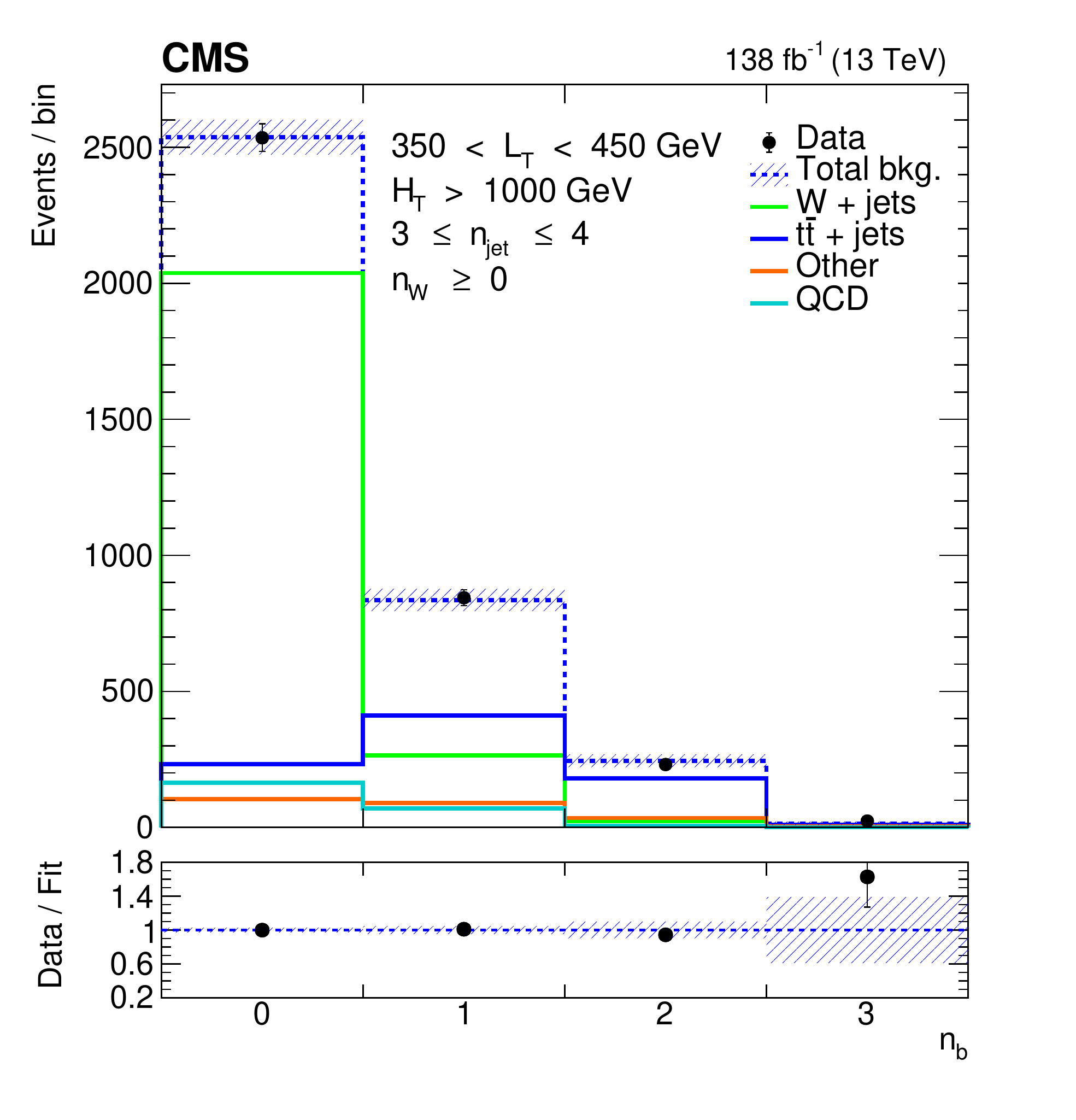}
\caption{Results of fits to the \nbtag multiplicity for control regions for the muon channel and with the requirements $3\leq\njet\leq4$, $250<\LT<350\GeV$, $500<\HT<750\GeV$, $\nwtag\geq1$, $\DF<1$ (left) and $3\leq\njet\leq4$, $350<\LT<450\GeV$, $\HT>1000\GeV$, $\nwtag\geq0$, $\DF<1$ (right).
The shaded area shows the fit uncertainty of the total background.}
\label{fig:btagMultiplicityFit_data}
\end{figure}

The fit of the \nbtag distribution is also performed in the SB to determine the fraction \fttSBCR, since the \ttbar contamination is significant and cannot be ignored.
Examples of these fits are shown in Fig.~\ref{fig:btagMultiplicityFit_data}.

The \ttbar yields are then subtracted in the numerator and denominator when determining \Rcs for the \Wjets estimate:
\begin{linenomath}\begin{equation}
    \RCSdatacorr\big(\threefourjets, \zerobtag, \onemuon\big) =
    \frac{
        \NSBSRdata - \fttSBCR \kappab \RCSdata\big(\fourfivejets, \onebtag\big) \NSBCRdata
    }{
        \Big(1 - \fttSBCR\Big) \NSBCRdata
    }\;.
\end{equation}\end{linenomath}
Similarly, a factor \kappaW, defined in Eq.~\eqref{eq:kappaw} and appearing in Eq.~\eqref{eq:NMBSRpredw}, corrects for a residual dependence of \RCSw on the jet multiplicity; its typical values are 0.7 to 1.1.
In addition, \kappaW also provides the extrapolation from the muon to the electron channel:
\begin{linenomath}\begin{equation} \label{eq:kappaw}
    \kappaW = \frac{
        \RCSw\big(\jetsasinmb, \zerobtag, n_{\Pell}=1\big)
    }{
        \Rcs_{\Wother}\big(\threefourjets, \zerobtag, \onemuon\big)
    }
    \text{, where }
    \RCSmc = \frac{\NSBSRmc}{\NSBCRmc}
    \text{ and }
    \text{MC} \in [\PW, \Wother].
\end{equation}\end{linenomath}
The final prediction of the \Wjets background is then given by:
\begin{linenomath}\begin{equation} \label{eq:NMBSRpredw}
    \NMBSRpredw = \underbrace{
        \kappaW \RCSdatacorr\big(\threefourjets, \zerobtag, \onemuon\big)
    }_{\text{transfer factor}}
    \underbrace{
        f_{\PW}^{\text{MB,CR}} \NMBCRdata
    }_{\text{\makebox[0pt]{\PW contribution in the control region}}}\;.
\end{equation}\end{linenomath}

\subsection{Dilepton control region correction}
\label{sec:dileptonnjet}

The background prediction is sensitive to the extrapolation of \Rcs from the low-\njet SB to the MB regions with higher jet multiplicities.
The \Rcs values differ significantly for events with only one genuine lepton compared to events with two genuine leptons (mainly dileptonic \ttbar), where one lepton is not identified or lost.
In the first case, the \Rcs values are of the order 0.01--0.02, while for dileptonic events the value is around 0.5.
In the latter case, the \MET in the event is not only caused by the neutrino of a leptonically decaying \PW boson, but also by the second genuine lepton that is not identified, mostly because it is a hadronically decaying \PGt lepton leading to more neutrinos in the event, or because it is out of acceptance.
This leads to more events in the high-\DF region and a significantly higher \Rcs.
In general, the prediction is not affected by the different \Rcs of the different processes, if the ratio of events with one genuine lepton to events with two genuine leptons (one lost or not identified) is the same for all \njet regions.

Any mismodeling in the simulation of these lost leptons would not be captured by the $\kappa$ factors in Eqs.~\eqref{eq:prediction} and \eqref{eq:NMBSRpredtt}.
We account for differences between simulation and data with additional event weights, separately for each \njet region, that are applied to genuine simulated dilepton events.
The high-purity dilepton events are transformed artificially into typical single-lepton events by removing the second lepton, as described in the following.
The dilepton control sample is selected by requiring two leptons of opposite charge.
In order to reduce the DY background in the \multib analysis, the invariant mass of same-flavor leptons is required to be more than 10\GeV away from the \PZ boson mass peak.
For the \zerob analysis, where the DY background is more important because of the zero \PQb tag requirement, we allow only two leptons of different flavor.
To simulate the feed-down of the dileptonic events into the single-lepton selection, one of the two leptons is removed from the event.
Since these ``lost leptons'' are mainly from $\PGt\to\text{hadrons}+\PGn$ decays, we replace the removed lepton with a jet with 2/3 of the original lepton's \pt to model the typical visible energy of a \PGt lepton, accounting for the missing momentum caused by the neutrino from the \PGt lepton decay.
In the next step \LT, \DF, and \HT values of the now ``single-lepton'' event (with the additional ``jet'') are recalculated.
In order to maximize the number of events, no \DF requirement is applied, and all events are used twice, with each reconstructed lepton being considered as the lost lepton.

In the events with one genuine lepton in the \zerob analysis, a change in the background composition (mainly \ttbar and \Wjets) could lead to a change of the correction factor.
The size of this additional change is hard to determine and it is desirable to disentangle these two effects.
In order to tackle this issue, we normalize these two backgrounds using weights extracted after performing the template fit on the \PQb tag multiplicity.

The correction factor, \wDL, is determined as a function of \njet for each event from a linear fit to the double ratio between data over MC yields for dilepton (transformed to ``single-lepton'') and single-lepton events of the form:
\begin{linenomath}\begin{equation}
    \wDL = a+b\big(\njet-\njetav\big),
\end{equation}\end{linenomath}
where $a$ is the constant, $b$ is the slope, and \njetav is the weighted mean.
The correction factor is applied as a weight to all simulated events that are flagged as dileptonic from generator level information.

\begin{figure}[!phtb]
\centering
\includegraphics[width=.4\textwidth]{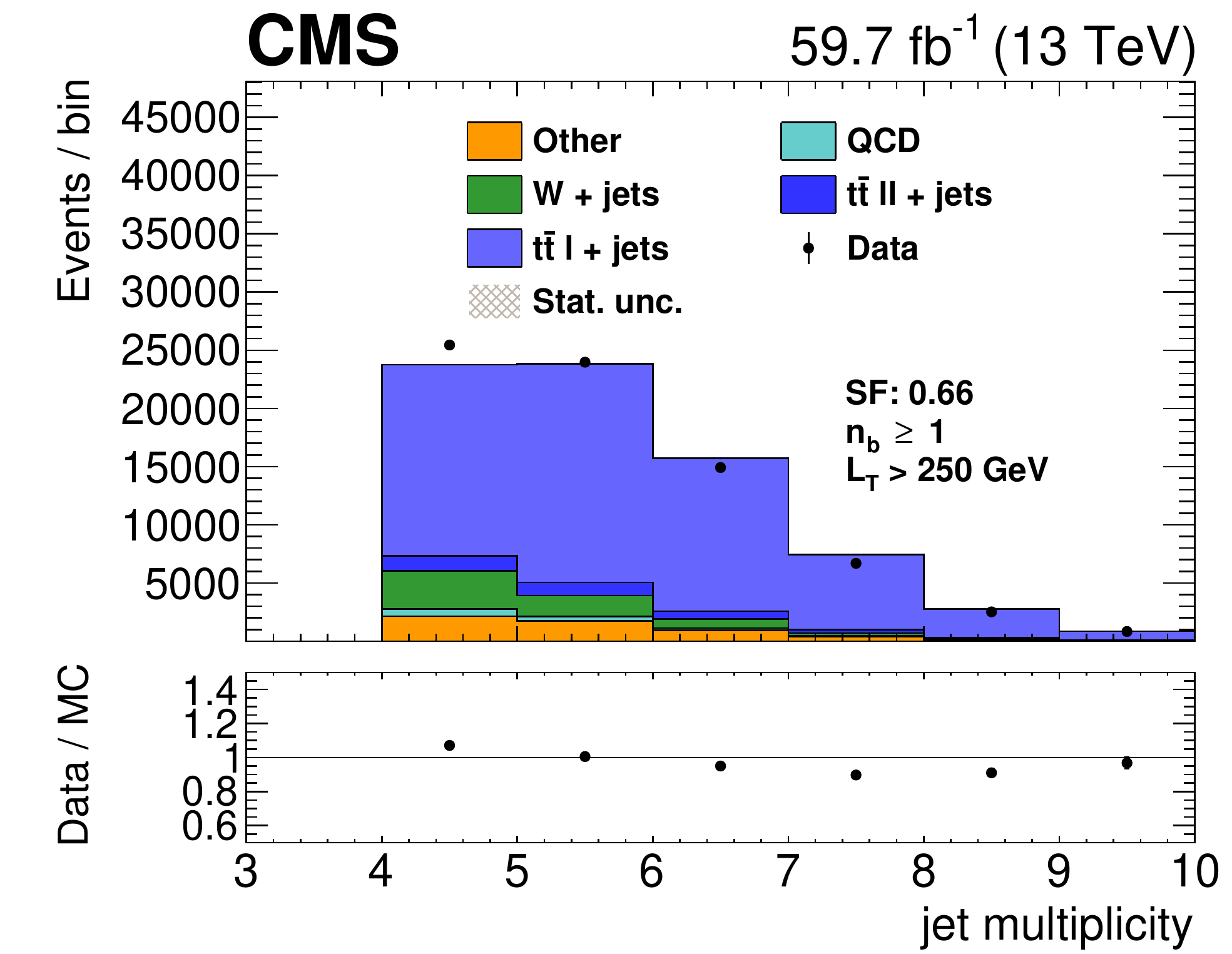}%
\hspace{.05\textwidth}%
\includegraphics[width=.4\textwidth]{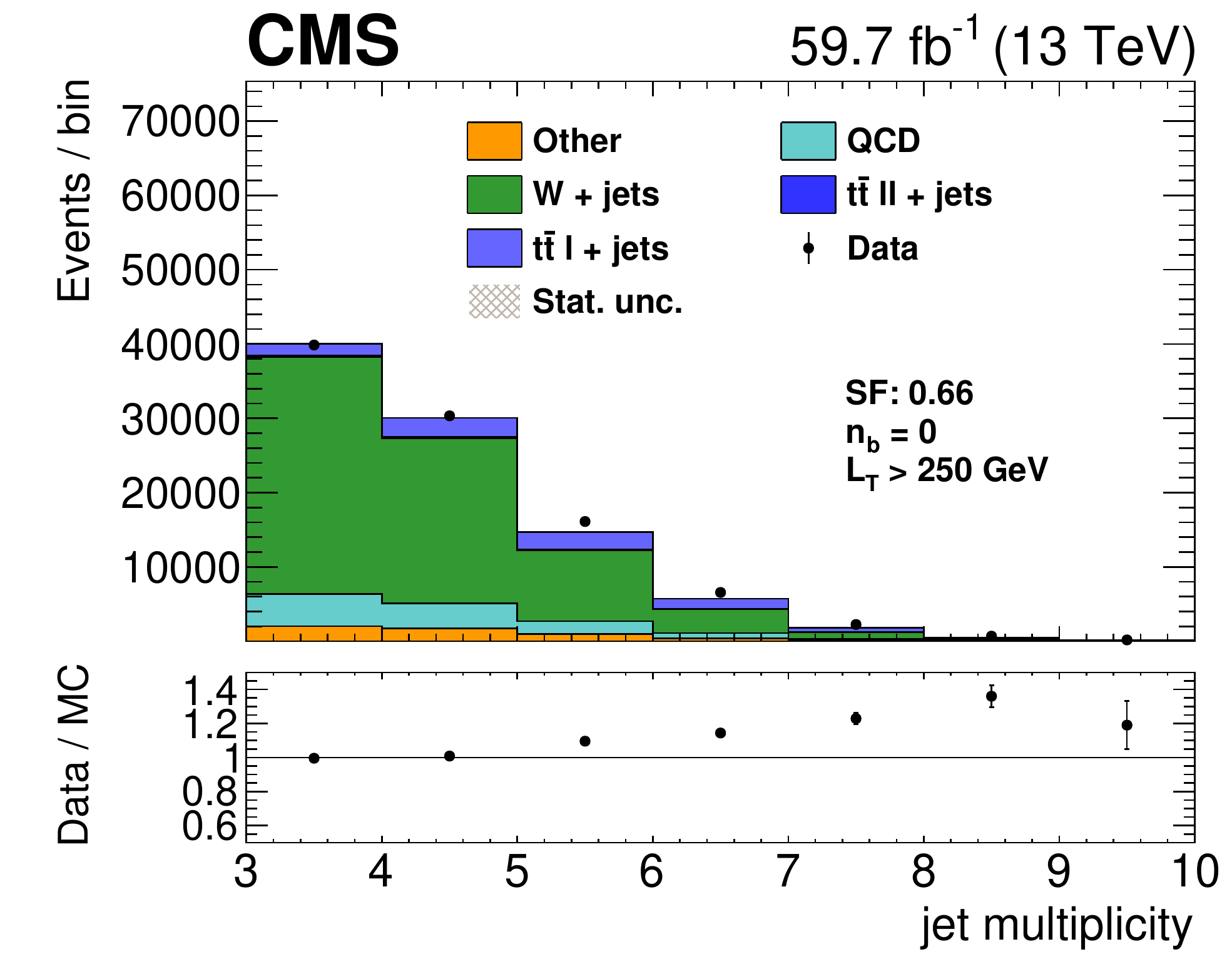}\\
\includegraphics[width=.4\textwidth]{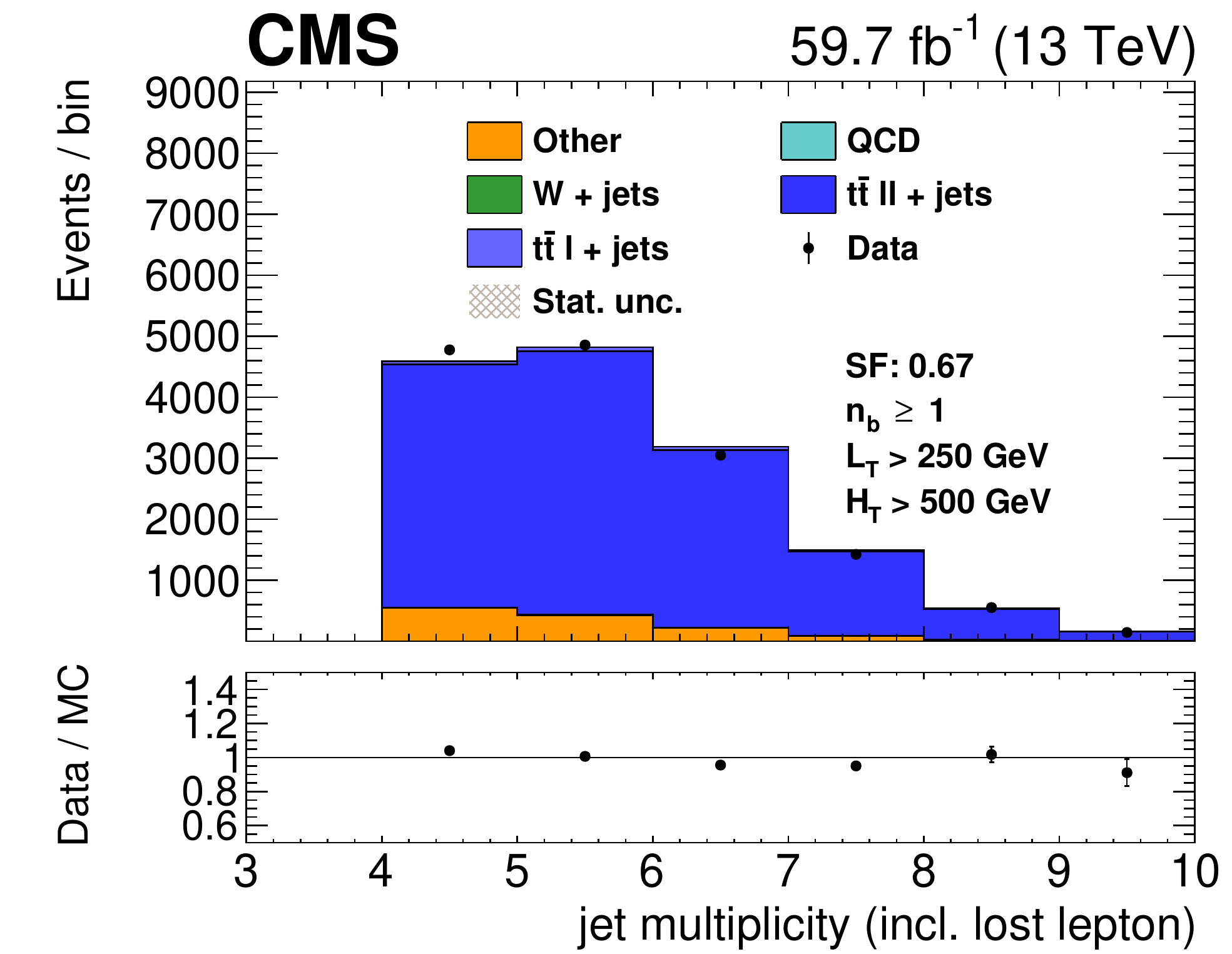}%
\hspace{.05\textwidth}%
\includegraphics[width=.4\textwidth]{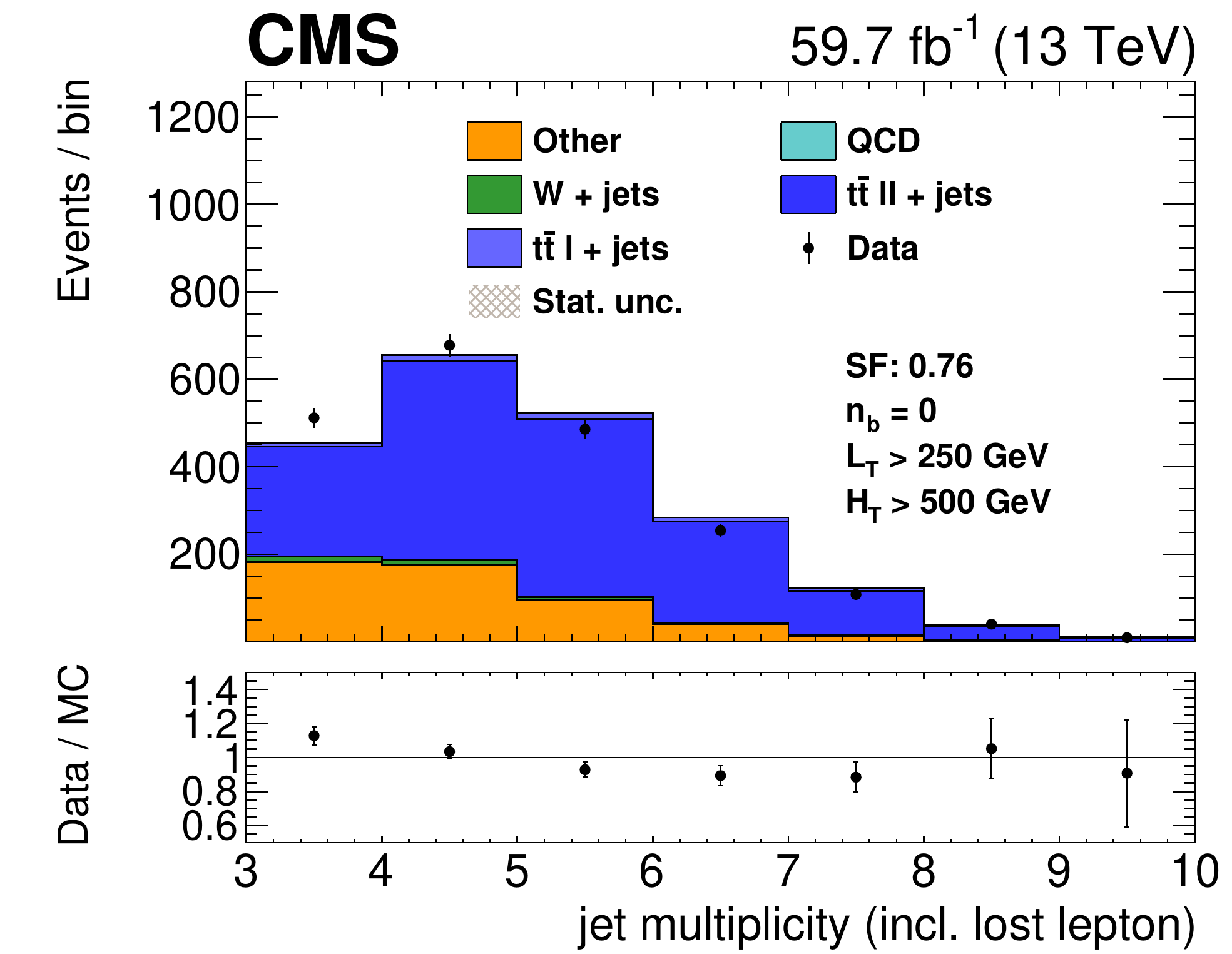}\\
\includegraphics[width=.4\textwidth]{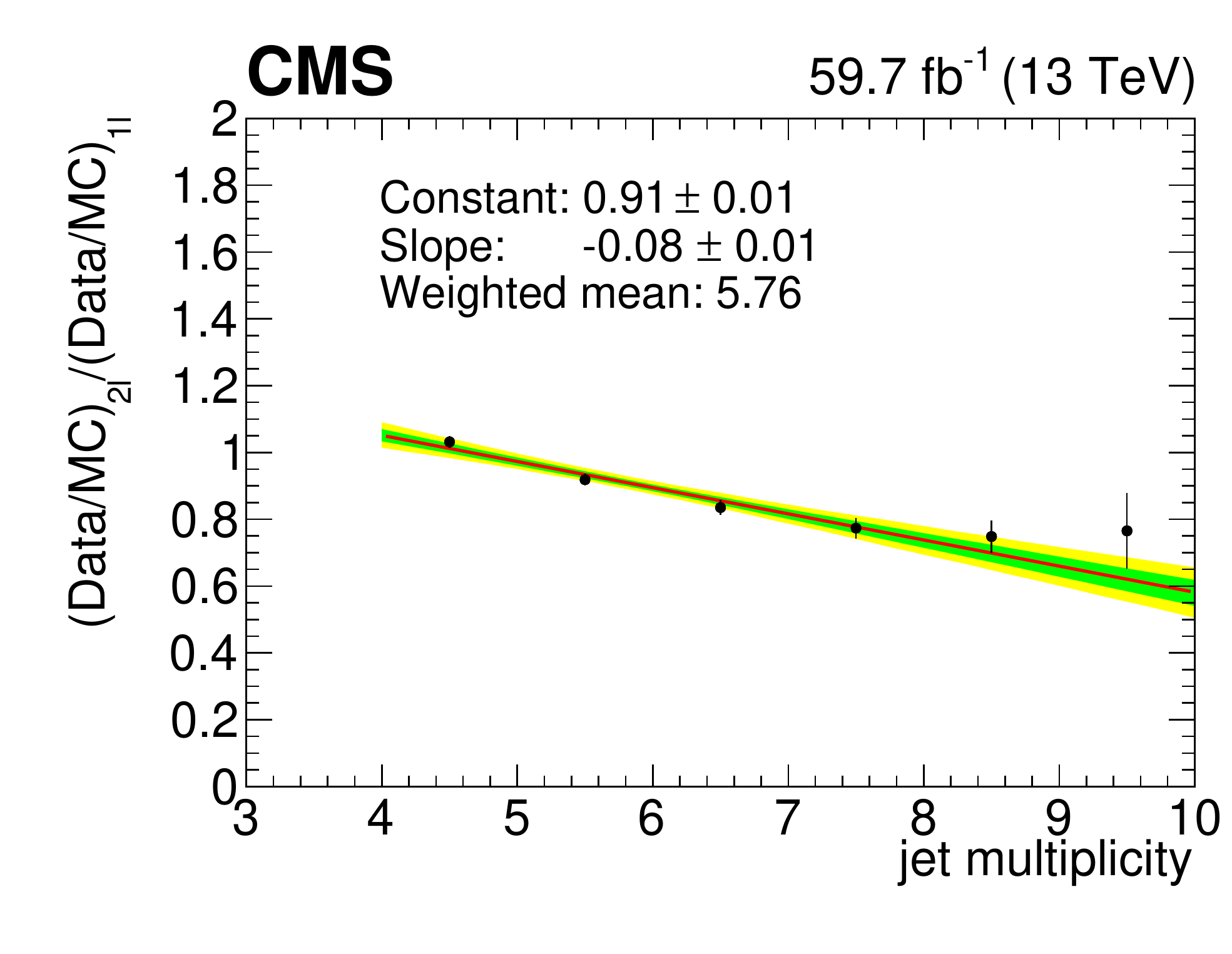}%
\hspace{.05\textwidth}%
\includegraphics[width=.4\textwidth]{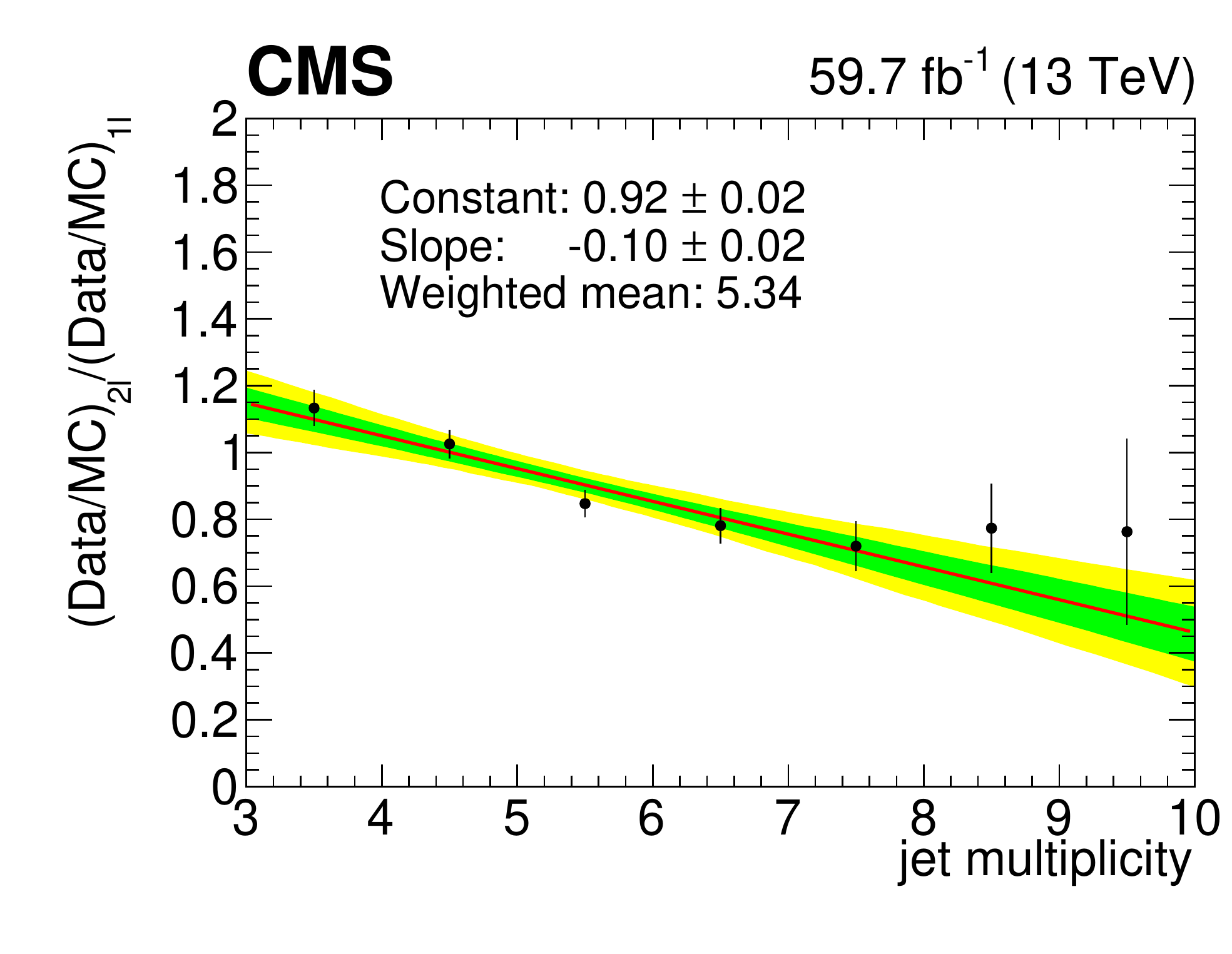}
\caption{The upper row shows the jet multiplicity distribution after the single-lepton baseline selection excluding the SRs for the \multib analysis (left) and for the \zerob analysis (right).
The middle row contains the dilepton CRs, again for the \multib analysis (left) and for the \zerob analysis (right).
The simulation is normalized to data with the SF mentioned in the plot.
The double ratio of the single-lepton and dilepton ratio between data and simulation together with fit results and their uncertainties is shown in the lower row for the \multib (left) and the \zerob (right) analysis.
The fits are performed for each data-taking year; 2018 is shown as an example.}
\label{fig:dilepton_njet}
\end{figure}

As an example, the evaluation using 2018 data is shown in Fig.~\ref{fig:dilepton_njet}.
The jet multiplicity distributions are separately shown for single-lepton (after the single-lepton baseline selection but excluding the SRs) and for dilepton events, both for the \multib and the \zerob analysis.
Additionally, the fitted double ratio is provided.

The uncertainties in the templates are considered as a source of systematic uncertainty in the analysis and are discussed in Section~\ref{sec:systematics}.

\subsection{Estimation of QCD multijet background}
\label{sec:qcd}

The QCD multijet events that pass the event selection typically have a reconstructed electron that originates from misidentified jets or from photon conversions in the inner detector.
This background contribution is estimated from the yield of ``anti-selected'' electron candidates that pass looser identification and isolation requirements and fail the tighter criteria for selected electrons.
Muons contribute naturally much less to this background, but are nevertheless studied in the same way.
The transfer factor \Fratio from the anti-selected to the selected lepton sample is extracted in a multijet-enriched control sample with zero \PQb-tagged jets and three or four other jets and therefore fewer prompt leptons.

\begin{figure}[!htb]
\centering
\includegraphics[width=.45\textwidth]{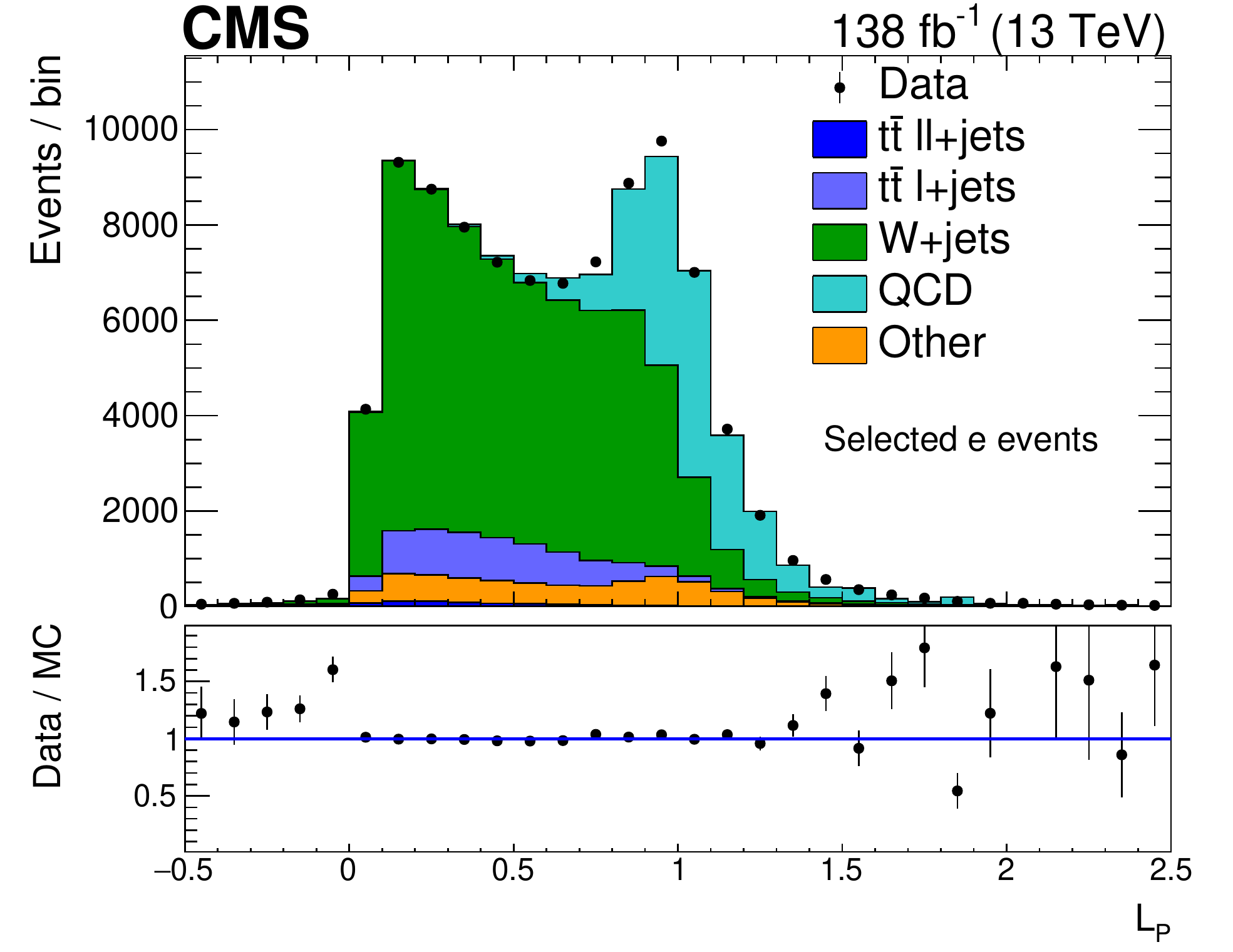}%
\hspace{.05\textwidth}%
\includegraphics[width=.45\textwidth]{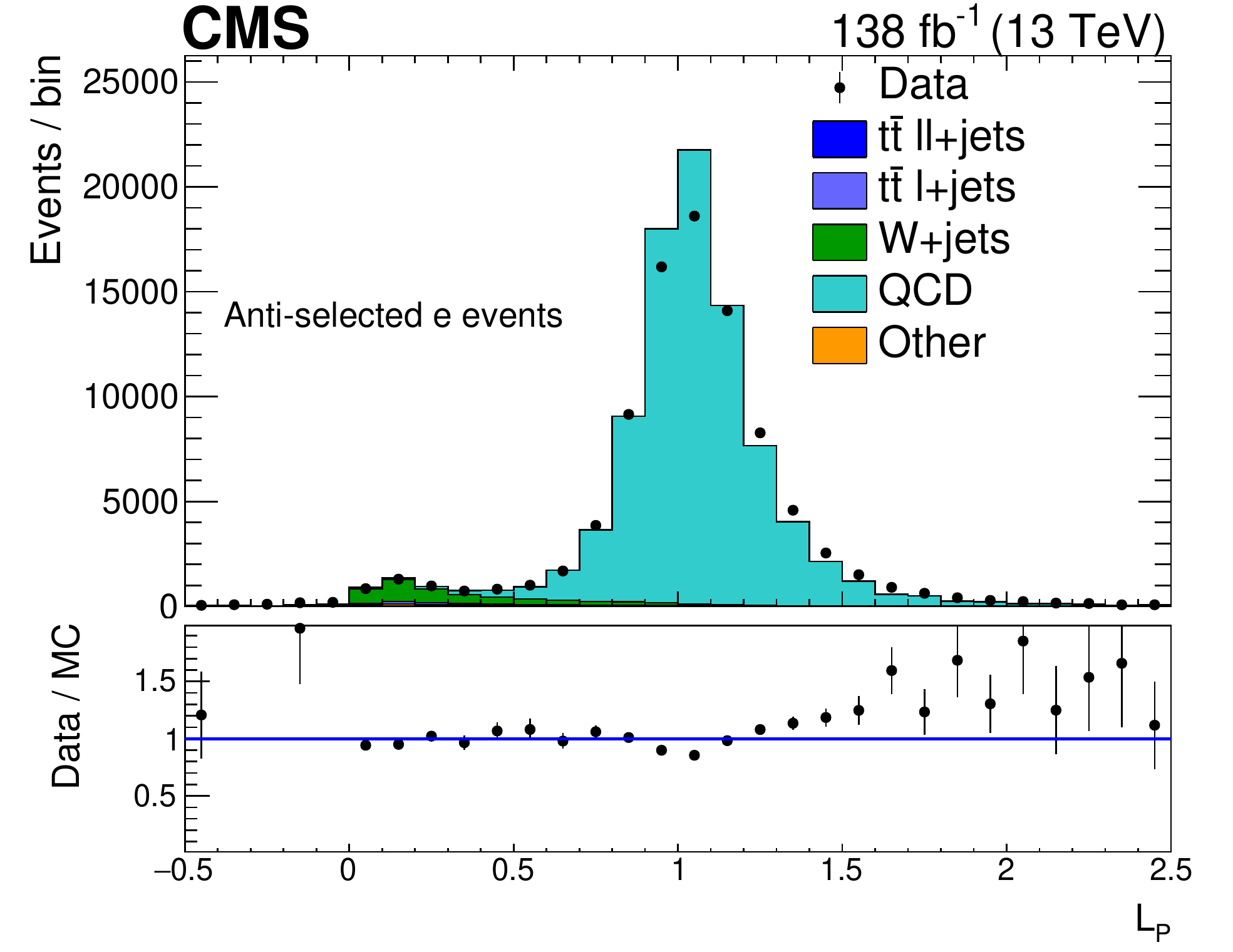}
\caption{The prefit \Lp distribution for selected (left) and anti-selected (right) electron candidates in the baseline QCD selection, with modified requirements of \threefourjets and \zerobtag.}
\label{fig:QCD}
\end{figure}

The estimation method applied here is very similar to the procedure developed in previous CMS analyses~\cite{CMS:2011kaj, CMS:2012vfw}.
It relies on the \Lp variable, which reflects the effective lepton polarization in the \PW decay, defined as:
\begin{linenomath}\begin{equation}
    \Lp = \frac{\pt^{\Pell}}{\pt^{\PW}} \cos(\DF).
\end{equation}\end{linenomath}
Here, \DF is again the angle between the transverse components of the momenta of lepton and reconstructed \PW boson, as defined in Eq.~\eqref{eq:dphi}.
According to the simulation, the selected lepton events comprise a mixture of EW and QCD backgrounds.
In contrast, the anti-selected electron events are clearly dominated by QCD background, as intended by the modified electron identification requirements.
As shown in Fig.~\ref{fig:QCD} for the electron channel, the EW background peaks around $\Lp=0$ and falls off towards higher values of \Lp, while the QCD background peaks around unity.
There is disagreement between the observed data and MC in the region of high \Lp, however the effect of this disagreement will be negligible for the fit to estimate the QCD yields.

Therefore, the number of EW and QCD events can be determined with template fits in \Lp to the selected and to the anti-selected lepton candidates.
The shape of the templates is taken from the corresponding simulated samples.
The ratio of QCD events in selected to anti-selected lepton events is then determined from data requiring zero \PQb tags and three or four jets:
\begin{linenomath}\begin{equation}\label{eq:fratio}
    \Fratio \big(\LT, \zerobtag, \threefourjets\big) = \frac{
        \Nsel^{\text{fit, data}} \big(\LT, \zerobtag, \threefourjets\big)
    }{
        \Nanti^{\text{fit, data}} \big(\LT, \zerobtag, \threefourjets\big)
    }\;.
\end{equation}\end{linenomath}
This ratio is calculated in bins of \LT, but inclusively in \HT, since the probability to misidentify jets as electrons is expected to be independent on the number of jets and \HT.
Typically, the \Fratio varies between 0.2 for smaller and 0.3 for large values of \LT.
This ratio is finally used to predict the QCD background for the SR bins with higher jet or \PQb-jet multiplicities bin by bin:
\begin{linenomath}\begin{equation}\label{eq:qcdpred}
    \Nsel^{\text{Pred}} \big(\njet, \nbtag\big) = \Fratio \big(\LT, \zerobtag, \threefourjets\big) \Nanti^{\text{data}} \big(\njet, \nbtag\big).
\end{equation}\end{linenomath}

\section{Systematic uncertainties}
\label{sec:systematics}

Our search results are subject to various systematic uncertainty sources related to the experimental apparatus and theoretical models.
The uncertainties can influence the background estimations and/or modify the signal predictions.
The impact of the uncertainties is evaluated individually for the \multib and the \zerob analysis, and also separately for the \ttbar and \Wjets background predictions in the \zerob analysis.
This is done by varying the yields of the MC simulation used to calculate the correction factors \kappaew or \kappab, \kappatt and \kappaW, for various uncertainty sources split by year.
The results are summarized in Tables~\ref{tab:sysTableSummary_mb} and \ref{tab:sysTableSummary_0b}, respectively.
In addition, the impact of uncertainties on the yield predictions for two representative signal points for each analysis are shown in Tables~\ref{tab:sysTableSummary_mb_sig} and \ref{tab:sysTableSummary_0b_sig}. For the \multib analysis, the \ptmiss uncertainty has a very high maximum value for T1tttt(2.2, 0.1) in one bin with low sensitivity to the signal.

\begin{table}[!ht]
\centering\renewcommand\arraystretch{1.1}
\topcaption{Summary of systematic uncertainties in the background prediction for the \multib analysis.
For each uncertainty source, the median, minimal (min), and maximal (max) impact on the total background prediction is shown in order of decreasing importance, where these quantities refer to the set of MB SR bins.}
\begin{tabular}{ccccccc}\hline
\multirow{2}{*}{Uncertainty source} & Total background \\[-1pt]
                                 & \medminmax      \\ \hline
Jet energy corrections           & 3.8 [0.2, 36.3] \\
QCD multijet                     & 3.8 [0.8, 71.0] \\
\ttV cross sections              & 2.8 [0.1, 22.6] \\
ISR modeling                     & 2.3 [0.4, 20.3] \\
Pileup modeling                  & 2.3 [0.1, 18.6] \\
Dileptonic correction            & 2.2 [0.4, 12.3] \\
\ttbar cross section             & 1.6 [0.1, 23.7] \\
\Wjets polarization              & 0.6 [0.1, 4.4]  \\
\PQb tagging (efficiency)        & 0.6 [0.1, 5.7]  \\
\Wjets cross section             & 0.4 [0.1, 7.7]  \\
\PQb tagging (misidentification) & 0.3 [0.1, 8.4]  \\
Lepton efficiency                & 0.2 [0.1, 1.6]  \\ \hline
\end{tabular}
\label{tab:sysTableSummary_mb}
\end{table}

\begin{table}[!ht]
\centering\renewcommand\arraystretch{1.1}
\topcaption{Summary of systematic uncertainties in the background prediction for the \zerob analysis.
For each uncertainty source, the median, minimal (min), and maximal (max) impact on the \ttbar, \Wjets, and total background prediction is shown in order of decreasing importance for the total background, where these quantities refer to the set of MB SR bins.}
\cmsTable{\begin{tabular}{ccccccc}\hline
\multirow{2}{*}{Uncertainty source} & \ttbar & \Wjets & Total background \\[-1pt]
                                 & \medminmax      & \medminmax      & \medminmax      \\ \hline
QCD multijet                     & \NA             & \NA             & 5.2 [1.5, 27.6] \\
\ttV cross sections              & 0.9 [0.2, 5.3]  & 0.3 [0.1, 2.1]  & 4.0 [1.0, 19.6] \\
Jet energy corrections           & 1.4 [0.1, 34.4] & 1.2 [0.1, 22.0] & 3.5 [0.5, 40.5] \\
Pileup modeling                  & 0.5 [0.1, 5.5]  & 0.6 [0.1, 4.8]  & 1.2 [0.1, 13.1] \\
Dileptonic correction            & 2.0 [0.2, 13.7] & 0.1 [0.1, 0.9]  & 0.8 [0.1, 4.7]  \\
\Wjets cross section             & 0.6 [0.1, 2.6]  & 1.5 [0.1, 13.7] & 0.7 [0.1, 4.5]  \\
\PQb tagging (efficiency)        & 0.3 [0.1, 2.7]  & 0.1 [0.1, 1.8]  & 0.6 [0.2, 4.6]  \\
\Wjets polarization              & 0.2 [0.1, 2.9]  & 0.8 [0.1, 7.6]  & 0.4 [0.1, 4.1]  \\
Lepton efficiency                & 0.1 [0.1, 1.4]  & 0.1 [0.1, 1.6]  & 0.4 [0.1, 2.3]  \\
\ttbar cross section             & 1.3 [0.1, 10.3] & \NA             & 0.3 [0.1, 3.2]  \\
Integrated luminosity            & \NA             & \NA             & 0.3 [0.1, 1.0]  \\
ISR modeling                     & 0.5 [0.1, 14.1] & \NA             & 0.1 [0.1, 4.4]  \\
\PQb tagging (misidentification) & 0.1 [0.1, 0.5]  & 0.1 [0.1, 0.3]  & 0.1 [0.1, 0.7]  \\ \hline
\end{tabular}}
\label{tab:sysTableSummary_0b}
\end{table}

\begin{table}[!ht]
\centering\renewcommand\arraystretch{1.1}
\topcaption{Summary of the main systematic uncertainties in the signal prediction for the \multib analysis, for two representative combinations of (gluino, neutralino) masses with large (2.2, 0.1)\TeV and small (1.8, 1.3)\TeV mass differences.
    For each uncertainty source, the median, minimal (min), and maximal (max) impact on the total background prediction is shown in order of decreasing importance for the T1tttt(1.8, 1.3)\TeV signal, where these quantities refer to the set of MB SR bins.}
\begin{tabular}{ccccccc}\hline
\multirow{2}{*}{Uncertainty source} & T1tttt(1.8, 1.3)\TeV & T1tttt(2.2, 0.1)\TeV \\[-1pt]
                                 & \medminmax        & \medminmax        \\ \hline
\PQt tagging                     & 10.0 [10.0, 10.0] & 10.0 [10.0, 10.0] \\
\ptmiss                          & 8.2 [1.3, 40.8 ]  & 1.6 [0.1, 61.2 ]     \\
Jet energy corrections           & 7.8 [0.1, 53.7 ]  & 5.2 [0.1, 50.0 ]     \\
\PQb tagging (efficiency)        & 5.1[0.1, 19.9]    & 6.5[0.1, 26.9]       \\
ISR modeling                     & 4.8 [0.1, 17.8 ]  & 7.0 [0.8, 30.8 ]     \\
Integrated luminosity            & 1.6 [1.6, 1.6]    & 1.6 [1.6, 1.6]       \\
\PQb tagging (misidentification) & 0.5 [0.1, 2.1 ]   & 0.4 [0.1, 7.2 ]      \\
\hline
\end{tabular}
\label{tab:sysTableSummary_mb_sig}
\end{table}

\begin{table}[!ht]
\centering\renewcommand\arraystretch{1.1}
\topcaption{Summary of the main systematic uncertainties in the signal prediction for the \zerob analysis, for two representative combinations of (gluino, neutralino) masses with large (2.2, 0.1)\TeV and small (1.8, 1.3)\TeV mass differences.
    For each uncertainty source, the median, minimal (min), and maximal (max) impact on the total background prediction is shown in order of decreasing importance for the T5qqqqWW(1.8, 1.3)\TeV signal, where these quantities refer to the set of MB SR bins.}
\begin{tabular}{ccccccc}\hline
\multirow{2}{*}{Uncertainty source} & T5qqqqWW(1.8, 1.3)\TeV & T5qqqqWW(2.2, 0.1)\TeV \\[-1pt]
                                 & \medminmax        & \medminmax        \\ \hline
\PW boson tagging                & 10.0 [10.0, 10.0] & 10.0 [10.0, 10.0] \\
Jet energy corrections           & 5.3 [0.8, 39.8]   & 5.0 [0.3, 50.0]   \\
\ptmiss                          & 4.4 [0.4, 23.1]   & 4.2 [0.1, 43.2]   \\
ISR modeling                     & 1.6 [0.1, 11.2]   & 1.6 [0.1, 15.0]   \\
Integrated luminosity            & 1.6 [1.6, 1.6]    & 1.6 [1.6, 1.6]    \\
\PQb tagging (efficiency)        & 0.5 [0.2, 5.9]    & 1.3 [0.2, 5.9]    \\
\PQb tagging (misidentification) & 0.3 [0.1, 0.9]    & 0.4 [0.1, 1.1]    \\ \hline
\end{tabular}
\label{tab:sysTableSummary_0b_sig}
\end{table}

One common large systematic uncertainty is given by the jet energy corrections, which are varied within their uncertainty~\cite{CMS:2016lmd} as a function of jet \pt and $\eta$, and these changes are propagated to all observables.
The SF related to the efficiencies for identifying leptons as well as \PQb quark jets, and the misidentification of the \PQc quark, light quark, or gluon jets are scaled up and down according to their uncertainties in the efficiency.
The uncertainty in the pileup is determined by varying the inelastic \pp cross-section of 69\unit{mb} by $\pm$5\%~\cite{CMS:2018mlc}.
All these uncertainties apply to both the background prediction and the signal yield.

The integrated luminosities of the 2016, 2017, and 2018 data-taking periods are individually measured with uncertainties in the 1.2--2.5\% range~\cite{CMS:2021xjt, CMS:2018elu, CMS:2019jhq}, while the total integrated luminosity has an uncertainty of 1.6\%.

Changes in the polarization of \PW bosons can affect the \DF variable. Thus, events are reweighted using the factor $w=1+\alpha(1-\cos\thetast)^2$, where \thetast is
the angle between the charged lepton and \PW boson in the \PW boson rest frame.
In \Wjets events, we take $\alpha$ to be 0.1, guided by the theoretical uncertainty and measurements found in Refs.~\cite{Bern:2011ie, CMS:2015cyj, CMS:2011kaj, ATLAS:2012au}.
For \ttbar{}+jets events, we take $\alpha=0.05$~\cite{ATLAS:2012nhi, ATLAS:2016ygv, CMS:2015cyp, ATLAS:2016fbc}.
For \Wjets events, where the initial state can have different polarizations for \PWp and \PWm bosons, we take as uncertainty the larger change in $\kappa$ resulting from reweighting only the \PWp bosons in the sample, and from reweighting all \PW bosons.

While the \Wjets and \ttbar backgrounds are estimated from data, a change in their relative contribution can lead to changes in the \Rcs at low jet multiplicities of the SB.
Therefore, the inclusive \Wjets and \ttbar cross sections are varied by 30\% above and below the nominal value~\cite{CMS:2017xrt} to account for possible biases in the estimation of the background composition, which only affects the calculation of the $\kappa$ factors.
The small contribution of \ttbar produced with an additional vector boson (\ttV) is varied by 100\% to account for the uncertainty in the theoretical prediction.
Uncertainties in the signal cross section are shown as explicit variations of the mass limits (Section~\ref{sec:int}).

The QCD uncertainty includes the statistical uncertainties in the anti-selected region.
As \Fratio is calculated for events with 3--4 jets, we apply an additional relative uncertainty that is larger for SR bins with higher jet multiplicities.
An uncertainty of 30\% for events with 6--8 jets, or 50\% for events with least 9 jets in the \multib analysis is applied on the QCD background estimate.
For the \zerob analysis, we take an uncertainty of 15\% for events with 5 jets, 30\% for events with 6--7 jets, and 50\% for events with at least 8 jets.

Since we consider a signal with multiple top quarks (\PW bosons) in the \multib (\zerob) analysis, the related tagging uncertainties have been investigated and are found to be consistent with those of Ref.~\cite{CMS:DP-2020-025}.
The taggers are described in detail in Section~\ref{sec:obj}.
The background estimation is not sensitive to details of the \PQt and \PW boson tagging performance, therefore a systematic uncertainty is only assigned for the signal efficiency.
The systematic uncertainties in the \PQt tagging efficiency and mistagging rate are estimated as follows:
The relative yields of events with different \PQt tag multiplicities are used to extract an overall efficiency and mistagging rate.
A difference of 5\% is observed, and the systematic uncertainty is taken to be twice this quantity, namely 10\%.
For \PW boson tagging, the efficiency and mistagging rates are extracted from a full comparison of data and simulation.
A total uncertainty of 10\% is found to account for all differences between data and simulation.

The SF applied to correct the ISR in signal samples and 2016 \ttbar, with values typically around 0.94, is varied by 4--5\% (Section~\ref{sec:simulation}).

Because the signal samples use fast simulation, an additional \ptmiss uncertainty is taken into account.
The analysis is performed twice, once using the generated and once using the reconstructed \ptmiss for signal events.
A flat uncertainty equal to one-half the difference between the acceptances is applied.

Lastly, systematic uncertainties related to the dileptonic correction explained in Section~\ref{sec:dileptonnjet} have to be taken into account.
The systematic variations around that new central value are determined by varying the fit to the double ratio by the following uncertainties: the variation of the constant value $a$ is extracted as the quadratic sum of the deviation of the central value of $a$ from unity and by the uncertainty in $a$ that is extracted from the fit itself.
The variation of the slope $b$ is determined as the quadratic sum of the deviation of $b$ from zero and the uncertainty in $b$ as given by the fit.

\section{Results}

\begin{figure}[!htb]
\centering
\includegraphics[width=\textwidth]{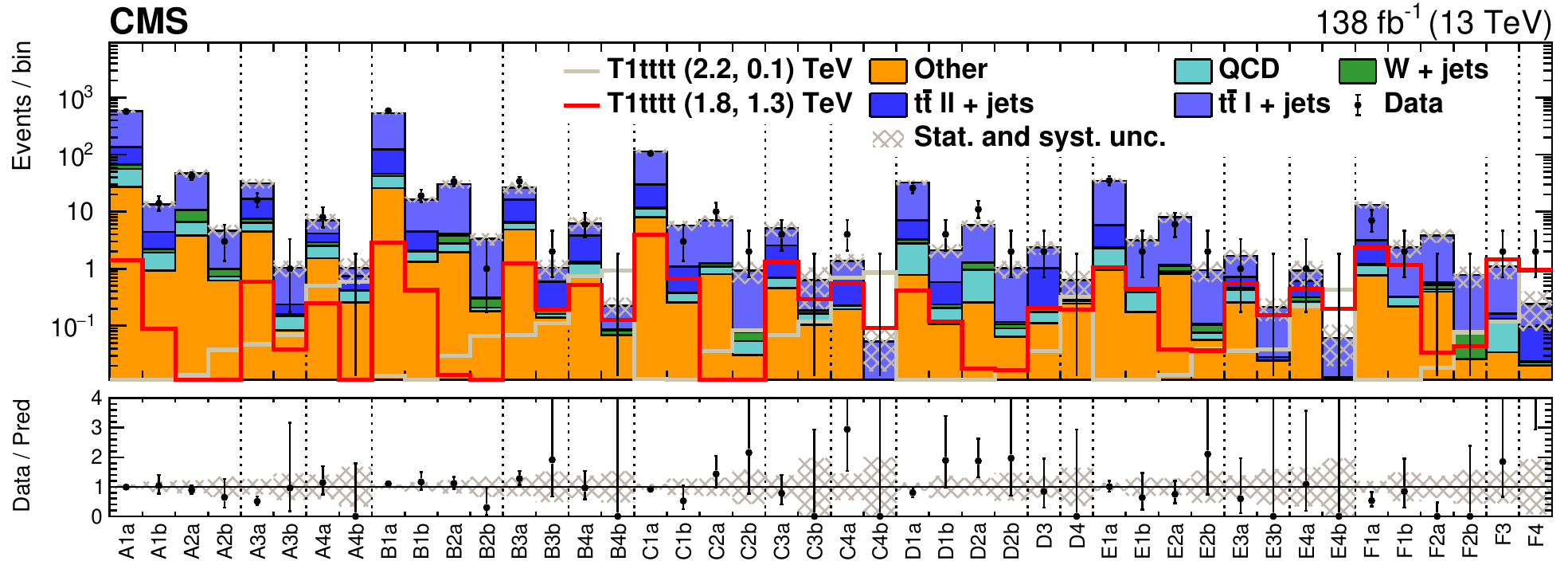}
\caption{Observed event yields in the MB SRs of the \multib analysis compared to signal and background predictions.
The relative fraction of the different SM EW background contributions determined in simulation is shown by the stacked, colored histograms, normalized so that their sum is equal to the background estimated using data control regions.
The QCD background is predicted using the \Lp method.
The signal is shown for two representative combinations of (gluino, neutralino) masses with large (2.2, 0.1)\TeV and small (1.8, 1.3)\TeV mass differences.
}
\label{fig:multib_SRMB}
\end{figure}

The observed data yields and the predicted background contributions for the \multib analysis are given in Table~\ref{tab:searchbins_mb} and shown in Fig.~\ref{fig:multib_SRMB} for the MB SR for the combination of all three data-taking years.
Good agreement is observed for almost all search bins, except for the last bin where two events are observed, against an expected value of $0.24 \pm 0.16$.
The observed limit is about one standard deviation lower than expected, because of the number of excess events in the last search bin.

\begin{figure}[!htb]
\centering
\includegraphics[width=\textwidth]{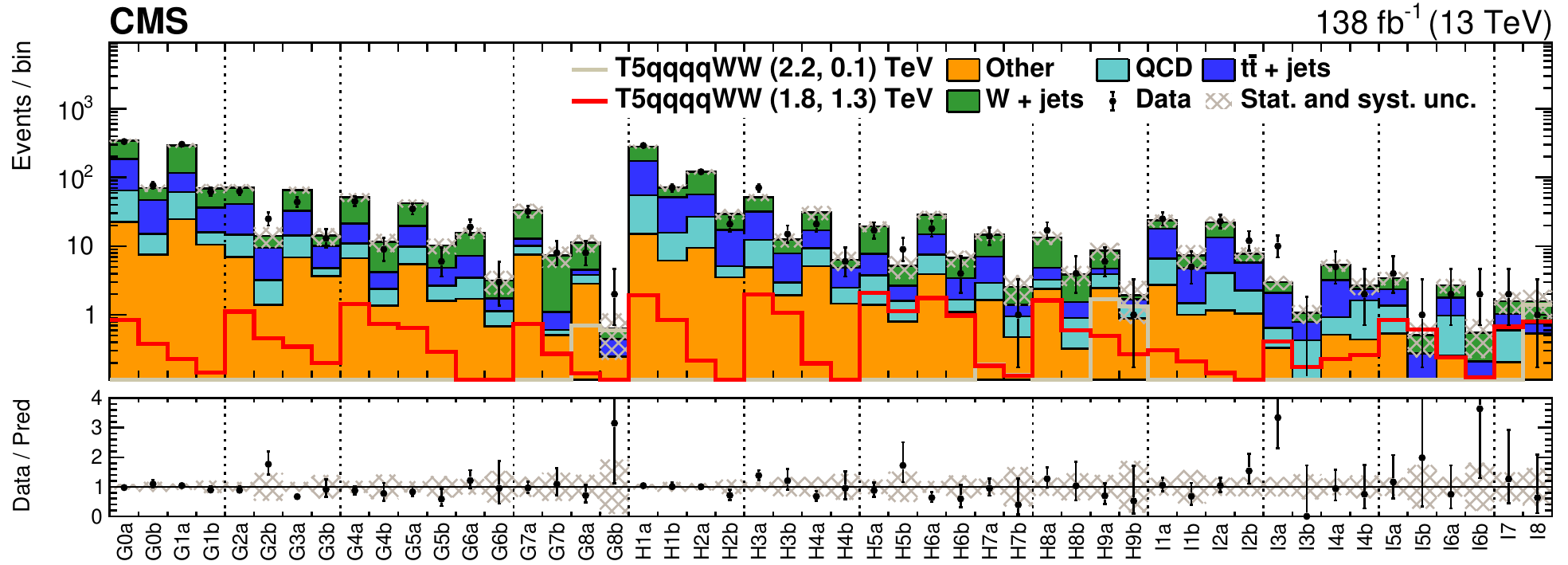}
\caption{Observed event yields in the MB SRs of the \zerob analysis compared to signal and background predictions.
The \Wjets, \ttbar, and QCD predictions are extracted from data control samples, while the other background contributions are estimated from simulation.
The signal is shown for two representative combinations of (gluino, neutralino) masses with large (2.2, 0.1)\TeV and small (1.8, 1.3)\TeV mass differences.}
\label{fig:SR_MB_zerob}
\end{figure}

Table~\ref{tab:searchbins_zb} and Fig.~\ref{fig:SR_MB_zerob} show the background predictions for the \zerob analysis and the data yields in the MB SR.
Here, we observe good agreement in almost all bins as well, and a deviation from the prediction is observed only for bins that are dominated by the background.
Namely, the bins G2b, H3a, and I3a, among the 50 \zerob bins, contain more observed events than expected by at least one standard deviation.

\section{Interpretation}
\label{sec:int}

To evaluate exclusion limits on the simplified SUSY models considered in this search, a likelihood function is defined that includes all SB and MB CRs and SRs.
For the \multib analysis, the measurement is performed for each set of four bins (SB CR, SB SR, MB CR and MB SR), as defined for the \Rcs method in Section~\ref{sec:background_mb}.
The number of entries in the bins are modeled by a product of Poisson distributions.
The \Rcs method for the background components of the Poisson mean parameters for the SRs is implemented by imposing the constraints given by Eq.~\eqref{eq:prediction} for the \multib analysis, or Eqs.~\eqref{eq:NMBSRpredtt} and \eqref{eq:NMBSRpredw} for the \zerob analysis.
Multijet background and rare processes are treated as independent contributions.
The systematic uncertainties enter the likelihood as nuisance parameters and are taken into account as a product of log-normal distributions.
For the \zerob analysis, the bins are defined in Section~\ref{sec:background_mb}.
The template fit on the \nbtag distributions is performed separately.
Thus, only the SB' CR, SB' SR, SB SR and MB SR search bins enter the likelihood.
The different constraints on the rate parameters implied by the extended \Rcs method are taken into account in a similar way and the systematic uncertainties are again modeled by log-normal distributions.
We set upper limits on the production cross section at 95\% confidence level (\CL).
These are estimated with the modified frequentist \CLs method~\cite{Junk:1999kv, Read:2002hq} using the asymptotic approximation~\cite{Cowan:2010js}.
The \CLs method is used with the test statistic $q_\mu = -2\ln\lambda_\mu$, where $\lambda_\mu$ refers to the ratio of the maximized likelihood for a given signal strength $\mu$ to the unconditional likelihood maximized for all parameters, including $\mu$.
The $\PSg\PSg$ pair production cross section is calculated at approximate NNLO and NNLL accuracy, and exclusion limits are set as a function of the (\mGlu, \mLSP) hypothesis.

\begin{figure}[!htb]
\centering
\includegraphics[width=.495\textwidth]{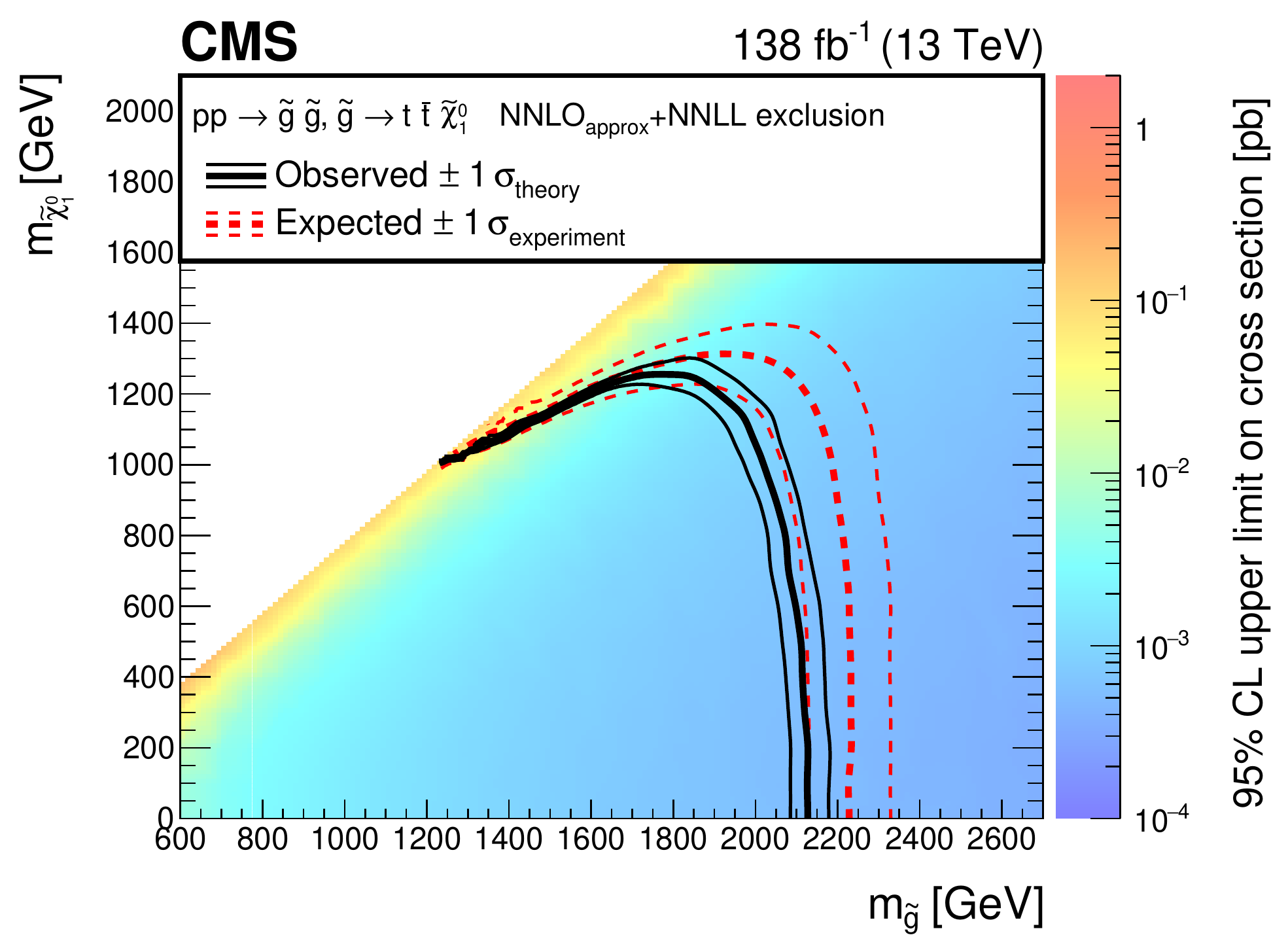}%
\hspace{.0075\textwidth}%
\includegraphics[width=.495\textwidth]{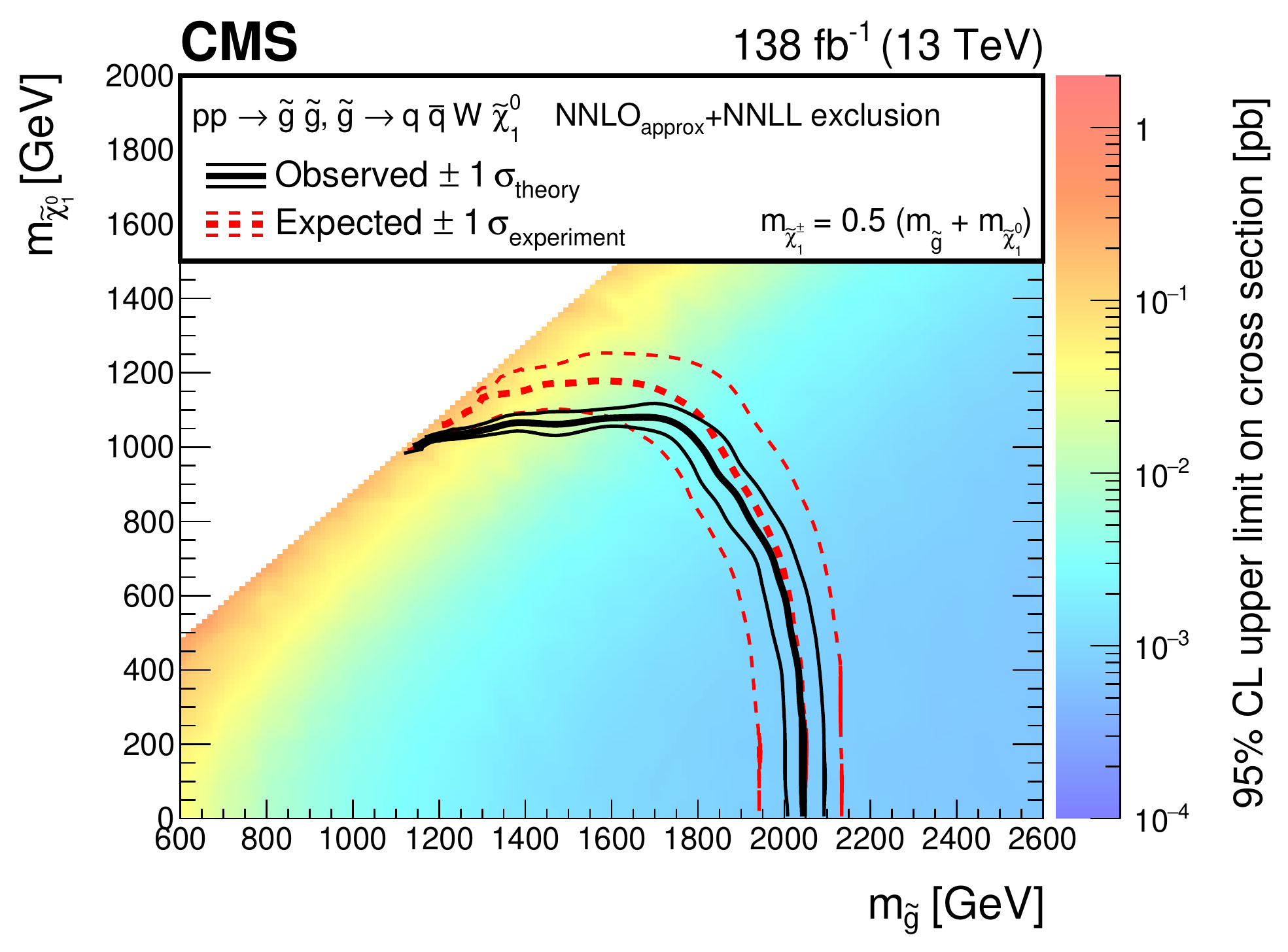}
\caption{Cross section limits at 95\% \CL for the T1tttt (left) and for the T5qqqqWW (right) model, as functions of the gluino and LSP masses, assuming a branching fraction of 100\%.
The mass of the intermediate chargino is taken to be halfway between the gluino and the neutralino masses.
The solid black (dashed red) lines correspond to the observed (expected) mass limits, with the thicker lines representing the central values and the thinner lines representing the $\pm1\sigma$ uncertainty bands related to the theoretical (experimental) uncertainties.}
\label{fig:limits}
\end{figure}

For the T1tttt model, which describes gluino pair production with each gluino decaying to a \ttbar pair and a \PSGczDo, the cross section limits are obtained using the \multib analysis.
They are shown in Fig.~\ref{fig:limits} (left) as functions of \mGlu and
\mLSP, assuming branching fractions of 100\%.
The observed limit is about one standard deviation lower than the expected one, which is caused by the observation of two events in the last bin, while only $0.24 \pm 0.16$ events are expected.

The results of the \zerob analysis are interpreted in the T5qqqqWW model, in which pair-produced gluinos decay to a (light) quark-antiquark pair and a chargino, which further decays to a \PW boson and the \PSGczDo.
The observed limit, shown in Fig.~\ref{fig:limits} (right), agrees with the expected limit over most of the mass range.

\section{Summary}

A search for supersymmetry has been performed using a sample of proton-proton collisions at $\sqrt{s}=13\TeV$ corresponding to an integrated luminosity of 138\fbinv, recorded by the CMS experiment in 2016--2018.
Events with a single charged lepton (electron or muon) and multiple jets are selected.
Top quark and \PW boson identification algorithms based on machine-learning techniques are employed to suppress the main background contributions in the analysis.
Various exclusive search regions are defined that differ in the number of jets, the number of jets identified as stemming from \PQb quarks, the number of hadronically decaying top quarks or \PW bosons, the scalar sum of all jet transverse momenta, and the scalar sum of the missing transverse momentum and the transverse momentum of the lepton.
By targeting final states with one lepton, this analysis represents a search for SUSY complementary to those without any leptons in their final states.

To reduce the main background processes from \ttbar and \Wjets production, the presence of a lepton produced in the leptonic decay of a \PW boson in the event is exploited.
Under the hypothesis that all of the missing transverse momentum in the event originates from the neutrino produced in a leptonic \PW boson decay, the \PW boson momentum is calculated.
The requirement of a large azimuthal angle between the directions of the lepton and of the reconstructed \PW boson decaying leptonically, notably reduces the background contributions.

The event yields observed in data are consistent with the expectations from the SM processes, which are estimated using control samples in data.
Exclusion limits on the supersymmetric particle masses in the context of two simplified models of gluino pair production are evaluated.

For the T1tttt simplified model, where each gluino decays to a top quark-antiquark pair and the lightest neutralino, the excluded gluino masses reach up to 2120\GeV, while the excluded neutralino masses reach up to 1250\GeV.
This result extends the exclusion limit on gluino (neutralino) masses from a previous CMS search~\cite{CMS:2017qth} by about 310 (150)\GeV.

The second simplified model, T5qqqqWW, also targets gluino pair production, but with decays to a light-flavor quark-antiquark pair and a chargino, which decays to a \PW boson and the lightest neutralino.
The chargino mass in this decay channel is assumed to be $m_{\PSGcpmDo}=0.5(\mGlu+\mLSP)$.
The excluded gluino masses reach up to 2050\GeV, while the excluded neutralino masses reach up to 1070\GeV.
This corresponds to an improvement on gluino (neutralino) masses by about 150 (120)\GeV in comparison with the previous result~\cite{CMS:2017qth}.

\begin{acknowledgments}
We congratulate our colleagues in the CERN accelerator departments for the excellent performance of the LHC and thank the technical and administrative staffs at CERN and at other CMS institutes for their contributions to the success of the CMS effort. In addition, we gratefully acknowledge the computing centers and personnel of the Worldwide LHC Computing Grid and other centers for delivering so effectively the computing infrastructure essential to our analyses. Finally, we acknowledge the enduring support for the construction and operation of the LHC, the CMS detector, and the supporting computing infrastructure provided by the following funding agencies: BMBWF and FWF (Austria); FNRS and FWO (Belgium); CNPq, CAPES, FAPERJ, FAPERGS, and FAPESP (Brazil); MES and BNSF (Bulgaria); CERN; CAS, MoST, and NSFC (China); MINCIENCIAS (Colombia); MSES and CSF (Croatia); RIF (Cyprus); SENESCYT (Ecuador); MoER, ERC PUT and ERDF (Estonia); Academy of Finland, MEC, and HIP (Finland); CEA and CNRS/IN2P3 (France); BMBF, DFG, and HGF (Germany); GSRI (Greece); NKFIH (Hungary); DAE and DST (India); IPM (Iran); SFI (Ireland); INFN (Italy); MSIP and NRF (Republic of Korea); MES (Latvia); LAS (Lithuania); MOE and UM (Malaysia); BUAP, CINVESTAV, CONACYT, LNS, SEP, and UASLP-FAI (Mexico); MOS (Montenegro); MBIE (New Zealand); PAEC (Pakistan); MES and NSC (Poland); FCT (Portugal); MESTD (Serbia); MCIN/AEI and PCTI (Spain); MOSTR (Sri Lanka); Swiss Funding Agencies (Switzerland); MST (Taipei); MHESI and NSTDA (Thailand); TUBITAK and TENMAK (Turkey); NASU (Ukraine); STFC (United Kingdom); DOE and NSF (USA).
    
\hyphenation{Rachada-pisek} Individuals have received support from the Marie-Curie program and the European Research Council and Horizon 2020 Grant, contract Nos.\ 675440, 724704, 752730, 758316, 765710, 824093, 884104, and COST Action CA16108 (European Union); the Leventis Foundation; the Alfred P.\ Sloan Foundation; the Alexander von Humboldt Foundation; the Belgian Federal Science Policy Office; the Fonds pour la Formation \`a la Recherche dans l'Industrie et dans l'Agriculture (FRIA-Belgium); the Agentschap voor Innovatie door Wetenschap en Technologie (IWT-Belgium); the F.R.S.-FNRS and FWO (Belgium) under the ``Excellence of Science -- EOS" -- be.h project n.\ 30820817; the Beijing Municipal Science \& Technology Commission, No. Z191100007219010; the Ministry of Education, Youth and Sports (MEYS) of the Czech Republic; the Hellenic Foundation for Research and Innovation (HFRI), Project Number 2288 (Greece); the Deutsche Forschungsgemeinschaft (DFG), under Germany's Excellence Strategy -- EXC 2121 ``Quantum Universe" -- 390833306, and under project number 400140256 - GRK2497; the Hungarian Academy of Sciences, the New National Excellence Program - \'UNKP, the NKFIH research grants K 124845, K 124850, K 128713, K 128786, K 129058, K 131991, K 133046, K 138136, K 143460, K 143477, 2020-2.2.1-ED-2021-00181, and TKP2021-NKTA-64 (Hungary); the Council of Science and Industrial Research, India; the Latvian Council of Science; the Ministry of Education and Science, project no. 2022/WK/14, and the National Science Center, contracts Opus 2021/41/B/ST2/01369 and 2021/43/B/ST2/01552 (Poland); the Funda\c{c}\~ao para a Ci\^encia e a Tecnologia, grant CEECIND/01334/2018 (Portugal); the National Priorities Research Program by Qatar National Research Fund; MCIN/AEI/10.13039/501100011033, ERDF ``a way of making Europe", and the Programa Estatal de Fomento de la Investigaci{\'o}n Cient{\'i}fica y T{\'e}cnica de Excelencia Mar\'{\i}a de Maeztu, grant MDM-2017-0765 and Programa Severo Ochoa del Principado de Asturias (Spain); the Chulalongkorn Academic into Its 2nd Century Project Advancement Project, and the National Science, Research and Innovation Fund via the Program Management Unit for Human Resources \& Institutional Development, Research and Innovation, grant B05F650021 (Thailand); the Kavli Foundation; the Nvidia Corporation; the SuperMicro Corporation; the Welch Foundation, contract C-1845; and the Weston Havens Foundation (USA).
\end{acknowledgments}

\bibliography{auto_generated}

\providecommand{\href}[2]{#2}\begingroup\raggedright\begin{thebibliography}{100}%
\makeatletter
\providecommand{\hrefCMSnoop }[0]{\@secondoftwo}%
\makeatother
\providecommand{\doi}{\texttt{doi:}\begingroup \urlstyle{tt}\Url}

\bibitem{Wess:1973kz}
\hrefCMSnoop {}{J.~Wess and B.~Zumino, ``A {Lagrangian} model invariant under
  supergauge transformations'',} \textit{ Phys. Lett. B} \textbf{ 49} (1974)
  52,
  \href{http://dx.doi.org/10.1016/0370-2693(74)90578-4}{\doi{10.1016/0370-2693(74)90578-4}}.

\bibitem{Fayet:1976cr}
\hrefCMSnoop {}{P.~Fayet and S.~Ferrara, ``Supersymmetry'',} \textit{ Phys.
  Rept.} \textbf{ 32} (1977) 249,
  \href{http://dx.doi.org/10.1016/0370-1573(77)90066-7}{\doi{10.1016/0370-1573(77)90066-7}}.

\bibitem{Barbieri:1982eh}
\hrefCMSnoop {}{R.~Barbieri, S.~Ferrara, and C.~A. Savoy, ``Gauge models with
  spontaneously broken local supersymmetry'',} \textit{ Phys. Lett. B} \textbf{
  119} (1982) 343,
  \href{http://dx.doi.org/10.1016/0370-2693(82)90685-2}{\doi{10.1016/0370-2693(82)90685-2}}.

\bibitem{Nilles:1983ge}
\hrefCMSnoop {}{H.~P. Nilles, ``Supersymmetry, supergravity and particle
  physics'',} \textit{ Phys. Rept.} \textbf{ 110} (1984) 1,
  \href{http://dx.doi.org/10.1016/0370-1573(84)90008-5}{\doi{10.1016/0370-1573(84)90008-5}}.

\bibitem{Haber:1984rc}
\hrefCMSnoop {}{H.~E. Haber and G.~L. Kane, ``The search for supersymmetry:
  probing physics beyond the standard model'',} \textit{ Phys. Rept.} \textbf{
  117} (1985) 75,
  \href{http://dx.doi.org/10.1016/0370-1573(85)90051-1}{\doi{10.1016/0370-1573(85)90051-1}}.

\bibitem{Martin:1997ns}
\hrefCMSnoop {}{S.~P. Martin, ``A supersymmetry primer'',} \textit{ Adv. Ser.
  Direct. High Energy Phys.} \textbf{ 21} (2010) 1,
  \href{http://dx.doi.org/10.1142/9789814307505_0001}{\doi{10.1142/9789814307505_0001}},
  \href{http://www.arXiv.org/abs/hep-ph/9709356}{\texttt{arXiv:hep-ph/9709356}}.

\bibitem{Farrar:1978xj}
\hrefCMSnoop {}{G.~R. Farrar and P.~Fayet, ``Phenomenology of the production,
  decay, and detection of new hadronic states associated with supersymmetry'',}
  \textit{ Phys. Lett. B} \textbf{ 76} (1978) 575,
  \href{http://dx.doi.org/10.1016/0370-2693(78)90858-4}{\doi{10.1016/0370-2693(78)90858-4}}.

\bibitem{Arkani-Hamed:2007gys}
N.~Arkani-Hamed\hrefCMSnoop {}{ {et~al.}, ``{MARMOSET}: The path from {LHC}
  data to the new standard model via on-shell effective theories'',} 2007.
  \href{http://www.arXiv.org/abs/hep-ph/0703088}{\texttt{arXiv:hep-ph/0703088}}.

\bibitem{Alwall:2008ag}
\hrefCMSnoop {}{J.~Alwall, P.~Schuster, and N.~Toro, ``Simplified models for a
  first characterization of new physics at the {LHC}'',} \textit{ Phys. Rev. D}
  \textbf{ 79} (2009) 075020,
  \href{http://dx.doi.org/10.1103/PhysRevD.79.075020}{\doi{10.1103/PhysRevD.79.075020}},
  \href{http://www.arXiv.org/abs/0810.3921}{\texttt{arXiv:0810.3921}}.

\bibitem{Alwall:2008va}
\hrefCMSnoop {}{J.~Alwall, M.-P. Le, M.~Lisanti, and J.~G. Wacker,
  ``Model-independent jets plus missing energy searches'',} \textit{ Phys. Rev.
  D} \textbf{ 79} (2009) 015005,
  \href{http://dx.doi.org/10.1103/PhysRevD.79.015005}{\doi{10.1103/PhysRevD.79.015005}},
  \href{http://www.arXiv.org/abs/0809.3264}{\texttt{arXiv:0809.3264}}.

\bibitem{LHCNewPhysicsWorkingGroup:2011mji}
\hrefCMSnoop {}{D.~Alves {et~al.}, ``Simplified models for {LHC} new physics
  searches'',} \textit{ J. Phys. G} \textbf{ 39} (2012) 105005,
  \href{http://dx.doi.org/10.1088/0954-3899/39/10/105005}{\doi{10.1088/0954-3899/39/10/105005}},
  \href{http://www.arXiv.org/abs/1105.2838}{\texttt{arXiv:1105.2838}}.

\bibitem{CMS:2013wdg}
\hrefCMSnoop {}{{CMS Collaboration}, ``Interpretation of searches for
  supersymmetry with simplified models'',} \textit{ Phys. Rev. D} \textbf{ 88}
  (2013) 052017,
  \href{http://dx.doi.org/10.1103/PhysRevD.88.052017}{\doi{10.1103/PhysRevD.88.052017}},
  \href{http://www.arXiv.org/abs/1301.2175}{\texttt{arXiv:1301.2175}}.

\bibitem{ATLAS:2021twp}
\hrefCMSnoop {}{{ATLAS Collaboration}, ``Search for squarks and gluinos in
  final states with one isolated lepton, jets, and missing transverse momentum
  at $\sqrt{s}={13\TeV}$ with the {ATLAS} detector'',} \textit{ Eur. Phys. J.
  C} \textbf{ 81} (2021) 600,
  \href{http://dx.doi.org/10.1140/epjc/s10052-021-09344-w}{\doi{10.1140/epjc/s10052-021-09344-w}},
  \href{http://www.arXiv.org/abs/2101.01629}{\texttt{arXiv:2101.01629}}.
  [Erratum: \DOI{10.1140/epjc/s10052-021-09748-8}].

\bibitem{ATLAS:2017xvf}
\hrefCMSnoop {}{{ATLAS Collaboration}, ``Search for squarks and gluinos in
  events with an isolated lepton, jets, and missing transverse momentum at
  $\sqrt{s}={13\TeV}$ with the {ATLAS} detector'',} \textit{ Phys. Rev. D}
  \textbf{ 96} (2017) 112010,
  \href{http://dx.doi.org/10.1103/PhysRevD.96.112010}{\doi{10.1103/PhysRevD.96.112010}},
  \href{http://www.arXiv.org/abs/1708.08232}{\texttt{arXiv:1708.08232}}.

\bibitem{ATLAS:2022ihe}
\hrefCMSnoop {}{{ATLAS Collaboration}, ``Search for supersymmetry in final
  states with missing transverse momentum and three or more {\PQb}-jets in
  139\fbinv of proton-proton collisions at $\sqrt{s}={13\TeV}$ with the {ATLAS}
  detector'',} \textit{ Eur. Phys. J. C} \textbf{ 83} (2023) 561,
  \href{http://dx.doi.org/10.1140/epjc/s10052-023-11543-6}{\doi{10.1140/epjc/s10052-023-11543-6}},
  \href{http://www.arXiv.org/abs/2211.08028}{\texttt{arXiv:2211.08028}}.

\bibitem{ATLAS:2016maz}
\hrefCMSnoop {}{{ATLAS Collaboration}, ``Search for gluinos in events with an
  isolated lepton, jets and missing transverse momentum at $\sqrt{s}={13\TeV}$
  with the {ATLAS} detector'',} \textit{ Eur. Phys. J. C} \textbf{ 76} (2016)
  565,
  \href{http://dx.doi.org/10.1140/epjc/s10052-016-4397-x}{\doi{10.1140/epjc/s10052-016-4397-x}},
  \href{http://www.arXiv.org/abs/1605.04285}{\texttt{arXiv:1605.04285}}.

\bibitem{ATLAS:2016gty}
\hrefCMSnoop {}{{ATLAS Collaboration}, ``Search for pair production of gluinos
  decaying via stop and sbottom in events with {\PQb}-jets and large missing
  transverse momentum in \pp collisions at $\sqrt{s}={13\TeV}$ with the {ATLAS}
  detector'',} \textit{ Phys. Rev. D} \textbf{ 94} (2016) 032003,
  \href{http://dx.doi.org/10.1103/PhysRevD.94.032003}{\doi{10.1103/PhysRevD.94.032003}},
  \href{http://www.arXiv.org/abs/1605.09318}{\texttt{arXiv:1605.09318}}.

\bibitem{CMS:2020cur}
\hrefCMSnoop {}{{CMS Collaboration}, ``Search for supersymmetry in \pp
  collisions at $\sqrt{s}={13\TeV}$ with 137\fbinv in final states with a
  single lepton using the sum of masses of large-radius jets'',} \textit{ Phys.
  Rev. D} \textbf{ 101} (2020) 052010,
  \href{http://dx.doi.org/10.1103/PhysRevD.101.052010}{\doi{10.1103/PhysRevD.101.052010}},
  \href{http://www.arXiv.org/abs/1911.07558}{\texttt{arXiv:1911.07558}}.

\bibitem{CMS:2017qth}
\hrefCMSnoop {}{{CMS Collaboration}, ``Search for supersymmetry in events with
  one lepton and multiple jets exploiting the angular correlation between the
  lepton and the missing transverse momentum in proton-proton collisions at
  $\sqrt{s}={13\TeV}$'',} \textit{ Phys. Lett. B} \textbf{ 780} (2018) 384,
  \href{http://dx.doi.org/10.1016/j.physletb.2018.03.028}{\doi{10.1016/j.physletb.2018.03.028}},
  \href{http://www.arXiv.org/abs/1709.09814}{\texttt{arXiv:1709.09814}}.

\bibitem{CMS:2016muu}
\hrefCMSnoop {}{{CMS Collaboration}, ``Search for supersymmetry in events with
  one lepton and multiple jets in proton-proton collisions at
  $\sqrt{s}={13\TeV}$'',} \textit{ Phys. Rev. D} \textbf{ 95} (2017) 012011,
  \href{http://dx.doi.org/10.1103/PhysRevD.95.012011}{\doi{10.1103/PhysRevD.95.012011}},
  \href{http://www.arXiv.org/abs/1609.09386}{\texttt{arXiv:1609.09386}}.

\bibitem{CMS:2016krz}
\hrefCMSnoop {}{{CMS Collaboration}, ``Search for supersymmetry in \pp
  collisions at $\sqrt{s}={13\TeV}$ in the single-lepton final state using the
  sum of masses of large-radius jets'',} \textit{ JHEP} \textbf{ 08} (2016)
  122,
  \href{http://dx.doi.org/10.1007/JHEP08(2016)122}{\doi{10.1007/JHEP08(2016)122}},
  \href{http://www.arXiv.org/abs/1605.04608}{\texttt{arXiv:1605.04608}}.

\bibitem{CMS:2018rst}
\hrefCMSnoop {}{{CMS Collaboration}, ``Inclusive search for supersymmetry in
  \pp collisions at $\sqrt{s}={13\TeV}$ using razor variables and boosted
  object identification in zero and one lepton final states'',} \textit{ JHEP}
  \textbf{ 03} (2019) 031,
  \href{http://dx.doi.org/10.1007/JHEP03(2019)031}{\doi{10.1007/JHEP03(2019)031}},
  \href{http://www.arXiv.org/abs/1812.06302}{\texttt{arXiv:1812.06302}}.

\bibitem{CMS:2017umd}
\hrefCMSnoop {}{{CMS Collaboration}, ``Search for supersymmetry in \pp
  collisions at $\sqrt{s}={13\TeV}$ in the single-lepton final state using the
  sum of masses of large-radius jets'',} \textit{ Phys. Rev. Lett.} \textbf{
  119} (2017) 151802,
  \href{http://dx.doi.org/10.1103/PhysRevLett.119.151802}{\doi{10.1103/PhysRevLett.119.151802}},
  \href{http://www.arXiv.org/abs/1705.04673}{\texttt{arXiv:1705.04673}}.

\bibitem{ATLAS:2020syg}
\hrefCMSnoop {}{{ATLAS Collaboration}, ``Search for squarks and gluinos in
  final states with jets and missing transverse momentum using 139\fbinv of
  $\sqrt{s}={13\TeV}$ \pp collision data with the {ATLAS} detector'',} \textit{
  JHEP} \textbf{ 02} (2021) 143,
  \href{http://dx.doi.org/10.1007/JHEP02(2021)143}{\doi{10.1007/JHEP02(2021)143}},
  \href{http://www.arXiv.org/abs/2010.14293}{\texttt{arXiv:2010.14293}}.

\bibitem{ATLAS:2017mjy}
\hrefCMSnoop {}{{ATLAS Collaboration}, ``Search for squarks and gluinos in
  final states with jets and missing transverse momentum using 36\fbinv of
  $\sqrt{s}={13\TeV}$ \pp collision data with the {ATLAS} detector'',} \textit{
  Phys. Rev. D} \textbf{ 97} (2018) 112001,
  \href{http://dx.doi.org/10.1103/PhysRevD.97.112001}{\doi{10.1103/PhysRevD.97.112001}},
  \href{http://www.arXiv.org/abs/1712.02332}{\texttt{arXiv:1712.02332}}.

\bibitem{ATLAS:2015gky}
\hrefCMSnoop {}{{ATLAS Collaboration}, ``Summary of the searches for squarks
  and gluinos using $\sqrt{s}={8\TeV}$ \pp collisions with the {ATLAS}
  experiment at the {LHC}'',} \textit{ JHEP} \textbf{ 10} (2015) 054,
  \href{http://dx.doi.org/10.1007/JHEP10(2015)054}{\doi{10.1007/JHEP10(2015)054}},
  \href{http://www.arXiv.org/abs/1507.05525}{\texttt{arXiv:1507.05525}}.

\bibitem{ATLAS:2020xgt}
\hrefCMSnoop {}{{ATLAS Collaboration}, ``Search for new phenomena in final
  states with large jet multiplicities and missing transverse momentum using
  $\sqrt{s}={13\TeV}$ proton-proton collisions recorded by {ATLAS} in {Run~2}
  of the {LHC}'',} \textit{ JHEP} \textbf{ 10} (2020) 062,
  \href{http://dx.doi.org/10.1007/JHEP10(2020)062}{\doi{10.1007/JHEP10(2020)062}},
  \href{http://www.arXiv.org/abs/2008.06032}{\texttt{arXiv:2008.06032}}.

\bibitem{CMS:2017tec}
\hrefCMSnoop {}{{CMS Collaboration}, ``Search for physics beyond the standard
  model in events with two leptons of same sign, missing transverse momentum,
  and jets in proton-proton collisions at $\sqrt{s}={13\TeV}$'',} \textit{ Eur.
  Phys. J. C} \textbf{ 77} (2017) 578,
  \href{http://dx.doi.org/10.1140/epjc/s10052-017-5079-z}{\doi{10.1140/epjc/s10052-017-5079-z}},
  \href{http://www.arXiv.org/abs/1704.07323}{\texttt{arXiv:1704.07323}}.

\bibitem{CMS:2013xio}
\hrefCMSnoop {}{{CMS Collaboration}, ``Search for new physics in events with
  same-sign dileptons and jets in \pp collisions at $\sqrt{s}={8\TeV}$'',}
  \textit{ JHEP} \textbf{ 01} (2014) 163,
  \href{http://dx.doi.org/10.1007/JHEP01(2014)163}{\doi{10.1007/JHEP01(2014)163}},
  \href{http://www.arXiv.org/abs/1311.6736}{\texttt{arXiv:1311.6736}}.
  [Erratum: \DOI{10.1007/JHEP01(2015)014}].

\bibitem{CMS:2016ohy}
\hrefCMSnoop {}{{CMS Collaboration}, ``Search for supersymmetry in the multijet
  and missing transverse momentum final state in \pp collisions at {13\TeV}'',}
  \textit{ Phys. Lett. B} \textbf{ 758} (2016) 152,
  \href{http://dx.doi.org/10.1016/j.physletb.2016.05.002}{\doi{10.1016/j.physletb.2016.05.002}},
  \href{http://www.arXiv.org/abs/1602.06581}{\texttt{arXiv:1602.06581}}.

\bibitem{CMS:2017abv}
\hrefCMSnoop {}{{CMS Collaboration}, ``Search for supersymmetry in multijet
  events with missing transverse momentum in proton-proton collisions at
  {13\TeV}'',} \textit{ Phys. Rev. D} \textbf{ 96} (2017) 032003,
  \href{http://dx.doi.org/10.1103/PhysRevD.96.032003}{\doi{10.1103/PhysRevD.96.032003}},
  \href{http://www.arXiv.org/abs/1704.07781}{\texttt{arXiv:1704.07781}}.

\bibitem{CMS:2019ybf}
\hrefCMSnoop {}{{CMS Collaboration}, ``Searches for physics beyond the standard
  model with the {\mTii} variable in hadronic final states with and without
  disappearing tracks in proton-proton collisions at $\sqrt{s}={13\TeV}$'',}
  \textit{ Eur. Phys. J. C} \textbf{ 80} (2020) 3,
  \href{http://dx.doi.org/10.1140/epjc/s10052-019-7493-x}{\doi{10.1140/epjc/s10052-019-7493-x}},
  \href{http://www.arXiv.org/abs/1909.03460}{\texttt{arXiv:1909.03460}}.

\bibitem{CMS:2019zmd}
\hrefCMSnoop {}{{CMS Collaboration}, ``Search for supersymmetry in
  proton-proton collisions at {13\TeV} in final states with jets and missing
  transverse momentum'',} \textit{ JHEP} \textbf{ 10} (2019) 244,
  \href{http://dx.doi.org/10.1007/JHEP10(2019)244}{\doi{10.1007/JHEP10(2019)244}},
  \href{http://www.arXiv.org/abs/1908.04722}{\texttt{arXiv:1908.04722}}.

\bibitem{CMS:2017qxu}
\hrefCMSnoop {}{{CMS Collaboration}, ``Search for supersymmetry in
  proton-proton collisions at {13\TeV} using identified top quarks'',} \textit{
  Phys. Rev. D} \textbf{ 97} (2018) 012007,
  \href{http://dx.doi.org/10.1103/PhysRevD.97.012007}{\doi{10.1103/PhysRevD.97.012007}},
  \href{http://www.arXiv.org/abs/1710.11188}{\texttt{arXiv:1710.11188}}.

\bibitem{CMS:2017okm}
\hrefCMSnoop {}{{CMS Collaboration}, ``Search for new phenomena with the
  {\mTii} variable in the all-hadronic final state produced in proton-proton
  collisions at $\sqrt{s}={13\TeV}$'',} \textit{ Eur. Phys. J. C} \textbf{ 77}
  (2017) 710,
  \href{http://dx.doi.org/10.1140/epjc/s10052-017-5267-x}{\doi{10.1140/epjc/s10052-017-5267-x}},
  \href{http://www.arXiv.org/abs/1705.04650}{\texttt{arXiv:1705.04650}}.

\bibitem{hepdata}
\hrefCMSnoop {}{}{HEPData} record for this analysis, 2022.
\newblock
  \href{http://dx.doi.org/10.17182/hepdata.135454}{\doi{10.17182/hepdata.135454}}.

\bibitem{CMS:2020uim}
\hrefCMSnoop {}{{CMS Collaboration}, ``Electron and photon reconstruction and
  identification with the {CMS} experiment at the {CERN} {LHC}'',} \textit{
  JINST} \textbf{ 16} (2021) P05014,
  \href{http://dx.doi.org/10.1088/1748-0221/16/05/P05014}{\doi{10.1088/1748-0221/16/05/P05014}},
  \href{http://www.arXiv.org/abs/2012.06888}{\texttt{arXiv:2012.06888}}.

\bibitem{CMS:2018rym}
\hrefCMSnoop {}{{CMS Collaboration}, ``Performance of the {CMS} muon detector
  and muon reconstruction with proton-proton collisions at
  $\sqrt{s}={13\TeV}$'',} \textit{ JINST} \textbf{ 13} (2018) P06015,
  \href{http://dx.doi.org/10.1088/1748-0221/13/06/P06015}{\doi{10.1088/1748-0221/13/06/P06015}},
  \href{http://www.arXiv.org/abs/1804.04528}{\texttt{arXiv:1804.04528}}.

\bibitem{CMS:2014pgm}
\hrefCMSnoop {}{{CMS Collaboration}, ``Description and performance of track and
  primary-vertex reconstruction with the {CMS} tracker'',} \textit{ JINST}
  \textbf{ 9} (2014) P10009,
  \href{http://dx.doi.org/10.1088/1748-0221/9/10/P10009}{\doi{10.1088/1748-0221/9/10/P10009}},
  \href{http://www.arXiv.org/abs/1405.6569}{\texttt{arXiv:1405.6569}}.

\bibitem{CMS:2020cmk}
\hrefCMSnoop {}{{CMS Collaboration}, ``Performance of the {CMS} {\Lone} trigger
  in proton-proton collisions at $\sqrt{s}={13\TeV}$'',} \textit{ JINST}
  \textbf{ 15} (2020) P10017,
  \href{http://dx.doi.org/10.1088/1748-0221/15/10/P10017}{\doi{10.1088/1748-0221/15/10/P10017}},
  \href{http://www.arXiv.org/abs/2006.10165}{\texttt{arXiv:2006.10165}}.

\bibitem{CMS:2016ngn}
\hrefCMSnoop {}{{CMS Collaboration}, ``The {CMS} trigger system'',} \textit{
  JINST} \textbf{ 12} (2017) P01020,
  \href{http://dx.doi.org/10.1088/1748-0221/12/01/P01020}{\doi{10.1088/1748-0221/12/01/P01020}},
  \href{http://www.arXiv.org/abs/1609.02366}{\texttt{arXiv:1609.02366}}.

\bibitem{CMS:2008xjf}
\hrefCMSnoop {}{{CMS Collaboration}, ``The {CMS} experiment at the {CERN}
  {LHC}'',} \textit{ JINST} \textbf{ 3} (2008) S08004,
  \href{http://dx.doi.org/10.1088/1748-0221/3/08/S08004}{\doi{10.1088/1748-0221/3/08/S08004}}.

\bibitem{Alwall:2014hca}
J.~Alwall\hrefCMSnoop {}{ {et~al.}, ``The automated computation of tree-level
  and next-to-leading order differential cross sections, and their matching to
  parton shower simulations'',} \textit{ JHEP} \textbf{ 07} (2014) 079,
  \href{http://dx.doi.org/10.1007/JHEP07(2014)079}{\doi{10.1007/JHEP07(2014)079}},
  \href{http://www.arXiv.org/abs/1405.0301}{\texttt{arXiv:1405.0301}}.

\bibitem{Frederix:2012ps}
\hrefCMSnoop {}{R.~Frederix and S.~Frixione, ``Merging meets matching in
  {\MCATNLO}'',} \textit{ JHEP} \textbf{ 12} (2012) 061,
  \href{http://dx.doi.org/10.1007/JHEP12(2012)061}{\doi{10.1007/JHEP12(2012)061}},
  \href{http://www.arXiv.org/abs/1209.6215}{\texttt{arXiv:1209.6215}}.

\bibitem{Nason:2004rx}
\hrefCMSnoop {}{P.~Nason, ``A new method for combining {NLO} {QCD} with shower
  {Monte Carlo} algorithms'',} \textit{ JHEP} \textbf{ 11} (2004) 040,
  \href{http://dx.doi.org/10.1088/1126-6708/2004/11/040}{\doi{10.1088/1126-6708/2004/11/040}},
  \href{http://www.arXiv.org/abs/hep-ph/0409146}{\texttt{arXiv:hep-ph/0409146}}.

\bibitem{Frixione:2007vw}
\hrefCMSnoop {}{S.~Frixione, P.~Nason, and C.~Oleari, ``Matching {NLO} {QCD}
  computations with parton shower simulations: the {\POWHEG} method'',}
  \textit{ JHEP} \textbf{ 11} (2007) 070,
  \href{http://dx.doi.org/10.1088/1126-6708/2007/11/070}{\doi{10.1088/1126-6708/2007/11/070}},
  \href{http://www.arXiv.org/abs/0709.2092}{\texttt{arXiv:0709.2092}}.

\bibitem{Alioli:2010xd}
\hrefCMSnoop {}{S.~Alioli, P.~Nason, C.~Oleari, and E.~Re, ``A general
  framework for implementing {NLO} calculations in shower {Monte Carlo}
  programs: the {\POWHEG} \textsc{box}'',} \textit{ JHEP} \textbf{ 06} (2010)
  043,
  \href{http://dx.doi.org/10.1007/JHEP06(2010)043}{\doi{10.1007/JHEP06(2010)043}},
  \href{http://www.arXiv.org/abs/1002.2581}{\texttt{arXiv:1002.2581}}.

\bibitem{Alioli:2009je}
\hrefCMSnoop {}{S.~Alioli, P.~Nason, C.~Oleari, and E.~Re, ``{NLO} single-top
  production matched with shower in {\POWHEG}: $s$- and $t$-channel
  contributions'',} \textit{ JHEP} \textbf{ 09} (2009) 111,
  \href{http://dx.doi.org/10.1088/1126-6708/2009/09/111}{\doi{10.1088/1126-6708/2009/09/111}},
  \href{http://www.arXiv.org/abs/0907.4076}{\texttt{arXiv:0907.4076}}.
  [Erratum: \DOI{10.1007/JHEP02(2010)011}].

\bibitem{Re:2010bp}
\hrefCMSnoop {}{E.~Re, ``Single-top ${\PW\PQt}$-channel production matched with
  parton showers using the {\POWHEG} method'',} \textit{ Eur. Phys. J. C}
  \textbf{ 71} (2011) 1547,
  \href{http://dx.doi.org/10.1140/epjc/s10052-011-1547-z}{\doi{10.1140/epjc/s10052-011-1547-z}},
  \href{http://www.arXiv.org/abs/1009.2450}{\texttt{arXiv:1009.2450}}.

\bibitem{Melia:2011tj}
\hrefCMSnoop {}{T.~Melia, P.~Nason, R.~R{\"o}ntsch, and G.~Zanderighi,
  ``${\PWp\PWm}$, ${\PW\PZ}$ and ${\PZ\PZ}$ production in the {\POWHEG}
  \textsc{box}'',} \textit{ JHEP} \textbf{ 11} (2011) 078,
  \href{http://dx.doi.org/10.1007/JHEP11(2011)078}{\doi{10.1007/JHEP11(2011)078}},
  \href{http://www.arXiv.org/abs/1107.5051}{\texttt{arXiv:1107.5051}}.

\bibitem{Nason:2013ydw}
\hrefCMSnoop {}{P.~Nason and G.~Zanderighi, ``${\PWp\PWm}$, ${\PW\PZ}$ and
  ${\PZ\PZ}$ production in the {\POWHEG-}\textsc{box-v2}'',} \textit{ Eur.
  Phys. J. C} \textbf{ 74} (2014) 2702,
  \href{http://dx.doi.org/10.1140/epjc/s10052-013-2702-5}{\doi{10.1140/epjc/s10052-013-2702-5}},
  \href{http://www.arXiv.org/abs/1311.1365}{\texttt{arXiv:1311.1365}}.

\bibitem{Hartanto:2015uka}
\hrefCMSnoop {}{H.~B. Hartanto, B.~J{\"a}ger, L.~Reina, and D.~Wackeroth,
  ``{Higgs} boson production in association with top quarks in the {\POWHEG}
  \textsc{box}'',} \textit{ Phys. Rev. D} \textbf{ 91} (2015) 094003,
  \href{http://dx.doi.org/10.1103/PhysRevD.91.094003}{\doi{10.1103/PhysRevD.91.094003}},
  \href{http://www.arXiv.org/abs/1501.04498}{\texttt{arXiv:1501.04498}}.

\bibitem{Sjostrand:2014zea}
T.~Sj{\"o}strand\hrefCMSnoop {}{ {et~al.}, ``An introduction to
  {\PYTHIA8.2}'',} \textit{ Comput. Phys. Commun.} \textbf{ 191} (2015) 159,
  \href{http://dx.doi.org/10.1016/j.cpc.2015.01.024}{\doi{10.1016/j.cpc.2015.01.024}},
  \href{http://www.arXiv.org/abs/1410.3012}{\texttt{arXiv:1410.3012}}.

\bibitem{Gavin:2012sy}
\hrefCMSnoop {}{S.~Quackenbush, R.~Gavin, Y.~Li, and F.~Petriello, ``{\PW}
  physics at the {LHC} with {\FEWZ2.1}'',} \textit{ Comput. Phys. Commun.}
  \textbf{ 184} (2013) 209,
  \href{http://dx.doi.org/10.1016/j.cpc.2012.09.005}{\doi{10.1016/j.cpc.2012.09.005}},
  \href{http://www.arXiv.org/abs/1201.5896}{\texttt{arXiv:1201.5896}}.

\bibitem{Gavin:2010az}
\hrefCMSnoop {}{R.~Gavin, Y.~Li, F.~Petriello, and S.~Quackenbush,
  ``{\FEWZ2.0}: A code for hadronic {\PZ} production at next-to-next-to-leading
  order'',} \textit{ Comput. Phys. Commun.} \textbf{ 182} (2011) 2388,
  \href{http://dx.doi.org/10.1016/j.cpc.2011.06.008}{\doi{10.1016/j.cpc.2011.06.008}},
  \href{http://www.arXiv.org/abs/1011.3540}{\texttt{arXiv:1011.3540}}.

\bibitem{Li:2012wna}
\hrefCMSnoop {}{Y.~Li and F.~Petriello, ``Combining {QCD} and electroweak
  corrections to dilepton production in the framework of the {\FEWZ} simulation
  code'',} \textit{ Phys. Rev. D} \textbf{ 86} (2012) 094034,
  \href{http://dx.doi.org/10.1103/PhysRevD.86.094034}{\doi{10.1103/PhysRevD.86.094034}},
  \href{http://www.arXiv.org/abs/1208.5967}{\texttt{arXiv:1208.5967}}.

\bibitem{Gehrmann:2014fva}
T.~Gehrmann\hrefCMSnoop {}{ {et~al.}, ``${\PWp\PWm}$ production at hadron
  colliders in next to next to leading order {QCD}'',} \textit{ Phys. Rev.
  Lett.} \textbf{ 113} (2014) 212001,
  \href{http://dx.doi.org/10.1103/PhysRevLett.113.212001}{\doi{10.1103/PhysRevLett.113.212001}},
  \href{http://www.arXiv.org/abs/1408.5243}{\texttt{arXiv:1408.5243}}.

\bibitem{Campbell:2011bn}
\hrefCMSnoop {}{J.~M. Campbell, R.~K. Ellis, and C.~Williams, ``Vector boson
  pair production at the {LHC}'',} \textit{ JHEP} \textbf{ 07} (2011) 018,
  \href{http://dx.doi.org/10.1007/JHEP07(2011)018}{\doi{10.1007/JHEP07(2011)018}},
  \href{http://www.arXiv.org/abs/1105.0020}{\texttt{arXiv:1105.0020}}.

\bibitem{Campbell:2015qma}
\hrefCMSnoop {}{J.~M. Campbell, R.~K. Ellis, and W.~T. Giele, ``A
  multi-threaded version of {\MCFM}'',} \textit{ Eur. Phys. J. C} \textbf{ 75}
  (2015) 246,
  \href{http://dx.doi.org/10.1140/epjc/s10052-015-3461-2}{\doi{10.1140/epjc/s10052-015-3461-2}},
  \href{http://www.arXiv.org/abs/1503.06182}{\texttt{arXiv:1503.06182}}.

\bibitem{Beneke:2011mq}
\hrefCMSnoop {}{M.~Beneke, P.~Falgari, S.~Klein, and C.~Schwinn, ``Hadronic
  top-quark pair production with {NNLL} threshold resummation'',} \textit{
  Nucl. Phys. B} \textbf{ 855} (2012) 695,
  \href{http://dx.doi.org/10.1016/j.nuclphysb.2011.10.021}{\doi{10.1016/j.nuclphysb.2011.10.021}},
  \href{http://www.arXiv.org/abs/1109.1536}{\texttt{arXiv:1109.1536}}.

\bibitem{Cacciari:2011hy}
M.~Cacciari\hrefCMSnoop {}{ {et~al.}, ``Top-pair production at hadron colliders
  with next-to-next-to-leading logarithmic soft-gluon resummation'',} \textit{
  Phys. Lett. B} \textbf{ 710} (2012) 612,
  \href{http://dx.doi.org/10.1016/j.physletb.2012.03.013}{\doi{10.1016/j.physletb.2012.03.013}},
  \href{http://www.arXiv.org/abs/1111.5869}{\texttt{arXiv:1111.5869}}.

\bibitem{Baernreuther:2012ws}
\hrefCMSnoop {}{P.~B{\"a}rnreuther, M.~Czakon, and A.~Mitov,
  ``Percent-level-precision physics at the {Tevatron}: next-to-next-to-leading
  order {QCD} corrections to $\qqbar\to\ttbar+{\PX}$'',} \textit{ Phys. Rev.
  Lett.} \textbf{ 109} (2012) 132001,
  \href{http://dx.doi.org/10.1103/PhysRevLett.109.132001}{\doi{10.1103/PhysRevLett.109.132001}},
  \href{http://www.arXiv.org/abs/1204.5201}{\texttt{arXiv:1204.5201}}.

\bibitem{Czakon:2012zr}
\hrefCMSnoop {}{M.~Czakon and A.~Mitov, ``{NNLO} corrections to top-pair
  production at hadron colliders: the all-fermionic scattering channels'',}
  \textit{ JHEP} \textbf{ 12} (2012) 054,
  \href{http://dx.doi.org/10.1007/JHEP12(2012)054}{\doi{10.1007/JHEP12(2012)054}},
  \href{http://www.arXiv.org/abs/1207.0236}{\texttt{arXiv:1207.0236}}.

\bibitem{Czakon:2012pz}
\hrefCMSnoop {}{M.~Czakon and A.~Mitov, ``{NNLO} corrections to top pair
  production at hadron colliders: the quark-gluon reaction'',} \textit{ JHEP}
  \textbf{ 01} (2013) 080,
  \href{http://dx.doi.org/10.1007/JHEP01(2013)080}{\doi{10.1007/JHEP01(2013)080}},
  \href{http://www.arXiv.org/abs/1210.6832}{\texttt{arXiv:1210.6832}}.

\bibitem{Czakon:2013goa}
\hrefCMSnoop {}{M.~Czakon, P.~Fiedler, and A.~Mitov, ``Total top-quark
  pair-production cross section at hadron colliders through ${O(\alpS^4)}$'',}
  \textit{ Phys. Rev. Lett.} \textbf{ 110} (2013) 252004,
  \href{http://dx.doi.org/10.1103/PhysRevLett.110.252004}{\doi{10.1103/PhysRevLett.110.252004}},
  \href{http://www.arXiv.org/abs/1303.6254}{\texttt{arXiv:1303.6254}}.

\bibitem{Czakon:2011xx}
\hrefCMSnoop {}{M.~Czakon and A.~Mitov, ``\textsc{top++}: A program for the
  calculation of the top-pair cross-section at hadron colliders'',} \textit{
  Comput. Phys. Commun.} \textbf{ 185} (2014) 2930,
  \href{http://dx.doi.org/10.1016/j.cpc.2014.06.021}{\doi{10.1016/j.cpc.2014.06.021}},
  \href{http://www.arXiv.org/abs/1112.5675}{\texttt{arXiv:1112.5675}}.

\bibitem{Beenakker:2016lwe}
W.~Beenakker\hrefCMSnoop {}{ {et~al.}, ``{NNLL}-fast: predictions for coloured
  supersymmetric particle production at the {LHC} with threshold and {Coulomb}
  resummation'',} \textit{ JHEP} \textbf{ 12} (2016) 133,
  \href{http://dx.doi.org/10.1007/JHEP12(2016)133}{\doi{10.1007/JHEP12(2016)133}},
  \href{http://www.arXiv.org/abs/1607.07741}{\texttt{arXiv:1607.07741}}.

\bibitem{Borschensky:2014cia}
C.~Borschensky\hrefCMSnoop {}{ {et~al.}, ``Squark and gluino production cross
  sections in ${\Pp\Pp}$ collisions at $\sqrt{s}=13$, 14, 33 and {100\TeV}'',}
  \textit{ Eur. Phys. J. C} \textbf{ 74} (2014) 3174,
  \href{http://dx.doi.org/10.1140/epjc/s10052-014-3174-y}{\doi{10.1140/epjc/s10052-014-3174-y}},
  \href{http://www.arXiv.org/abs/1407.5066}{\texttt{arXiv:1407.5066}}.

\bibitem{Beenakker:1996ch}
\hrefCMSnoop {}{W.~Beenakker, R.~H{\"o}pker, M.~Spira, and P.~M. Zerwas,
  ``Squark and gluino production at hadron colliders'',} \textit{ Nucl. Phys.
  B} \textbf{ 492} (1997) 51,
  \href{http://dx.doi.org/10.1016/S0550-3213(97)80027-2}{\doi{10.1016/S0550-3213(97)80027-2}},
  \href{http://www.arXiv.org/abs/hep-ph/9610490}{\texttt{arXiv:hep-ph/9610490}}.

\bibitem{Kulesza:2008jb}
\hrefCMSnoop {}{A.~Kulesza and L.~Motyka, ``Threshold resummation for
  squark-antisquark and gluino-pair production at the {LHC}'',} \textit{ Phys.
  Rev. Lett.} \textbf{ 102} (2009) 111802,
  \href{http://dx.doi.org/10.1103/PhysRevLett.102.111802}{\doi{10.1103/PhysRevLett.102.111802}},
  \href{http://www.arXiv.org/abs/0807.2405}{\texttt{arXiv:0807.2405}}.

\bibitem{Kulesza:2009kq}
\hrefCMSnoop {}{A.~Kulesza and L.~Motyka, ``Soft gluon resummation for the
  production of gluino-gluino and squark-antisquark pairs at the {LHC}'',}
  \textit{ Phys. Rev. D} \textbf{ 80} (2009) 095004,
  \href{http://dx.doi.org/10.1103/PhysRevD.80.095004}{\doi{10.1103/PhysRevD.80.095004}},
  \href{http://www.arXiv.org/abs/0905.4749}{\texttt{arXiv:0905.4749}}.

\bibitem{Beenakker:2009ha}
W.~Beenakker\hrefCMSnoop {}{ {et~al.}, ``Soft-gluon resummation for squark and
  gluino hadroproduction'',} \textit{ JHEP} \textbf{ 12} (2009) 041,
  \href{http://dx.doi.org/10.1088/1126-6708/2009/12/041}{\doi{10.1088/1126-6708/2009/12/041}},
  \href{http://www.arXiv.org/abs/0909.4418}{\texttt{arXiv:0909.4418}}.

\bibitem{Beenakker:2011sf}
W.~Beenakker\hrefCMSnoop {}{ {et~al.}, ``{NNLL} resummation for
  squark-antisquark pair production at the {LHC}'',} \textit{ JHEP} \textbf{
  01} (2012) 076,
  \href{http://dx.doi.org/10.1007/JHEP01(2012)076}{\doi{10.1007/JHEP01(2012)076}},
  \href{http://www.arXiv.org/abs/1110.2446}{\texttt{arXiv:1110.2446}}.

\bibitem{Beenakker:2013mva}
W.~Beenakker\hrefCMSnoop {}{ {et~al.}, ``Towards {NNLL} resummation: hard
  matching coefficients for squark and gluino hadroproduction'',} \textit{
  JHEP} \textbf{ 10} (2013) 120,
  \href{http://dx.doi.org/10.1007/JHEP10(2013)120}{\doi{10.1007/JHEP10(2013)120}},
  \href{http://www.arXiv.org/abs/1304.6354}{\texttt{arXiv:1304.6354}}.

\bibitem{Beenakker:2014sma}
W.~Beenakker\hrefCMSnoop {}{ {et~al.}, ``{NNLL} resummation for squark and
  gluino production at the {LHC}'',} \textit{ JHEP} \textbf{ 12} (2014) 023,
  \href{http://dx.doi.org/10.1007/JHEP12(2014)023}{\doi{10.1007/JHEP12(2014)023}},
  \href{http://www.arXiv.org/abs/1404.3134}{\texttt{arXiv:1404.3134}}.

\bibitem{Beenakker:1997ut}
W.~Beenakker\hrefCMSnoop {}{ {et~al.}, ``Stop production at hadron
  colliders'',} \textit{ Nucl. Phys. B} \textbf{ 515} (1998) 3,
  \href{http://dx.doi.org/10.1016/S0550-3213(98)00014-5}{\doi{10.1016/S0550-3213(98)00014-5}},
  \href{http://www.arXiv.org/abs/hep-ph/9710451}{\texttt{arXiv:hep-ph/9710451}}.

\bibitem{Beenakker:2010nq}
W.~Beenakker\hrefCMSnoop {}{ {et~al.}, ``Supersymmetric top and bottom squark
  production at hadron colliders'',} \textit{ JHEP} \textbf{ 08} (2010) 098,
  \href{http://dx.doi.org/10.1007/JHEP08(2010)098}{\doi{10.1007/JHEP08(2010)098}},
  \href{http://www.arXiv.org/abs/1006.4771}{\texttt{arXiv:1006.4771}}.

\bibitem{Beenakker:2016gmf}
W.~Beenakker\hrefCMSnoop {}{ {et~al.}, ``{NNLL} resummation for stop
  pair-production at the {LHC}'',} \textit{ JHEP} \textbf{ 05} (2016) 153,
  \href{http://dx.doi.org/10.1007/JHEP05(2016)153}{\doi{10.1007/JHEP05(2016)153}},
  \href{http://www.arXiv.org/abs/1601.02954}{\texttt{arXiv:1601.02954}}.

\bibitem{Mangano:2006rw}
\hrefCMSnoop {}{M.~L. Mangano, M.~Moretti, F.~Piccinini, and M.~Treccani,
  ``Matching matrix elements and shower evolution for top-pair production in
  hadronic collisions'',} \textit{ JHEP} \textbf{ 01} (2007) 013,
  \href{http://dx.doi.org/10.1088/1126-6708/2007/01/013}{\doi{10.1088/1126-6708/2007/01/013}},
  \href{http://www.arXiv.org/abs/hep-ph/0611129}{\texttt{arXiv:hep-ph/0611129}}.

\bibitem{CMS:2015wcf}
\hrefCMSnoop {}{{CMS Collaboration}, ``Event generator tunes obtained from
  underlying event and multiparton scattering measurement'',} \textit{ Eur.
  Phys. J. C} \textbf{ 76} (2016) 155,
  \href{http://dx.doi.org/10.1140/epjc/s10052-016-3988-x}{\doi{10.1140/epjc/s10052-016-3988-x}},
  \href{http://www.arXiv.org/abs/1512.00815}{\texttt{arXiv:1512.00815}}.

\bibitem{CMS:2019csb}
\hrefCMSnoop {}{{CMS Collaboration}, ``Extraction and validation of a new set
  of {CMS} {\PYTHIA8} tunes from underlying-event measurements'',} \textit{
  Eur. Phys. J. C} \textbf{ 80} (2020) 4,
  \href{http://dx.doi.org/10.1140/epjc/s10052-019-7499-4}{\doi{10.1140/epjc/s10052-019-7499-4}},
  \href{http://www.arXiv.org/abs/1903.12179}{\texttt{arXiv:1903.12179}}.

\bibitem{Ball:2015}
\hrefCMSnoop {}{{NNPDF} Collaboration, ``Parton distributions for the {LHC}
  {Run II}'',} \textit{ JHEP} \textbf{ 04} (2015) 040,
  \href{http://dx.doi.org/10.1007/JHEP04(2015)040}{\doi{10.1007/JHEP04(2015)040}},
  \href{http://www.arXiv.org/abs/1410.8849}{\texttt{arXiv:1410.8849}}.

\bibitem{Ball:2017nwa}
\hrefCMSnoop {}{{NNPDF} Collaboration, ``Parton distributions from
  high-precision collider data'',} \textit{ Eur. Phys. J. C} \textbf{ 77}
  (2017) 663,
  \href{http://dx.doi.org/10.1140/epjc/s10052-017-5199-5}{\doi{10.1140/epjc/s10052-017-5199-5}},
  \href{http://www.arXiv.org/abs/1706.00428}{\texttt{arXiv:1706.00428}}.

\bibitem{Agostinelli:2002hh}
\hrefCMSnoop {}{{GEANT4} Collaboration, ``{\GEANTfour}---a simulation
  toolkit'',} \textit{ Nucl. Instrum. Meth. A} \textbf{ 506} (2003) 250,
  \href{http://dx.doi.org/10.1016/S0168-9002(03)01368-8}{\doi{10.1016/S0168-9002(03)01368-8}}.

\bibitem{Abdullin:2011zz}
\hrefCMSnoop {}{{CMS Collaboration}, S.~Abdullin {et~al.}, ``The fast
  simulation of the {CMS} detector at {LHC}'',} in \textit{ {Proc. 18th Int.
  Conf. on Computing in High Energy and Nuclear Phys. (CHEP 2010): Taipei,
  Taiwan, October 18--22, 2010}}.
\newblock 2011.
\newblock [J. Phys. Conf. Ser. 331 (2011) 032049].
  \href{http://dx.doi.org/10.1088/1742-6596/331/3/032049}{\doi{10.1088/1742-6596/331/3/032049}}.

\bibitem{Giammanco:2014bza}
\hrefCMSnoop {}{A.~Giammanco, ``The fast simulation of the {CMS} experiment'',}
  in \textit{ {Proc. 20th Int. Conf. on Computing in High Energy and Nuclear
  Phys. (CHEP 2013): Amsterdam, The Netherlands, October 14--18, 2013}}.
\newblock 2014.
\newblock [J. Phys. Conf. Ser. 513 (2014) 022012].
  \href{http://dx.doi.org/10.1088/1742-6596/513/2/022012}{\doi{10.1088/1742-6596/513/2/022012}}.

\bibitem{CMS:2017yfk}
\hrefCMSnoop {}{{CMS Collaboration}, ``Particle-flow reconstruction and global
  event description with the {CMS} detector'',} \textit{ JINST} \textbf{ 12}
  (2017) P10003,
  \href{http://dx.doi.org/10.1088/1748-0221/12/10/P10003}{\doi{10.1088/1748-0221/12/10/P10003}},
  \href{http://www.arXiv.org/abs/1706.04965}{\texttt{arXiv:1706.04965}}.

\bibitem{CMS-TDR-15-02}
\href {http://cds.cern.ch/record/2020886}{{CMS Collaboration}, ``Technical
  proposal for the {Phase-II} upgrade of the {Compact Muon Solenoid}'',} CMS
  Technical Proposal CERN-LHCC-2015-010, CMS-TDR-15-02, 2015.

\bibitem{Rehermann:2010vq}
\hrefCMSnoop {}{K.~Rehermann and B.~Tweedie, ``Efficient identification of
  boosted semileptonic top quarks at the {LHC}'',} \textit{ JHEP} \textbf{ 03}
  (2011) 059,
  \href{http://dx.doi.org/10.1007/JHEP03(2011)059}{\doi{10.1007/JHEP03(2011)059}},
  \href{http://www.arXiv.org/abs/1007.2221}{\texttt{arXiv:1007.2221}}.

\bibitem{Cacciari:2008gp}
\hrefCMSnoop {}{M.~Cacciari, G.~P. Salam, and G.~Soyez, ``The anti-\kt jet
  clustering algorithm'',} \textit{ JHEP} \textbf{ 04} (2008) 063,
  \href{http://dx.doi.org/10.1088/1126-6708/2008/04/063}{\doi{10.1088/1126-6708/2008/04/063}},
  \href{http://www.arXiv.org/abs/0802.1189}{\texttt{arXiv:0802.1189}}.

\bibitem{Cacciari:2011ma}
\hrefCMSnoop {}{M.~Cacciari, G.~P. Salam, and G.~Soyez, ``{\FASTJET} user
  manual'',} \textit{ Eur. Phys. J. C} \textbf{ 72} (2012) 1896,
  \href{http://dx.doi.org/10.1140/epjc/s10052-012-1896-2}{\doi{10.1140/epjc/s10052-012-1896-2}},
  \href{http://www.arXiv.org/abs/1111.6097}{\texttt{arXiv:1111.6097}}.

\bibitem{CMS:2020ebo}
\hrefCMSnoop {}{{CMS Collaboration}, ``Pileup mitigation at {CMS} in {13\TeV}
  data'',} \textit{ JINST} \textbf{ 15} (2020) P09018,
  \href{http://dx.doi.org/10.1088/1748-0221/15/09/P09018}{\doi{10.1088/1748-0221/15/09/P09018}},
  \href{http://www.arXiv.org/abs/2003.00503}{\texttt{arXiv:2003.00503}}.

\bibitem{Bertolini:2014bba}
\hrefCMSnoop {}{D.~Bertolini, P.~Harris, M.~Low, and N.~Tran, ``Pileup per
  particle identification'',} \textit{ JHEP} \textbf{ 10} (2014) 059,
  \href{http://dx.doi.org/10.1007/JHEP10(2014)059}{\doi{10.1007/JHEP10(2014)059}},
  \href{http://www.arXiv.org/abs/1407.6013}{\texttt{arXiv:1407.6013}}.

\bibitem{CMS:2016lmd}
\hrefCMSnoop {}{{CMS Collaboration}, ``Jet energy scale and resolution in the
  {CMS} experiment in \pp collisions at {8\TeV}'',} \textit{ JINST} \textbf{
  12} (2017) P02014,
  \href{http://dx.doi.org/10.1088/1748-0221/12/02/P02014}{\doi{10.1088/1748-0221/12/02/P02014}},
  \href{http://www.arXiv.org/abs/1607.03663}{\texttt{arXiv:1607.03663}}.

\bibitem{CMS:2017wtu}
\hrefCMSnoop {}{{CMS Collaboration}, ``Identification of heavy-flavour jets
  with the {CMS} detector in \pp collisions at {13\TeV}'',} \textit{ JINST}
  \textbf{ 13} (2018) P05011,
  \href{http://dx.doi.org/10.1088/1748-0221/13/05/P05011}{\doi{10.1088/1748-0221/13/05/P05011}},
  \href{http://www.arXiv.org/abs/1712.07158}{\texttt{arXiv:1712.07158}}.

\bibitem{CMS:2020poo}
\hrefCMSnoop {}{{CMS Collaboration}, ``Identification of heavy, energetic,
  hadronically decaying particles using machine-learning techniques'',}
  \textit{ JINST} \textbf{ 15} (2020) P06005,
  \href{http://dx.doi.org/10.1088/1748-0221/15/06/P06005}{\doi{10.1088/1748-0221/15/06/P06005}},
  \href{http://www.arXiv.org/abs/2004.08262}{\texttt{arXiv:2004.08262}}.

\bibitem{CMS:2017mbm}
\hrefCMSnoop {}{{CMS Collaboration}, ``Search for direct production of
  supersymmetric partners of the top quark in the all-jets final state in
  proton-proton collisions at $\sqrt{s}={13\TeV}$'',} \textit{ JHEP} \textbf{
  10} (2017) 005,
  \href{http://dx.doi.org/10.1007/JHEP10(2017)005}{\doi{10.1007/JHEP10(2017)005}},
  \href{http://www.arXiv.org/abs/1707.03316}{\texttt{arXiv:1707.03316}}.

\bibitem{CMS:2019ctu}
\hrefCMSnoop {}{{CMS Collaboration}, ``Performance of missing transverse
  momentum reconstruction in proton-proton collisions at $\sqrt{s}={13\TeV}$
  using the {CMS} detector'',} \textit{ JINST} \textbf{ 14} (2019) P07004,
  \href{http://dx.doi.org/10.1088/1748-0221/14/07/P07004}{\doi{10.1088/1748-0221/14/07/P07004}},
  \href{http://www.arXiv.org/abs/1903.06078}{\texttt{arXiv:1903.06078}}.

\bibitem{Lester:1999tx}
\hrefCMSnoop {}{C.~G. Lester and D.~J. Summers, ``Measuring masses of
  semi-invisibly decaying particle pairs produced at hadron colliders'',}
  \textit{ Phys. Lett. B} \textbf{ 463} (1999) 99,
  \href{http://dx.doi.org/10.1016/S0370-2693(99)00945-4}{\doi{10.1016/S0370-2693(99)00945-4}},
  \href{http://www.arXiv.org/abs/hep-ph/9906349}{\texttt{arXiv:hep-ph/9906349}}.

\bibitem{CMS:2011kaj}
\hrefCMSnoop {}{{CMS Collaboration}, ``Measurement of the polarization of {\PW}
  bosons with large transverse momenta in {\PW}+jets events at the {LHC}'',}
  \textit{ Phys. Rev. Lett.} \textbf{ 107} (2011) 021802,
  \href{http://dx.doi.org/10.1103/PhysRevLett.107.021802}{\doi{10.1103/PhysRevLett.107.021802}},
  \href{http://www.arXiv.org/abs/1104.3829}{\texttt{arXiv:1104.3829}}.

\bibitem{CMS:2012vfw}
\hrefCMSnoop {}{{CMS Collaboration}, ``Search for supersymmetry in \pp
  collisions at $\sqrt{s}={7\TeV}$ in events with a single lepton, jets, and
  missing transverse momentum'',} \textit{ Eur. Phys. J. C} \textbf{ 73} (2013)
  2404,
  \href{http://dx.doi.org/10.1140/epjc/s10052-013-2404-z}{\doi{10.1140/epjc/s10052-013-2404-z}},
  \href{http://www.arXiv.org/abs/1212.6428}{\texttt{arXiv:1212.6428}}.

\bibitem{CMS:2018mlc}
\hrefCMSnoop {}{{CMS Collaboration}, ``Measurement of the inelastic
  proton-proton cross section at $\sqrt{s}={13\TeV}$'',} \textit{ JHEP}
  \textbf{ 07} (2018) 161,
  \href{http://dx.doi.org/10.1007/JHEP07(2018)161}{\doi{10.1007/JHEP07(2018)161}},
  \href{http://www.arXiv.org/abs/1802.02613}{\texttt{arXiv:1802.02613}}.

\bibitem{CMS:2021xjt}
\hrefCMSnoop {}{{CMS Collaboration}, ``Precision luminosity measurement in
  proton-proton collisions at $\sqrt{s}={13\TeV}$ in 2015 and 2016 at {CMS}'',}
  \textit{ Eur. Phys. J. C} \textbf{ 81} (2021) 800,
  \href{http://dx.doi.org/10.1140/epjc/s10052-021-09538-2}{\doi{10.1140/epjc/s10052-021-09538-2}},
  \href{http://www.arXiv.org/abs/2104.01927}{\texttt{arXiv:2104.01927}}.

\bibitem{CMS:2018elu}
\href {http://cds.cern.ch/record/2621960}{{CMS Collaboration}, ``{CMS}
  luminosity measurement for the 2017 data-taking period at
  $\sqrt{s}={13\TeV}$'',} CMS Physics Analysis Summary CMS-PAS-LUM-17-004,
  2018.

\bibitem{CMS:2019jhq}
\href {http://cds.cern.ch/record/2676164}{{CMS Collaboration}, ``{CMS}
  luminosity measurement for the 2018 data-taking period at
  $\sqrt{s}={13\TeV}$'',} CMS Physics Analysis Summary CMS-PAS-LUM-18-002,
  2019.

\bibitem{Bern:2011ie}
Z.~Bern\hrefCMSnoop {}{ {et~al.}, ``Left-handed {\PW} bosons at the {LHC}'',}
  \textit{ Phys. Rev. D} \textbf{ 84} (2011) 034008,
  \href{http://dx.doi.org/10.1103/PhysRevD.84.034008}{\doi{10.1103/PhysRevD.84.034008}},
  \href{http://www.arXiv.org/abs/1103.5445}{\texttt{arXiv:1103.5445}}.

\bibitem{CMS:2015cyj}
\hrefCMSnoop {}{{CMS Collaboration}, ``Angular coefficients of {\PZ} bosons
  produced in \pp collisions at $\sqrt{s}={8\TeV}$ and decaying to {\MM} as a
  function of transverse momentum and rapidity'',} \textit{ Phys. Lett. B}
  \textbf{ 750} (2015) 154,
  \href{http://dx.doi.org/10.1016/j.physletb.2015.08.061}{\doi{10.1016/j.physletb.2015.08.061}},
  \href{http://www.arXiv.org/abs/1504.03512}{\texttt{arXiv:1504.03512}}.

\bibitem{ATLAS:2012au}
\hrefCMSnoop {}{{ATLAS Collaboration}, ``Measurement of the polarisation of
  {\PW} bosons produced with large transverse momentum in \pp collisions at
  $\sqrt{s}={7\TeV}$ with the {ATLAS} experiment'',} \textit{ Eur. Phys. J. C}
  \textbf{ 72} (2012) 2001,
  \href{http://dx.doi.org/10.1140/epjc/s10052-012-2001-6}{\doi{10.1140/epjc/s10052-012-2001-6}},
  \href{http://www.arXiv.org/abs/1203.2165}{\texttt{arXiv:1203.2165}}.

\bibitem{ATLAS:2012nhi}
\hrefCMSnoop {}{{ATLAS Collaboration}, ``Measurement of the {\PW} boson
  polarization in top quark decays with the {ATLAS} detector'',} \textit{ JHEP}
  \textbf{ 06} (2012) 088,
  \href{http://dx.doi.org/10.1007/JHEP06(2012)088}{\doi{10.1007/JHEP06(2012)088}},
  \href{http://www.arXiv.org/abs/1205.2484}{\texttt{arXiv:1205.2484}}.

\bibitem{ATLAS:2016ygv}
\hrefCMSnoop {}{{ATLAS Collaboration}, ``Measurement of the inelastic
  proton-proton cross section at $\sqrt{s}={13\TeV}$ with the {ATLAS} detector
  at the {LHC}'',} \textit{ Phys. Rev. Lett.} \textbf{ 117} (2016) 182002,
  \href{http://dx.doi.org/10.1103/PhysRevLett.117.182002}{\doi{10.1103/PhysRevLett.117.182002}},
  \href{http://www.arXiv.org/abs/1606.02625}{\texttt{arXiv:1606.02625}}.

\bibitem{CMS:2015cyp}
\hrefCMSnoop {}{{CMS Collaboration}, ``Measurement of top quark polarisation in
  $t$-channel single top quark production'',} \textit{ JHEP} \textbf{ 04}
  (2016) 073,
  \href{http://dx.doi.org/10.1007/JHEP04(2016)073}{\doi{10.1007/JHEP04(2016)073}},
  \href{http://www.arXiv.org/abs/1511.02138}{\texttt{arXiv:1511.02138}}.

\bibitem{ATLAS:2016fbc}
\hrefCMSnoop {}{{ATLAS Collaboration}, ``Measurement of the {\PW} boson
  polarisation in \ttbar events from \pp collisions at $\sqrt{s}={8\TeV}$ in
  the lepton+jets channel with {ATLAS}'',} \textit{ Eur. Phys. J. C} \textbf{
  77} (2017) 264,
  \href{http://dx.doi.org/10.1140/epjc/s10052-017-4819-4}{\doi{10.1140/epjc/s10052-017-4819-4}},
  \href{http://www.arXiv.org/abs/1612.02577}{\texttt{arXiv:1612.02577}}.
  [Erratum: \DOI{10.1140/epjc/s10052-018-6520-7}].

\bibitem{CMS:2017xrt}
\hrefCMSnoop {}{{CMS Collaboration}, ``Measurement of the \ttbar production
  cross section using events with one lepton and at least one jet in \pp
  collisions at $\sqrt{s}={13\TeV}$'',} \textit{ JHEP} \textbf{ 09} (2017) 051,
  \href{http://dx.doi.org/10.1007/JHEP09(2017)051}{\doi{10.1007/JHEP09(2017)051}},
  \href{http://www.arXiv.org/abs/1701.06228}{\texttt{arXiv:1701.06228}}.

\bibitem{CMS:DP-2020-025}
\href {https://cds.cern.ch/record/2718978}{{CMS Collaboration}, ``{\PW} and top
  tagging scale factors for {Run} 2 data'',} CMS Detector Performance Note
  CMS-DP-2020-025, 2020.

\bibitem{Junk:1999kv}
\hrefCMSnoop {}{T.~Junk, ``Confidence level computation for combining searches
  with small statistics'',} \textit{ Nucl. Instrum. Meth. A} \textbf{ 434}
  (1999) 435,
  \href{http://dx.doi.org/10.1016/S0168-9002(99)00498-2}{\doi{10.1016/S0168-9002(99)00498-2}},
  \href{http://www.arXiv.org/abs/hep-ex/9902006}{\texttt{arXiv:hep-ex/9902006}}.

\bibitem{Read:2002hq}
\hrefCMSnoop {}{A.~Read, ``Presentation of search results: the {\CLs}
  technique'',} \textit{ J. Phys. G} \textbf{ 28} (2002) 2693,
  \href{http://dx.doi.org/10.1088/0954-3899/28/10/313}{\doi{10.1088/0954-3899/28/10/313}}.

\bibitem{Cowan:2010js}
\hrefCMSnoop {}{G.~Cowan, K.~Cranmer, E.~Gross, and O.~Vitells, ``Asymptotic
  formulae for likelihood-based tests of new physics'',} \textit{ Eur. Phys. J.
  C} \textbf{ 71} (2011) 1554,
  \href{http://dx.doi.org/10.1140/epjc/s10052-011-1554-0}{\doi{10.1140/epjc/s10052-011-1554-0}},
  \href{http://www.arXiv.org/abs/1007.1727}{\texttt{arXiv:1007.1727}}.
  [Erratum: \DOI{10.1140/epjc/s10052-013-2501-z}].

\end{thebibliography}\endgroup
\cleardoublepage \appendix\section{The CMS Collaboration \label{app:collab}}\begin{sloppypar}\hyphenpenalty=5000\widowpenalty=500\clubpenalty=5000
\cmsinstitute{Yerevan Physics Institute, Yerevan, Armenia}
{\tolerance=6000
A.~Tumasyan\cmsAuthorMark{1}\cmsorcid{0009-0000-0684-6742}
\par}
\cmsinstitute{Institut f\"{u}r Hochenergiephysik, Vienna, Austria}
{\tolerance=6000
W.~Adam\cmsorcid{0000-0001-9099-4341}, J.W.~Andrejkovic, T.~Bergauer\cmsorcid{0000-0002-5786-0293}, S.~Chatterjee\cmsorcid{0000-0003-2660-0349}, K.~Damanakis\cmsorcid{0000-0001-5389-2872}, M.~Dragicevic\cmsorcid{0000-0003-1967-6783}, A.~Escalante~Del~Valle\cmsorcid{0000-0002-9702-6359}, P.S.~Hussain\cmsorcid{0000-0002-4825-5278}, M.~Jeitler\cmsAuthorMark{2}\cmsorcid{0000-0002-5141-9560}, N.~Krammer\cmsorcid{0000-0002-0548-0985}, L.~Lechner\cmsorcid{0000-0002-3065-1141}, D.~Liko\cmsorcid{0000-0002-3380-473X}, I.~Mikulec\cmsorcid{0000-0003-0385-2746}, P.~Paulitsch, F.M.~Pitters, J.~Schieck\cmsAuthorMark{2}\cmsorcid{0000-0002-1058-8093}, R.~Sch\"{o}fbeck\cmsorcid{0000-0002-2332-8784}, D.~Schwarz\cmsorcid{0000-0002-3821-7331}, S.~Templ\cmsorcid{0000-0003-3137-5692}, W.~Waltenberger\cmsorcid{0000-0002-6215-7228}, C.-E.~Wulz\cmsAuthorMark{2}\cmsorcid{0000-0001-9226-5812}
% ...
\par}
\vskip\cmsinstskip
\dag:~Deceased\\
$^{1}$Also at Yerevan State University, Yerevan, Armenia\\
$^{2}$Also at TU Wien, Vienna, Austria\\
% ...
\end{sloppypar}
\end{document}